\documentclass[structabstract]{aa}  
\usepackage{natbib}
\usepackage{graphicx}
\usepackage{subfig}
 \usepackage{txfonts}
 \usepackage{pdflscape}
 \usepackage{rotating}
 \usepackage{longtable}

\usepackage{color}
  {}

\begin{document}

\title{Colors of barlenses: evidence for connecting them to boxy/peanut bulges}
 

   \author{M.~Herrera-Endoqui\inst{1}
          \and 
          H.~Salo\inst{1}
          \and
          E.~Laurikainen\inst{1}
          \and
          J.H.~Knapen\inst{2,3}
          }

   \institute{Astronomy Research Unit, University of Oulu, FI-90014, Finland \\
              \email{martin.herreraendoqui@oulu.fi}
         \and 
Instituto de Astrof\'isica de Canarias, E-38200 La Laguna, Tenerife, Spain
\and Departamento    de Astrof\'isica, Universidad de La Laguna, E-38205 La Laguna, Tenerife, Spain
}
             
\titlerunning{Barlenses}
\authorrunning{Herrera-Endoqui et al.}

   \date{Received: accepted:}

 
  \abstract
%
{}
{ We aim to study the colors and orientations of structures in low and intermediate
  inclination barred galaxies. We test the hypothesis
  that barlenses, roundish central components embedded in bars, could
  form a part of the bar in a similar manner to boxy/peanut bulges in
  the edge-on view.}
{A sample of 79 barlens galaxies was selected from the Spitzer Survey
  of Stellar Structure in Galaxies (S$^4$G) and the Near IR S0 galaxy
  Survey (NIRS0S), based on previous morphological classifications at
  3.6 $\mu$m and 2.2 $\mu$m wavelengths.  For these galaxies the
  sizes, ellipticities, and orientations of barlenses were measured,
  parameters which were used to define the barlens regions in the
  color measurements. In particular, the orientations of
    barlenses were studied with respect to those of the ``thin bars''
    and the line-of-nodes of the disks.  For a subsample of 47 galaxies color
  index maps were constructed using the Sloan Digital Sky Survey (SDSS)
  images in five optical bands, {\it u}, {\it g}, {\it r}, {\it i},
  and {\it z}.  Colors of bars, barlenses, disks, and central regions
  of the galaxies were measured using two different approaches and
  color-color diagrams sensitive to metallicity, stellar surface
  gravity, and short lived stars were constructed. Color differences
  between the structure components were also calculated for each
  individual galaxy, and presented in histogram form.  }
{We find that the colors of barlenses are very similar to those of the
  surrounding bars, indicating that most probably they form part
   of the bar. We also find that barlenses have orientations closer to the
   disk line-of-nodes than to the thin bars, which is
   consistent with the idea that they are vertically thick, in a
   similar manner as the boxy/peanut structures in more inclined
   galaxies. Typically, the colors of barlenses are also similar
  to those of normal E/S0 galaxies.  Galaxy by galaxy studies also
  show that in spiral galaxies very dusty barlenses also exist,
  along with barlenses with rejuvenated stellar populations. The
  central regions of galaxies are found to be on average redder than
  bars or barlenses, although galaxies with bluer
  central peaks also exist. }  {}
%
%
\maketitle
%

\section{Introduction}


\begin{figure*}
\begin{centering}
\includegraphics[scale=.7]{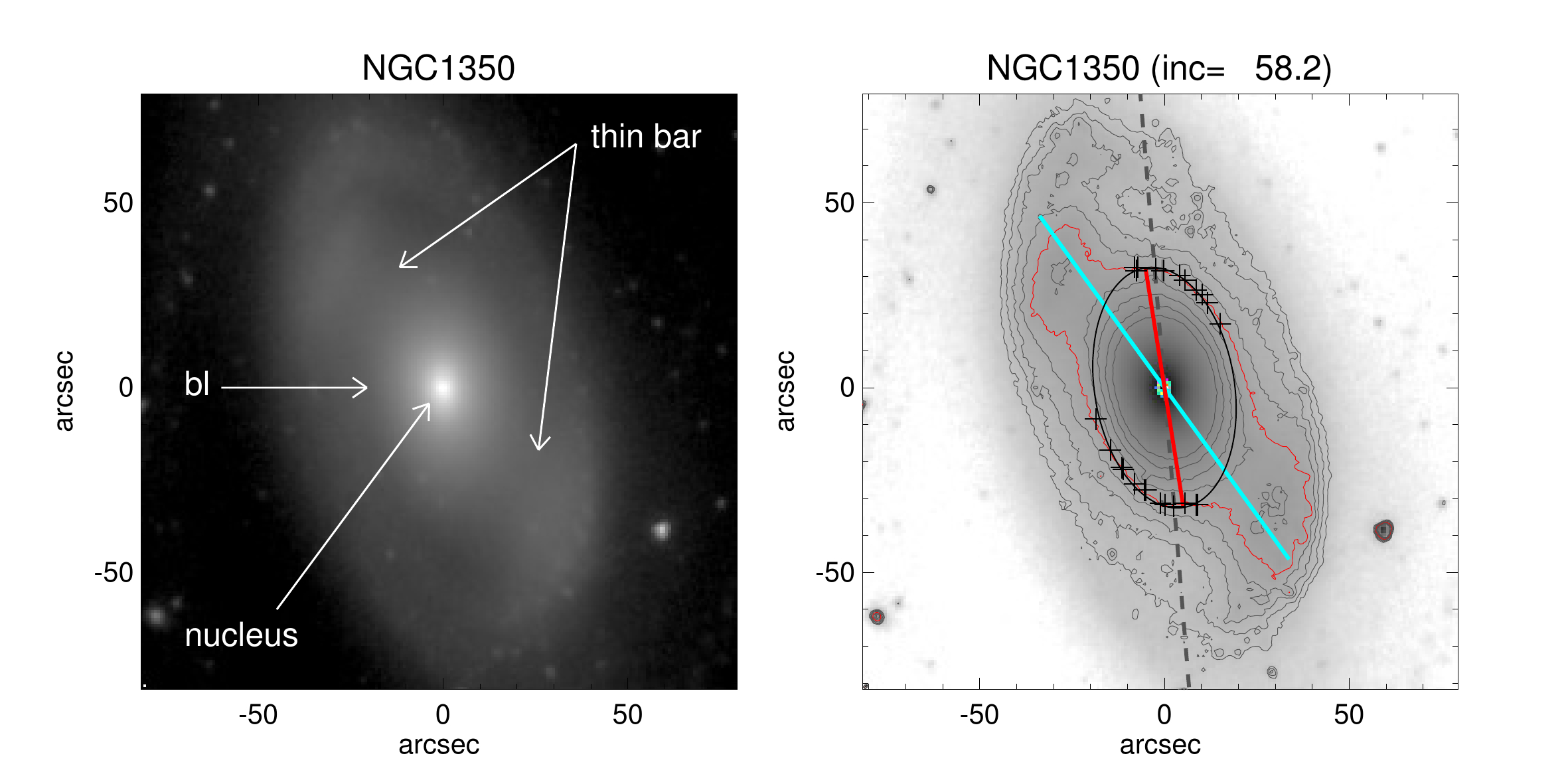}
 \caption{Example barlens galaxy, NGC 1350. {Left panel}
   shows the bar region of the 3.6 $\mu$m image. Marked are the
   structure components measured in this study: 
   ``bl'' refers to barlens, ``thin bar'' is the bar region
   outside the barlens, and ``nucleus'' refers to possible small bulge
   in the central regions. {Right panel} shows the contours of the
   same image. Lines indicating the orientations of
   the bar (blue line), barlens (red line), and the disk line-of-nodes
   (dashed line) are over-plotted.  Black crosses show the points selected to measure the
   size and orientation of the barlens by fitting an ellipse to these points. }
\label{bl_example_1350}
\end{centering}
\end{figure*}

Galactic bulges is an active research topic, not least because of the
recent discoveries concerning the Milky Way bulge, which is now
known to form part of the Milky Way bar \citep{weiland1994}. 
Photometrically bulges are defined as flux above the
underlying disk, and observers often divide them into ``classical
bulges'', which are highly relaxed, velocity dispersion supported
spheroidal components, and rotationally supported pseudo-bulges, which
originate from the disk material \citep{kormendy1983}. The formation of a
pseudo-bulge is associated to the evolution of a bar, which can
vertically thicken the central bar regions, for example by bar
buckling effects \citep{combes1990,raha1991,pfenniger1991},
leading to boxy- or peanut-shaped bulges \citep[see][]{atha2005}.
Bars can also trigger gas inflows, which, via star formation,
can collect stars to the central regions of the galaxies, manifested
as central disks (i.e., ``disky pseudo-bulges'').  Classical and
pseudo-bulges are often distinguished via low S\'ersic indices
($n\sim$ 1 means a pseudo-bulge) and low bulge-to-total flux ratios,
but this approach has shown out to be oversimplified 
\citep[see review by][]{kormendy2016}.
More sophisticated multi-parameter methods, including
kinematics, stellar populations, and metallicities, have been
developed, of which an update is given by \citet{fisher2016}.
However, all these methods have been developed for separating
classical bulges from inner disks (i.e., disky pseudo-bulges), without
paying attention to the vertically thick inner bar components, which
might be visible at a large range of galaxy inclinations.

In fact, there has been a long-standing ambiguity in interpreting
bulges at low and high galaxy inclinations in the nearby
Universe. While in the edge-on view many bulges have boxy or peanut
shape isophotes and are thereby associated with bars \citep{lutticke2000a,bureau2006},
in less inclined galaxies the central
flux concentrations in a large majority of galaxies have been
interpreted in terms of classical bulges 
\citep[see the discussion by][]{laurisalo2016}.
Isophotal analysis \citep{atha2006,erwin2013} have shown that boxy/peanut
bulges, based on their isophotal orientation with respect to the bar
and the galaxy line-of-nodes, can be identified even at fairly low
galaxy inclinations.  In a few galaxies boxy/peanut bulges have been
identified also kinematically \citep{mendezabreu2008,mendezabreu2014} in
almost face-on view, but such observations are difficult to make.

As a solution to the puzzle of how boxy/peanut bulges appear in nearly
face-on galaxies, \citet{lauri2014} suggested that they manifest as
barlenses, that is, central lens-like morphologies embedded in bars. Based
on multi-component structural decompositions they showed that among
the Milky Way mass galaxies such vertically thick bar components might
contain most of the bulge mass in the local Universe.  Theoretical
evidence for the idea that barlenses and boxy/peanut bulges might be
one and the same component was given by \citet{atha2015}, where
detailed comparisons of simulation models with such properties as size
and ellipticity of a barlens were made. If the above interpretation is
correct, we would expect barlenses at intermediate galaxy inclinations
to appear in isophotal analysis in a similar manner as the boxy/peanut
structures. 
{Moreover, the colors of barlenses can be compared with those of  
the bars outside the barlens regions in order to look for similarities. 
However, the colors of barlenses have not been 
studied yet, and also, there are no theoretical models in which the colors
were directly predicted.  Our hypothesis in this study is that if
barlenses (in a similar manner as boxy/peanut bulges) are formed of
the same material as the rest of the bar, their colors should be
similar to those of the rest of the bar.}

Barlenses were recognized and coded into morphological classification
by \citet{lauri2011} for the Near IR S0 galaxy Survey
(NIRS0S), and by \citet{buta2015} at 3.6 $\mu$m for the
{\it Spitzer} Survey of Stellar Structure in Galaxies 
\citep[S$^4$G;][]{sheth2010}.
Nuclear bars, rings and lenses are embedded in
barlenses in many galaxies, but their sizes are always smaller than those of barlenses.
An example of a barlens galaxy, NGC 1350, is shown in Fig. 1 (left
panel), where the different structure components in the bar region are
illustrated. The bar region outside the barlens is called the ``thin bar''
(or simply bar), considering that it is assumed to be
vertically thin.  Depending on the level of gas accretion and the
subsequent star formation histories of barlens galaxies, the central
components can have variable colors.  However, any classical bulges
that might appear in these galaxies are expected to be dominated by
old stellar populations.  The spheroidals are dynamically hot and
therefore any accretion of external gas into these components would be
difficult.





\begin{figure*}
\begin{centering}
\includegraphics[scale=.7]{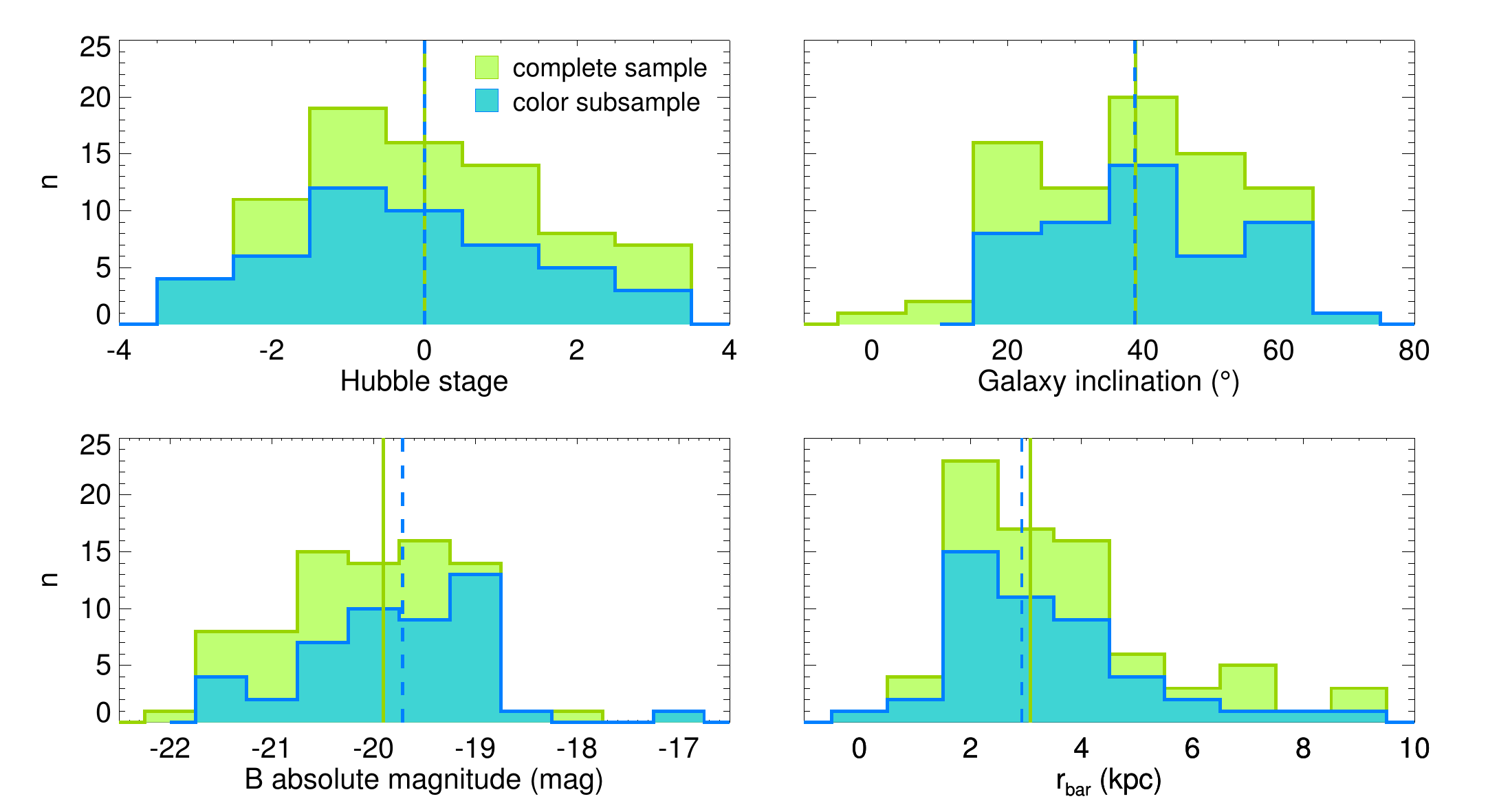}
\caption{Distributions of morphological type, inclination, absolute B magnitude, and bar size, 
 for the whole sample of barlens galaxies (green histogram), and for the subsample of galaxies with
 SDSS imaging (blue histogram). The green solid and blue dashed vertical lines show the median 
 values of the distributions for the complete and color subsample, respectively.}
\label{sample_distros}
\end{centering}
\end{figure*}

\section{Sample and data}

\subsection{Sample}

Our sample consists of a set of barlens galaxies from two
near-infrared galaxy surveys.  The {\it Spitzer} Survey of Stellar
Structure in Galaxies \citep[][hereafter S$^4$G]{sheth2010} which
contains 2352 galaxies observed at 3.6 and 4.5 $\mu$m with the
Infrared Array Camera \citep[IRAC;][]{fazio2004} on board the {\it
  Spitzer} Space Telescope. This survey is limited in volume (distance
d $<$ 40 Mpc, Galactic latitude $|$b$|$ $>$ 30 deg), magnitude
(B$_{corr}$ $<$ 15.5 mag, corrected for internal extinction) and size
(D$_{25}$ $>$ 1 arcmin). It covers all Hubble types and disk
inclinations. From this survey we selected all galaxies with barlenses
in the classification by \citet{buta2015}. {In that study no
  inclination limit was used, however the barlens galaxies have in 
  general inclination $\lesssim$ 60$^{\circ}$. Only in one case the inclination
  is $\sim$ 71$^{\circ}$}.  The other survey is the
Near-Infrared S0 Galaxy survey \citep[hereafter NIRS0S;][]{lauri2011},
which consists of $\sim$ 200 early-type disk galaxies observed in the
K$_{\rm s}$-band, {with total magnitude B$_{\rm T} \leq$ 12.5 mag. The inclination limit of this sample is
  65$^{\circ}$.}  Altogether, S$^4$G and NIRS0S contain 79 barlens
galaxies. If not otherwise mentioned that forms our sample in all the
following. This sample will be used to study the orientation of
barlenses with respect to those of the bar and the disk
line-of-nodes. Of these galaxies 22 appear both in NIRS0S and S$^4$G,
and NIRS0S has 11 barlens galaxies which do not form part of S$^4$G.
The barlens galaxy sample is the same as used by \citet{lauri2014}.

\subsection{Infrared data}

The 3.6 $\mu$m images from S$^4$G typically reach the surface
brightnesses of 26.5 (AB) mag arcsec$^{-2}$, which translates to
roughly 28 mag arcsec$^{-2}$ in the B-band. These images have a pixel 
resolution of 0$.\!\!^{\prime\prime}$75 and full-width half-maximum (FWHM) $\approx 2$ \arcsec. 
The NIRS0S images at 2.2
$\mu$m typically reach B-band surface brightnesses of 27 mag
arcsec$^{-2}$, with pixel resolution in the range 0$.\!\!^{\prime\prime}$23 $-$ 0$.\!\!^{\prime\prime}$61
(typically 0$.\!\!^{\prime\prime}$25). 
When 3.6 $\mu$m images exist they have a priority in
our analysis.  The complete list of barlens galaxies appear in Table
1 (see Appendix C): shown are the morphological types, orientations (PA$_{\rm disk}$)
and inclinations (i$_{\rm disk}$) of the disks.
For S$^4$G galaxies the
morphological types are from \citet{buta2015}, 
the orientations and inclinations of the disks are from \citet{salo2015},
and the parameters of bars from \citet{herrera2015}. 
For NIRS0S galaxies the morphological types and bar and disk parameters are taken from the 
original paper \citep{lauri2011}. 
In these references the bar parameters were obtained 
by both visual and ellipse fitting methods, however, only the visual length 
(r$_{\rm bar}$) and orientation (PA$_{\rm bar}$) are used here. 
The disk parameters were obtained using isophotal 
ellipse fits, from which the ellipticity and orientations of the outer isophotes 
of the disks are derived.
The distances are taken from archive data in the respective survey
\citep[in the case of S$^4$G galaxies see][]{munoz2015}.
These are redshift independent 
distances obtained from NED\footnote{The NASA/IPAC Extragalactic Database (NED) is
operated by the Jet Propulsion Laboratory, California Institute of Technology, under contract
with the National Aeronautics and Space Administration.} 
when available and otherwise calculated assuming $H_0$ = 71 km s$^{-1}$ Mpc$^{-1}$.

\subsection{Optical data}

In order to obtain the optical colors of the structure components we
use the {\it Sloan} Digital Sky Survey \citep[][hereafter SDSS]{york2000,gunn1998} 
images in five bands: {\it u, g, r, i,} and {\it z}. The reduced
images are taken from \citet{knapen2014} who used images from SDSS
DR7 and SDSS-III DR8 produced from the SDSS photometric pipeline, but
using their own sky subtraction. These mosaics are already
sky-subtracted and the different bands are aligned with each other. Such mosaiced images exist
for all those S$^4$G galaxies for which SDSS data is available, including
47 barlens galaxies. This set of galaxies will be referred to as the ``color 
subsample'' in the rest of the paper. The pixel scale in the SDSS images is 0$.\!\!^{\prime\prime}$396.
These images can be used to derive surface brightness profiles
down to $\mu{\rm _r} \sim$ 27 mag arcsec$^{-2}$ \citep{pohlen2006},
equivalent to 27.5 mag arcsec$^{-2}$ in B-band.\\

The basic properties, that is, the Hubble
stages, galaxy inclinations, optical galaxy magnitudes, and bar sizes,
of the complete sample and the color subsample are shown in Fig. 2. The Hubble stage distribution is
previously shown also by \citet{lauri2014} for the complete
sample. 
The vertical lines represent the median values of the distributions.
For the complete sample these values are Hubble stage T = 0, inclination inc = 38.9$^\circ$, 
B absolute magnitude is -19.9 mag, 
and r$_{\rm bar}$ = 3.1 kpc. In the case of the color subsample 
the values are T = 0, inc = 38.8$^\circ$, B absolute magnitude is -19.7 mag, 
and r$_{\rm bar}$ = 2.9 kpc.
We note that the distributions and median properties of the color subsample are
very similar to those of the complete sample.


\begin{figure*}
\begin{centering}
\includegraphics[scale=.5]{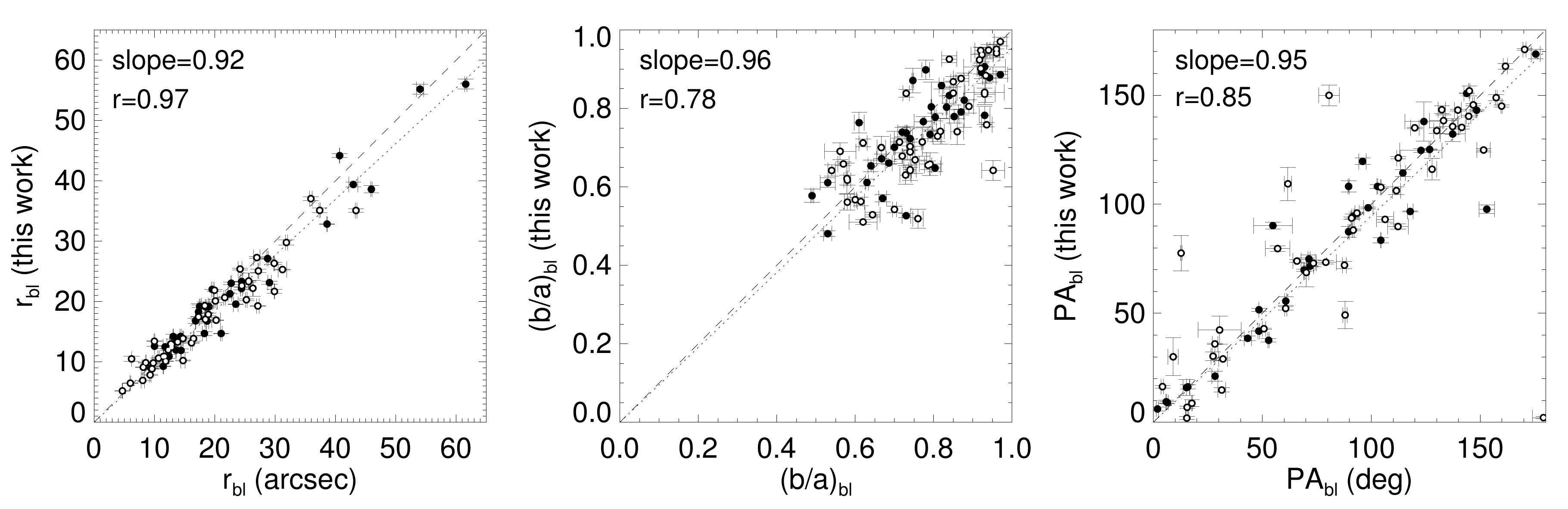}
 \caption{Sizes (r$_{\rm bl}$), axial ratios ((b/a)$_{\rm bl}$) and
   position angles (PA$_{\rm bl}$) of barlenses are compared with those 
  previosuly obtained by \citet{herrera2015} or  \citet{lauri2011}. Dotted
   lines indicate linear fits between the old and new measurements,
   whereas the dashed line stands for a unit slope. The labels indicate the slope
   and correlation coefficient of the linear fit.
   The galaxies that belong to the color subsample are indicated with an open symbol.}
\label{bl_compare_old_new}
\end{centering}
\end{figure*}

\section{Measuring the sizes and orientations of barlenses}
 
\subsection{Sizes}
Apparent sizes and shapes of all the barlenses in our sample 
were measured using the infrared images, fitting ellipses 
to their outer isophotes. 
For the galaxies in our sample the sizes have been previously measured
by \citet{herrera2015} or \citet{lauri2011}, but using a more visual
approach. The difference between the two approaches is that while in
the previous measurements the edges of barlenses were visually
estimated at a certain magnitude scale, in this study they were defined
following the outermost isophotes of barlenses indicated by their 
surface brightnesses (see Table 1).  Points were marked on
top of the contour, but avoiding the zone dominated by the bar major
axis flux.  Then an ellipse was automatically fit to those points,
which gives the size (r$_{\rm bl}$), minor-to-major axis ratio
((b/a)$_{\rm bl}$), and orientation (PA$_{\rm bl}$) of the barlens.
An example of our measurements is shown in Fig. 1 (right
panel), and the derived barlens parameters are compared with those of
the previous measurements in Fig. 3. 
In Table 1 we give the measurements of the barlens parameters
and their uncertainties,
which were calculated from the set of three measurements (unc=stdev/$\surd$N,
where N=3 is the number of measurements). The uncertainties in the orientation 
parameters of the disks were taken from the respective papers. 
In the case of bars, the uncertainties were calculated
in the same manner described above but with N=2, 
using the parameters obtained with the visual and ellipse fitting methods
in the respective catalogs.
Fig. 3 indicates that there are no significant systematic difference
between the two measurements.  In the position angles and b/a-values
the scatter easily appears in measurement of structures having low
ellipticities, which is the case also with many barlenses. From hereon
we use the isophotal measurements obtained in the current study, which
were used also to define the zones in our color index measurements (see
Sect. 4.2).

\begin{figure}
\begin{centering}
\includegraphics[scale=.7]{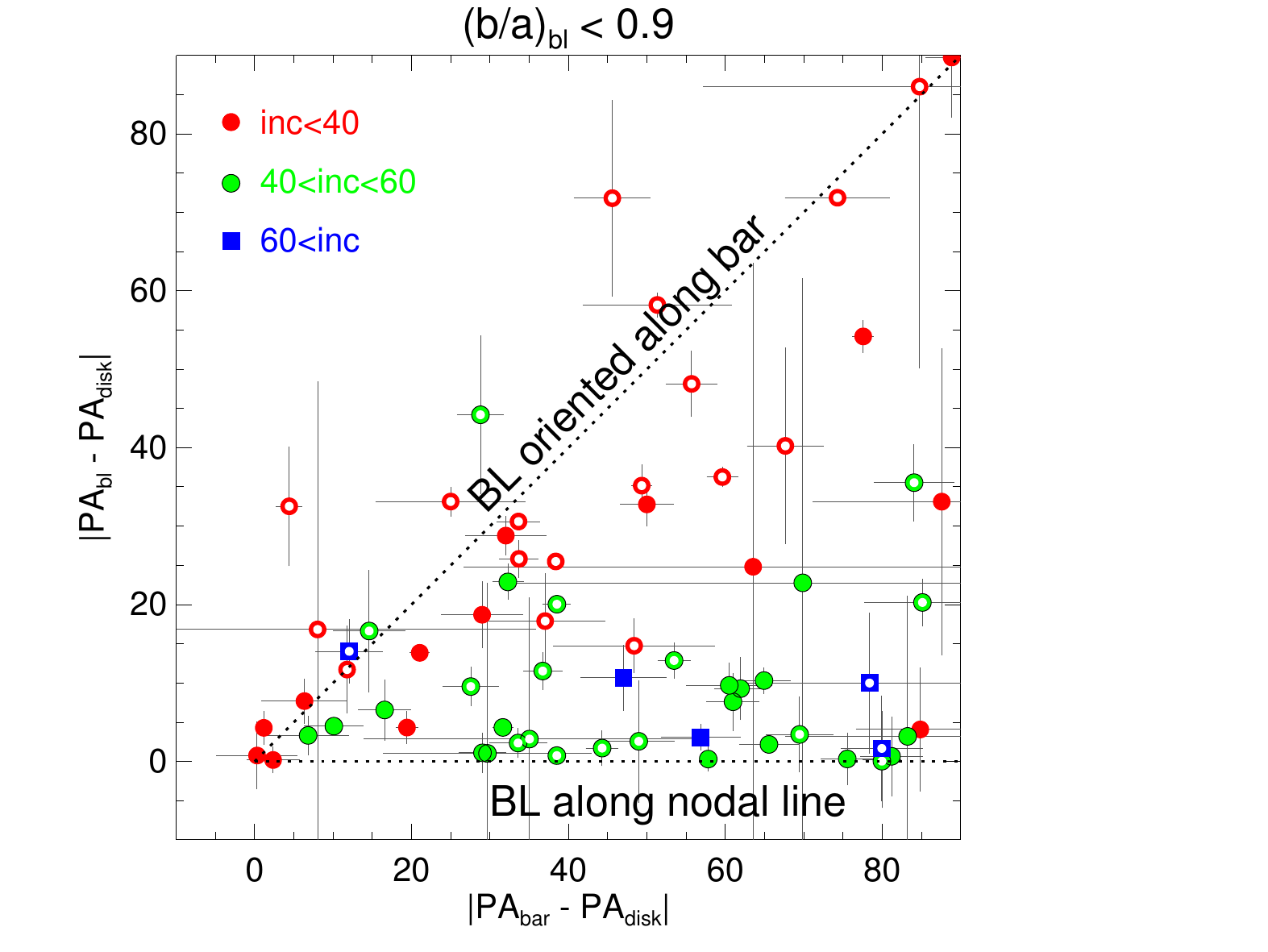}
\caption{Orientations of barlenses compared with the
orientations of bars and those of the lines-of-nodes (disks). The y-axis 
is the difference between the orientations of the barlens
and the disk, whereas the x-axis is the difference between
the orientations of the bar and the disk. The differences are given 
in degrees. Shown are only galaxies where the barlens axial ratio (b/a)$_{\rm bl}<$ 0.9.
The galaxies that belong to the color subsample are indicated with an open symbol.}
\label{bl_bar_pa}
\end{centering}
\end{figure}

\begin{figure*}
\begin{centering}
\includegraphics[trim = 0mm 8mm 0mm 23mm, clip, scale=.95]{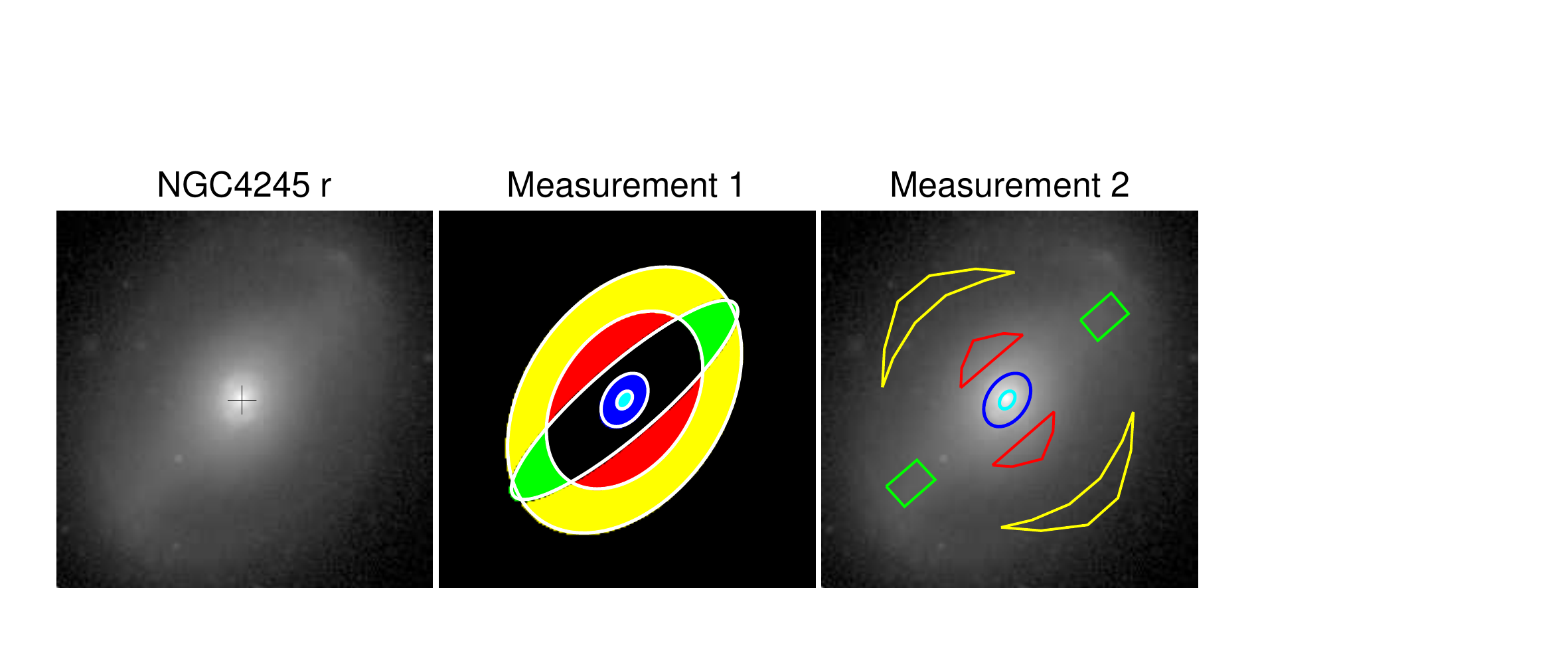}
 \caption{Illustration of the two methods used for defining the
   regions for the measurements of the component colors. In the left panel the {\it r}-band
   image of NGC 4245 is shown. In the middle panel we show the automatically defined
   regions for the central region (``nuc1'' and ``nuc2''; light and dark
   blue, respectively), for the barlens (``blc''; red), the bar (green), and
   the disk (yellow), see the text for details. In the right panel the manually
   selected regions for the barlens, the bar and the disk are shown (central regions are
   the same as in the middle panel).  All images are in the same scale.}
\label{bl_methods}
\end{centering}
\end{figure*}

\subsection{Orientations}

The orientations of barlenses (PA$_{\rm bl}$) are compared with the
orientations of bars (PA$_{\rm bar}$) and the lines-of-nodes
(PA$_{\rm disk}$) in Fig. 4. All the orientations were measured in the sky-plane
counter-clockwise from the north.  The main idea for making this
comparison is the following \citep[see also][]{erwin2013}: if
barlenses indeed are vertically thick components in a similar manner
as boxy/peanut/X-shaped structures of bars (the strongest boxy/peanuts
have X-shape morphology), we would expect that to be manifested 
in their orientations. Namely, it has been demonstrated by 
simulation models \citep{bettoni1994,atha2006}
that when the vertically thick boxy/peanut, embedded in a thin
bar, is seen in the non-edge-on view, due to an inclination effect
the two bar components would have slightly different orientations: the
orientation of the boxy/peanut would always be closer to the
line-of-nodes than the orientation of the thin bar would be.

In Fig. 4 the galaxies are arbitrarily divided into three galaxy 
inclination bins (inc $<$ 40$^\circ$, 40$^\circ$ $<$ inc $<$ 60$^\circ$ and 
inc $>$ 60$^\circ$) in order to observe the behavior of the 
barlens orientation with respect to galaxy inclination. 
The x-axis shows the orientation difference between the
bar and the disk, and the y-axis how much the major-axis barlens
orientation deviates from that of the disk.  In the diagonal the
barlens orientations are along the bar major-axis, and in the horizontal
line along the line-of-nodes.  Taking into account the measurement
uncertainties shown in Fig. 3, it appears that barlenses are
oriented between the diagonal and the horizontal line, which is
consistent with the idea that they are vertically more extended than
the thin bar component. At the lowest galaxy inclinations
(inc $<$ 40$^{\circ}$) the
barlens orientation is more closely along the thin bar, following
from their small intrinsic elongation along the thin bar major axis.
For higher inclination galaxies (inc $>$ 60$^{\circ}$)
both the thin bar and the barlens
are oriented closer to the line-of-nodes.  Contrary to barlenses, none
of the boxy/peanut bulges studied by \citet{erwin2013}
showed an alignment with the thin bar.  This is expected since the
galaxies they studied appeared at galaxy inclinations of
inc = 45$^{\circ}$ $-$ 65$^{\circ}$, where the orientations are expected to
fall between the line-of-nodes and the orientation of the thin bar.


\section{Obtaining the colors using the SDSS images}

\subsection{Making the color images}

Color maps were constructed in {\it (u-g), (g-r), (r-i)}, and
{\it (i-z)}-indices for the color subsample. 
These indices were later used to make color-color diagrams indicative
of the different stellar populations and metallicities. 
The images were rotated so that the bar major axis, determined from
the 3.6 or 2.2 $\mu$m images, became horizontal. We trust the available sky
subtractions, because the original uncertainties that
appeared during the SDSS data release were checked by \citet{knapen2014}.

However, before making the color images we needed to be sure that the
two images used to make them have the same resolution.  For
that purpose the Gaussian FWHMs were measured using the IRAF\footnote{IRAF
is distributed by the National Optical Astronomy Observatories,
which are operated by the Association of Universities for Research
in Astronomy, Inc., under cooperative agreement with the National
Science Foundation.} 
task ``imexamine'', applied to several field stars in each band for every
galaxy. The mean values for the FWHMs are 3.28, 3.07, 2.76, 2.60 and
2.68 pixels in {\it u, g, r, i} and {\it z} bands, respectively. 
These values are equivalent to 1$.\!\!^{\prime\prime}$30, 1$.\!\!^{\prime\prime}$22,
1$.\!\!^{\prime\prime}$09, 1$.\!\!^{\prime\prime}$03, and 1$.\!\!^{\prime\prime}$06,
respectively.
As discussed by \citet{bergvall2010}
non-Gaussian extended tails at very low surface
brightnesses appear, in particular in the {\it i}-band.  However, although
these tails might be important at low surface brightnesses, such
effects are not important in our study where only high surface
brightness structures are studied.  For each color image the higher
resolution image was convolved with a Gaussian point spread function (PSF) so that the
convolved image had the same FWHM as the lower resolution image. The
convolved images were then converted to magnitude-units and corrected
for Galactic extinction using the re-calibrated \citet{schlegel1998} 
extinctions in the infrared, made by \citet{schlafly2011},
assuming a reddening law with R$_{\rm V}$ = 3.1.

\subsection{Measuring the colors of the structure components}

The convolved images were used to measure the colors of the different
structure components. Instead of using directly the color index maps
the fluxes in each filter were extracted for the regions defining the
structure components. The fluxes in the two bands were then
converted to colors applying also the wavelength specific Galactic extinction
correction. The colors of the structure components were then measured
with two different methods:
\vspace{0.25cm}

\noindent Approach 1 (see Fig. 5, middle panel):
\vspace{0.2cm}


\noindent  Colors of the two bar components, the disks, and the central
peaks, were measured using an automatic approach defining the
measured regions in the following manner:
\vspace{0.1cm}

\noindent {\it nuc1 (central region, light blue color):} an elliptical
region around the galaxy center was selected having 
position angle and axial ratio of
as those of the barlens, and outer radius r = 0.1 $\times$ r$_{\rm bl}$.
The region selected this way is
smaller than nuclear rings 
typically are, thus avoiding color contamination from them. 
Given this definition of nuc1, in 16 galaxies the nuc1 region is smaller
than the FWHM of the SDSS image (1.3\arcsec). Our central regions
defined by nuc1 go from unresolved to barely resolved nuclei, to much
larger regions. 
In principle this central region could be associated to
a small classical bulge, but as the colors alone cannot be used to
define bulges, we avoid using that concept.

\vspace{0.1cm}

\noindent {\it nuc2 (central region, dark blue color):} a similar
elliptical region around the galaxy center as above was selected, but
using a larger radius, r = 0.3 $\times$ r$_{\rm bl}$.  This radius was
large enough to cover possible nuclear rings.
\vspace{0.1cm}

\noindent {\it bl (barlens):} barlens region was taken to be an
elliptical zone inside the measured barlens radius. The shape and
orientation of this region was defined by the measured b/a and
the position angle of the barlens.  However, the region covering the
central region (nuc2) was excluded.
\vspace{0.1cm}

\noindent {\it blc (barlens, red color):} this barlens region was
otherwise the same as that of bl above, except that 
the region covering the thin bar (i.e., bar) inside the barlens
radius was excluded. The advantage to the previous barlens measurement is that this
approach avoids contamination of flux along the bar major axis in the
barlens region.
\vspace{0.1cm}

\noindent {\it bar (i.e., thin bar, green color):} was taken to be an
elliptical region inside r = 0.9 $\times$ r$_{\rm bar}$, and outside the
barlens. 
This way we avoid mixing the colors of inner rings with those
representing the bars themselves: bars are often surrounded by star forming
rings, superimposed with the bars. 
We used the measured position angle of the bar, and the axial
ratio b/a = 0.5 $\times$ (b/a)$_{\rm bar}$. However, as a lower limit
to b/a we used b/a = 0.2. For the bar this modified value of b/a was
used, because the value that comes out directly from ellipse fitting
overestimates the b/a of the thin bar component.  This is because
the elliptical isophotal shapes are contaminated by the fairly round
barlenses sitting inside the bars.
\vspace{0.1cm}

\noindent {\it disk (yellow color):} for the disk we used an elliptical
region inside the radius r = 0.9 $\times$ r$_{\rm bar}$, using the
ellipticity and position angle of the barlens. The regions
covering the thin bar (green) and everything inside the barlens
radius were excluded. The full bar radius was not used because we wanted to
avoid possible contamination of inner rings, which often show recent
star formation.
\vspace{0.25cm}

\noindent Approach 2 (see Fig. 5, right panel):
\vspace{0.2cm}

\noindent Specific regions were visually selected and drawn on top of
the {\it r}-band images. For a barlens two zones at opposite sides of the
bar major axis were selected (red color), avoiding the region of the
thin bar. The thin bar (green color) was measured in two zones
outside the barlens, and avoiding possible star forming inner rings
superimposed with the bar.  For the disk (yellow color) two zones were
selected approximately at the same radial distances where the edges
of the thin bar appear. 


\setcounter{table}{1}
\begin{table}
\scriptsize
\caption{Color index measurements for the structures of the galaxies in the 
color subsample.}
\label{}
\centering

\begin{tabular}{lcccc}
\hline 
\noalign{\smallskip}
            &    {\it (u-g)}  &   {\it (g-r)}  &   {\it (r-i)}  &   {\it (i-z)} \\
\noalign{\smallskip}
\hline  \\
\noalign{\smallskip}
Approach 1:  &     &   &   &   \\

\noalign{\smallskip}

nuc1:   &     &   &   &   \\
 N\tablefootmark{\dagger}      &         45 &     45 &     45 &     45 \\
mean     &      1.772 &  0.793 &  0.377 &  0.353 \\
median   &      1.845 &  0.811 &  0.397 &  0.313 \\
min      &      0.920 & -0.197 & -0.114 &  0.133 \\
max      &      2.769 &  1.364 &  0.689 &  1.093 \\
std/$\surd$N  &  0.047 &  0.029 &  0.017 &  0.024 \\
\\
nuc2:   &     &   &   &   \\
mean     &      1.718 &  0.783 &  0.384 &  0.319 \\
median   &      1.780 &  0.782 &  0.386 &  0.299 \\
min      &      0.950 &  0.427 &  0.247 &  0.137 \\
max      &      2.589 &  1.292 &  0.651 &  0.633 \\
std/$\surd$N &   0.042 &  0.017 &  0.009 &  0.012 \\
\\
bl:   &     &   &   &   \\
mean     &      1.708 &  0.760 &  0.395 &  0.271 \\
median   &      1.722 &  0.758 &  0.392 &  0.262 \\
min      &      1.429 &  0.660 &  0.323 &  0.153 \\
max      &      2.217 &  1.044 &  0.551 &  0.617 \\
std/$\surd$N &   0.021 &  0.009 &  0.005 &  0.009 \\
\\
blc:   &     &   &   &   \\
mean     &      1.683 &  0.745 &  0.396 &  0.256 \\
median   &      1.695 &  0.750 &  0.398 &  0.248 \\
min      &      1.341 &  0.638 &  0.327 &  0.151 \\
max      &      2.073 &  0.921 &  0.492 &  0.606 \\
std/$\surd$N  &  0.018 &  0.007 &  0.005 &  0.009 \\
\\
bar:   &     &   &   &   \\
mean     &      1.690 &  0.742 &  0.392 &  0.262 \\
median   &      1.707 &  0.753 &  0.396 &  0.260 \\
min      &      1.136 &  0.542 &  0.270 &  0.148 \\
max      &      2.245 &  1.051 &  0.560 &  0.614 \\
std/$\surd$N  &  0.025 &  0.012 &  0.007 &  0.009 \\
\\
disk:   &     &   &   &   \\
mean     &      1.622 &  0.718 &  0.399 &  0.229 \\
median   &      1.653 &  0.727 &  0.398 &  0.224 \\
min      &      1.183 &  0.564 &  0.293 &  0.145 \\
max      &      1.986 &  0.881 &  0.498 &  0.589 \\
std/$\surd$N  &  0.021 &  0.009 &  0.006 &  0.010 \\
\noalign{\smallskip}
\hline 
\noalign{\smallskip}

Approach 2:  &     &   &   &   \\

\noalign{\smallskip}
bl: &&&&\\
N \tablefootmark{\dagger}       &            45  &    45  &    45   &   45\\
mean     &      1.685 &  0.752 &  0.392 &  0.261\\
median   &      1.718 &  0.749 &  0.394 &  0.248\\
min      &      1.211 &  0.622 &  0.306 &  0.162\\
max      &      2.074 &  0.896 &  0.473 &  0.613\\
std/$\surd$N  &  0.022 &  0.008 &  0.005 &  0.009\\
\\
bar: &&&&\\
mean      &     1.685 &  0.734 &  0.386 &  0.261\\
median    &     1.700 &  0.738 &  0.390 &  0.255\\
min       &     0.991 &  0.465 &  0.205 &  0.133\\
max       &     2.267 &  1.043 &  0.561 &  0.599\\
std/$\surd$N  &  0.028 &  0.013 &  0.007 &  0.009\\ 
\\
disk: &&&&\\
mean      &     1.532 &  0.686 &  0.390 &  0.214\\
median    &     1.550 &  0.702 &  0.398 &  0.215\\
min       &     0.966 &  0.478 &  0.288 &  0.098\\
max       &     1.810 &  0.843 &  0.484 &  0.592\\
std/$\surd$N  &  0.027 &  0.011 &  0.007 &  0.011\\
\noalign{\smallskip}
\hline 
\end{tabular} 
\tablefoot{
\tablefoottext{\dagger}{Data for only 45 of the 47 galaxies in our
color subsample are presented because NGC 3384 and NGC 4448 are not used in the 
analysis (see text).}
}
\end{table}

\vspace{0.2cm} 

The measurements using the two approaches are given in Table 2, where
the mean and median, and minimum and maximum colors, for each
structure component are shown. Also given are the standard error of
the mean, obtained by dividing the sample standard deviation with
$\surd N$.  The advantage of approach 1 is that the
measurements can be carried out automatically. It also gives a robust
manner to include a maximum number of pixels for obtaining the colors
of barlenses and thin bars.  Approach 2 is more human
guided and makes possible to avoid any obviously visible and unwanted
star forming regions that might contaminate the color.  In spite of
these differences the obtained colors in the two measurements are
fairly similar.
In Table 2 we present data for only 45 of the 47
galaxies in the color subsample. The optical images of NGC 3384 contain 
artifacts perhaps produced at the mosaicking stage, thus affecting the measurement
of the colors. This can be observed as fringes in the resulting color maps
(see Appendix A). The case of NGC 4448 is an unfortunate combination of 
high inclination (inc = 71.2$^\circ$) and an almost end-on view of the bar, which does not allow an estimation 
of the colors of the thin bar using our methods. These two galaxies were
excluded from our quantitative analysis, nevertheless, their color maps and color
profiles are shown in Appendices A and B, respectively.

Any differences that appear between these two measurements can be used
as uncertainties of our measurements.  These differences are
  shown in Table 3.  For barlenses (bl in approach 1) the color
differences in {\it (g-r}), {\it (r-i)} and {\it (i-z)} between the
two measurements are less than or similar to 0.01 magnitudes,
and in {\it (u-g)} similar to 0.02 magnitudes.  This is the case
both for the mean and median values. For the thin bars the color
differences between the two measurements are less than or similar
  to 0.01 magnitudes in all the colors, both for the mean and
median values. For the disks the mean and median colors are similar
within 0.01 magnitudes in  {\it (r-i)} and {\it (i-z)}, and within
0.03 magnitudes in {\it (g-r)}. 
Their {\it (u-g)} color is on average 0.10 magnitudes
bluer in approach 2. Small differences in {\it (u-g)} color in the disk easily
appear, because it is very sensitive to recent star formation and
dust.

Besides the measurement-related uncertainty estimated from the
  difference of the two approaches, we need to consider the photometric
  uncertainty of the colors and the variance caused by the finite size number
  of the galaxies in our sample. Since we are dealing with the colors
  of the bright central parts of galaxies, the uncertainty of colors
  is dominated by the uncertainty of the magnitude calibration
  zeropoints. According to \citet{knapen2014} the photometric
  calibration of SDSS DR7 is reliable on the level of 2$\%$. Assuming
  that the zeropoint errors in different filters are independent
  indicates an uncertainty of 0.03 mag for the color of an individual
  galaxy. For a sample of 47 galaxies the uncertainty of the mean colors
  is $0.03/\sqrt{47} \approx 0.004$ mags, thus smaller than the
  typical mean difference (0.01 mag) between the two measurement
  approaches.  It is also much smaller than the standard error of the
  mean, calculated from the sample variance and the number of galaxies
  in the sample (Table 2): the largest error of the mean is 0.047 mags,
  for the {\it (u-g)} color of the nuc1 measurements.  We may thus conclude
  that any difference in the mean colors of the components at the
  level or exceeding about 0.05 mag is likely to be real.

We are aware that the color of the underlying disk can affect the
color of the bar.  What is sometimes done is to remove the
contribution of the disk to the other structure components by assuming
that the disk is exponential, which disk is then extrapolated into the
inner parts of the galaxies in each wavelength \citep[see for
  example][]{balcells1994}.  However, in case of barred galaxies that
kind of strategy is not very reliable, because the disk under the bar
is not necessarily exponential. Also, the absolute colors of measured
structure components are not critical in this study, because we are
particularly interested on the relative color differences of those
components.







\section{Analysis of the colors}

\subsection{Mean colors}

\subsubsection{Bars and the central peaks}

The optical bands {\it u, g, r, i}, and {\it z}, as given in the
  Sloan Digital Sky Survey photometric system, cover the wavelength
  range of 3000$-$11500 $\AA$ \citep[see][]{fukugita1996}, the
  effective wavelengths ($\lambda_{\rm eff}$) of the bands being 3500
  $\AA$ ({\it u}), 4800 $\AA$ ({\it g}), 6200 $\AA$ ({\it r}), 7600
  $\AA$ ({\it i}), and 9000 $\AA$ ({\it z}). This wavelength range is
  dominated by O5, B5, A0$-$AF, G0$-$G5, K and M stars. In this study
  we use the colors {\it (u-g)}, {\it (g-r)}, {\it (r-i)} and {\it
    (i-z)}. For these bands \citet{shimasaku2001} has done aperture
  photometry for normal galaxies covering the whole range of Hubble
  types. {They used galaxies brighter than 18 mag in the {\it g}-band, which is equivalent to 17.5 mag 
  in b-band}.
  They compared the obtained colors with those in the
  spectrophotometric atlas of galaxies by \citet{kennicutt1992}, 
  converted to the same photometric system. 

It was shown by \citet{shimasaku2001}, using stellar population
synthesis models, that fairly good solutions can be found for the
color-color diagrams {\it (g-r)} vs. {\it (u-g)}, and {\it (r-i)}
vs. {\it (g-r)}, but in {\it (i-z)} vs. {\it (r-i)} the observed
galaxy colors are shifted by 0.1$-$0.2 mags so that the galaxies in
{\it (i-z)} are bluer than predicted. Of course, the solutions are not
unique, because there is a degeneracy between the stellar ages and
metallicities.  In the above color-color diagrams this degeneracy has
been studied by \citet{lenz1998}. Using \citet{kurucz1991} models they
calculated the stellar effective temperatures (T$_{\rm eff}$) for two
different surface gravities representative of giant and dwarf stars,
and for metallicities of [M/H] = +1.0, 0.0 and -5.0. They showed that
the loci are quite thin for stars hotter than 4500 K, but that in the
red end (populated by M stars) T$_{\rm eff}$ depends on the different
stellar parameters. This is important in our study where the colors
are dominated by giant stars.


\setcounter{table}{2}
\begin{table}
\scriptsize
\caption{Color differences between the two approaches for the different 
structures.
}
\label{table_color_errors}
\centering
\begin{tabular}{cccccc}
\noalign{\smallskip}
\hline  \\
            & {\it (u-g)}  &  {\it (g-r)}  &  {\it (r-i)}  &  {\it (i-z)}  \\
\noalign{\smallskip}
\hline 
\\
Differences between &&&&\\
approaches\tablefootmark{a}: &&&&\\
\\
bl:      &&&&\\
$\Delta$(mean)    & 0.023 & 0.008 & 0.003 & 0.010 \\
$\Delta$(median)  & 0.004 & 0.009 & 0.002 & 0.014 \\

\\
bar:   &&&&\\
$\Delta$(mean)    & 0.005 & 0.008 & 0.006 & 0.001 \\
$\Delta$(median) & 0.007 & 0.015 & 0.006 & 0.005 \\ 

\\
disk:   &&&&\\
$\Delta$(mean)    & 0.090 & 0.032 & 0.009 & 0.015 \\
$\Delta$(median)  & 0.103 & 0.025 & 0.000& 0.009 \\
\\
\hline

\end{tabular} 
\tablefoot{  
\tablefoottext{a}{Color differences between the two approaches calculated
from the mean and median values given in Table 2.}}
\end{table}

In the
following the mean and median colors given in Table 2 for the measured
structure components are compared with the total colors of normal
galaxies in different galaxy surveys where the SDSS images have been
used.  These include (see Table 4) the mean colors of E galaxies by
\citet{shimasaku2001} {which comprises 87 E's out of their}
sample of 456 bright galaxies, and the
colors in the {\it Spitzer} Infrared Nearby Galaxies Survey \citep[SINGS, ][]{munoz2009},
which galaxies we have divided into two
morphological bins of bright galaxies (S0$-$Sb, and Sbc$-$late-type
galaxies). The sample by \citet[][ see their Table 1]{shimasaku2001} is
the most representative of normal galaxies. SINGS is biased to
a sample of 32 bright galaxies {(R absolute magnitude
in the range from -18.4 to -22.1 mag})
and it practically lacks elliptical galaxies. 

It appears that the colors of barlenses and thin bars, using all of
our color indices, are very similar with each other (see Table 2). In
all colors the differences in the mean and median values are within
the dispersion of our measurements. The colors of the two bar
components are also very similar with the mean colors of normal
elliptical galaxies as given by \citet[][their Table
  1]{shimasaku2001}.  The biggest difference appears in  {\it
    (u-g)} and {\it (g-r)} colors in which the bar components are on
average 0.1 magnitudes bluer than normal elliptical galaxies{
  (1.69 vs. 1.79 in {\it (u-g)} and 0.74 vs. 0.83 in {\it (g-r)}) : the
  differences are at least two-fold compared to the standard errors of the
  mean colors of bar components.
The central regions (nuc1) are slightly redder than the two bar
components particularly in {\it (u-g)} and {\it (i-z)}, where the differences in
the mean values are $\sim$ 0.09. 
The mean
colors of {\it (u-g)} = 1.77 and {\it (g-r)} = 0.79 for the central regions (nuc1) are
very similar to the colors of elliptical galaxies.
In SINGS all the mean colors of S0$-$Sb galaxies are
bluer than those in \citeauthor{shimasaku2001}

The fact that bars in our sample have on average similar red colors as
the elliptical galaxies, is a manifestation that they are dominated by
old stellar populations. This is important because it is well known
that bars with young stellar populations also exist. In a sample of 18
S0$-$Sbc galaxies \citet{gadotti2006} divided bars into two
categories, namely those dominated by old ({\it B-I} = 2.17 $\pm$ 0.12) and young
({\it B-I} = 1.49 $\pm$ 0.20) stellar populations. The colors of bars in their study were
measured at the end of the bar. We have six galaxies in common with
their sample, of which four belong to their category of ``old bars''. For
these galaxies we transformed our {\it u, g, r} and {\it i} SDSS colors into {\it (B-I)} colors using
the equations by Lupton (2005)\footnote{These equations are found
in the SDSS webpage containing transformations between SDSS magnitudes and other
systems: www.sdss.org/dr8/algorithms/sdssUBVRITransform.php}. The obtained {\it (B-I)} colors
are in agreement with those given by \citeauthor{gadotti2006} for five of
the galaxies, at maximum deviating $\sim$ 0.3 mag. Only for NGC
4394 the difference is as high as $\sim$ 0.7 mag (redder in this work),
probably due to a prominent inner ring that surrounds the bar, which
might affect the color at the outer edge. In fact, the mean {\it (B-I)} color 
for all our common galaxies is $\langle${\it (B-I)}$\rangle$ = 2.07, which is
in agreement with the value for bars dominated by old stellar populations
in \citeauthor{gadotti2006}.
They discussed that although
``young bars'' in terms of stellar ages preferentially appear in young
parent galaxies, that is not enough to explain the color differences
between the two bar categories.

\setcounter{table}{3}
\begin{table}
\scriptsize
\caption{Color indices for elliptical \citep{shimasaku2001} and SINGS galaxies  
\citep{munoz2009}.
}
\label{}
\centering

\begin{tabular}{cccccc}
\noalign{\smallskip}
\hline  \\
            & {\it (u-g)}  &  {\it (g-r)}  &  {\it (r-i)}  &  {\it (i-z)}  \\
\noalign{\smallskip}
\hline 
\\
Shimasaku et al. 
\\
E:
\\
{N} & {86}  & {87}  & {87} &  {87}  \\ 
mean &    1.79   & 0.83   &  0.41   &  0.27  \\
std  &    0.26   & 0.14   &  0.05   &  0.06  \\
\noalign{\smallskip}
\hline 
\noalign{\smallskip}
\noalign{\smallskip}
SINGS
\\
\\
S0$-$Sb: &&&&\\
N           &    13 &    13 &    13 &    12  \\
mean        & 1.408 & 0.670 & 0.374 & 0.180  \\
median      & 1.423 & 0.681 & 0.384 & 0.197  \\
min         & 0.903 & 0.518 & 0.238 & 0.031  \\
max         & 1.627 & 0.772 & 0.438 & 0.251  \\
std         & 0.187 & 0.076 & 0.054 & 0.059  \\
\\
Sbc$-$LTG: &&&&\\
N           &    17 &    17 &    17 &    17  \\
mean        & 1.108 & 0.532 & 0.298 & 0.146  \\
median      & 1.099 & 0.531 & 0.284 & 0.153  \\
min         & 0.788 & 0.384 & 0.223 &-0.025  \\
max         & 1.435 & 0.781 & 0.447 & 0.247  \\
std         & 0.184 & 0.107 & 0.058 & 0.067  \\
\noalign{\smallskip}
\hline 
 
\noalign{\smallskip}
\end{tabular} 
\end{table}

The colors and stellar populations of bulges have been extensively
investigated in the literature, but the colors of boxy/peanut bulges are
only rarely studied. One such study is that made by \citet{williams2012},
thus giving us a point of comparison with barlenses. They
made long-slit spectroscopy for 28 edge-on galaxies, of which 22 had
boxy/peanuts. The parent galaxies had Hubble types of S0$-$Sb, which is
also the Hubble type range where the barlenses in our sample appear.
They studied separately the central peaks within the radius of a few
arcseconds covering the PSF, and the boxy/peanuts outside that region.
They found that the stellar properties of the central peaks were
indistinguishable from those of typical E/S0 galaxies. However, that
was not the case for the boxy/peanuts: although their stellar
populations were not much different from those of the central peaks,
their metallicity gradients were steeper than for the E/S0 galaxies.


\subsubsection{The underlying disks}

One of the first attempts to compare the colors of bars and the
underlying disks comes from \citet{okamura1976}. They studied
five barred galaxies, concluding that although the spiral
arms are bluer than bars, measured using the {\it (B-V)} color index, the
underlying disks have similar colors with bars.  On the other hand,
\citet{elmegreen1985} studied 15 barred galaxies in {\it (B-I)}
color, and found that the spiral arm regions can be bluer than the bar
even if the spiral arms themselves were not particularly
blue. However, counter-examples were also found. In this study we find
that the disks have very similar colors with bars (thin bars) in
{\it (r-i)} (the difference in the mean color is $\Delta$ = 0.005 mag), 
but slightly bluer in the other bands: $\Delta$ = 0.11, 0.03  
and 0.04 magnitudes in {\it (u-g)}, {\it (g-r)} and {\it (i-z)},
respectively. This is of particular interest because the old stellar
populations making bars and disks are expected to be the same. 
The disks of the barlens galaxies discussed above might actually be 
even bluer if the colors were measured outside the bar radius as was done
by \citet{elmegreen1985}.

\begin{figure*}
\begin{centering}
\includegraphics[scale=.6]{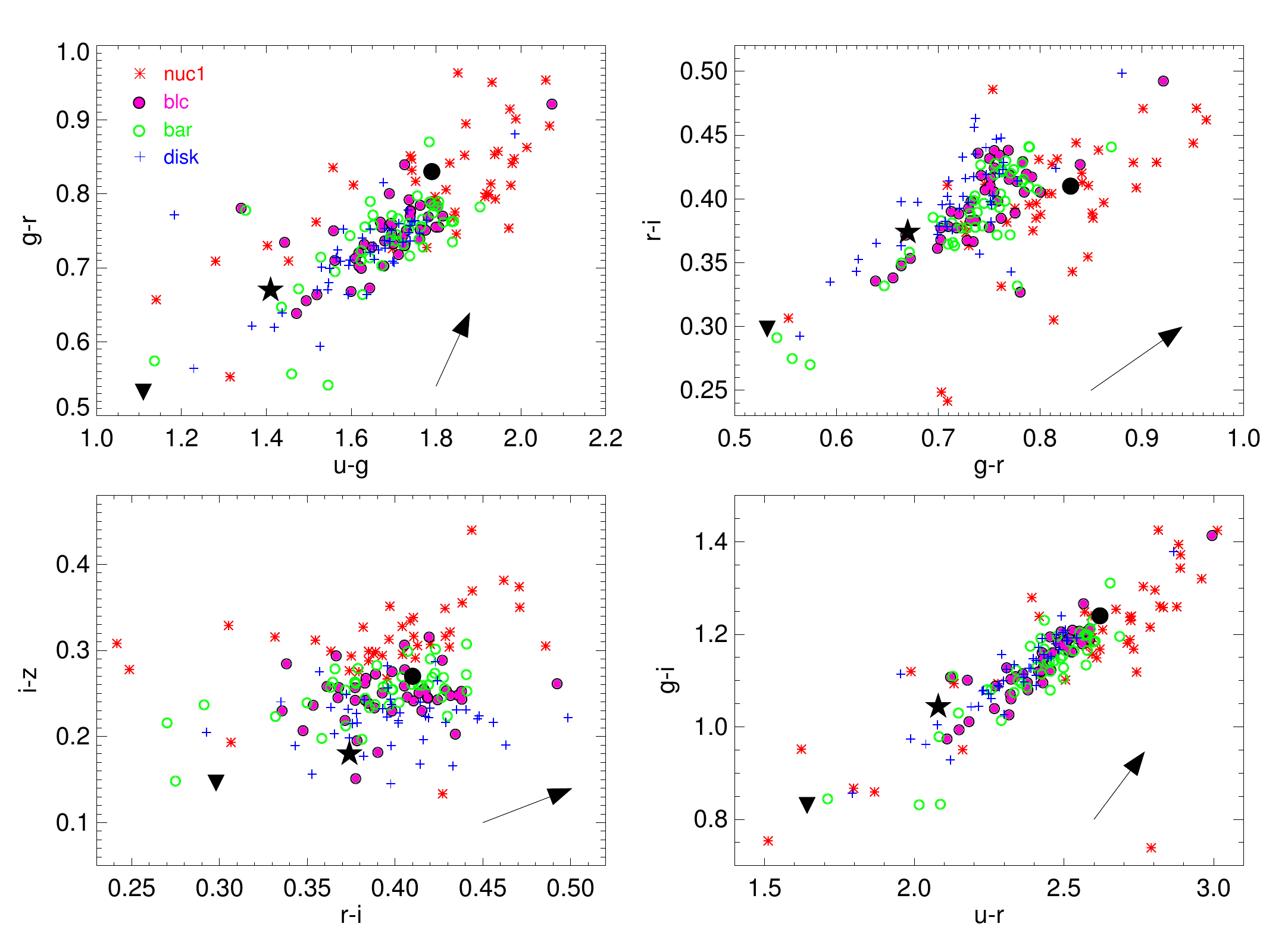}
\caption{Color-color diagrams discussed in the text. We
   use the color measurements obtained using the approach 1. The
   colors are corrected for Galactic extinction. The arrows indicate
   the reddening vectors (the length correspond to 0.1 mag extinction
   in the {\it r}-band). The black circle represents the mean color of the elliptical galaxies 
   from \citet{shimasaku2001}.
The black star and the black downward triangle represent the mean colors of  S0$-$Sb and 
Sbc$-$late-type galaxies from \citet{munoz2009},
respectively.}
\label{bl_color_color}
\end{centering}
\end{figure*}

\subsection{Color-color diagrams}

In Fig. 6 the color-color diagrams are plotted showing the central
peaks (red asterisks), barlenses (pink filled circles), 
thin bars (green open circles),
and disks (blue crosses). In the case of central 
peaks, only nuc1 regions are shown because nuc2 might 
contain contamination from the colors of nuclear rings.
In the two upper panels
plotted are {\it (g-r)} vs. {\it (u-g)}, and {\it (r-i)} vs. {\it (g-r)}, 
whereas in the lower panels {\it (i-z)} vs. {\it (r-i)} and {\it (g-i)} vs. 
{\it (u-r)} are shown. 
As a comparison we also plotted the mean values for the elliptical galaxies (black circle) 
in \citet{shimasaku2001} and the S0/a$-$Sb (black star) 
and Sbc$-$late-type galaxies (black downward triangle)
in \citet{munoz2009},  shown in Table 4.

Photometric separations of the stellar populations in these color
bands were discussed for example by \citet{fukugita1996}
and \citet{lenz1998}.  In all the color-color diagrams the reddening vectors are
also shown.  In Fig. 7 the tracks of the stellar evolutionary models
are shown in the same diagrams, and also the mean values for the different
types of galaxies from the literature as in Fig. 6. The models are from 
\citet{bruzual2003}, using their ``Padova 1994'' evolutionary tracks, 
the \citet{chabrier2003} Initial Mass Function (IMF), and STELIB spectral library \citep{leborgne2003}.
The tracks were calculated for different stellar ages and metallicities,
indicated in the figure. According to \citet{terlevich2002} typical
metallicities for the galaxies in Hubble types of S0$-$Sa are between  
[Fe/H] = -0.45 $-$ 0.50. 

Between the {\it u} and {\it g} bands the stellar spectra contain the Balmer jump,
which is sensitive to stellar surface gravity.  Therefore the first
diagram (upper left panel) is good for separating the main sequence
stars from giant stars. As most of the line-blanketing from heavy
elements also occurs at short wavelengths, among the selected diagrams
this is perhaps also the most sensitive to metallicity. In the
Fig. 3 by \citet{lenz1998}, for a given {\it (g-r)}
color the more metal rich stars have a higher {\it (u-g)} color index.
The second diagram (upper right panel) separates the stars according
to their effective temperatures in a more clear manner, although
metallicity also still plays some role.  At the color range useful in
this study, the third diagram (lower left panel) is at some level
also sensitive to the metallicity of giants stars. The {\it (u-r)} color index in
the forth diagram (lower right panel) is sensitive to short lived
stars, in a similar manner as the equivalent width of H$\alpha$
(EW(H$\alpha$)) \citep[see][]{casado2015}.

\begin{figure*}
\begin{centering}
\hskip 0.3cm\includegraphics[scale=.6]{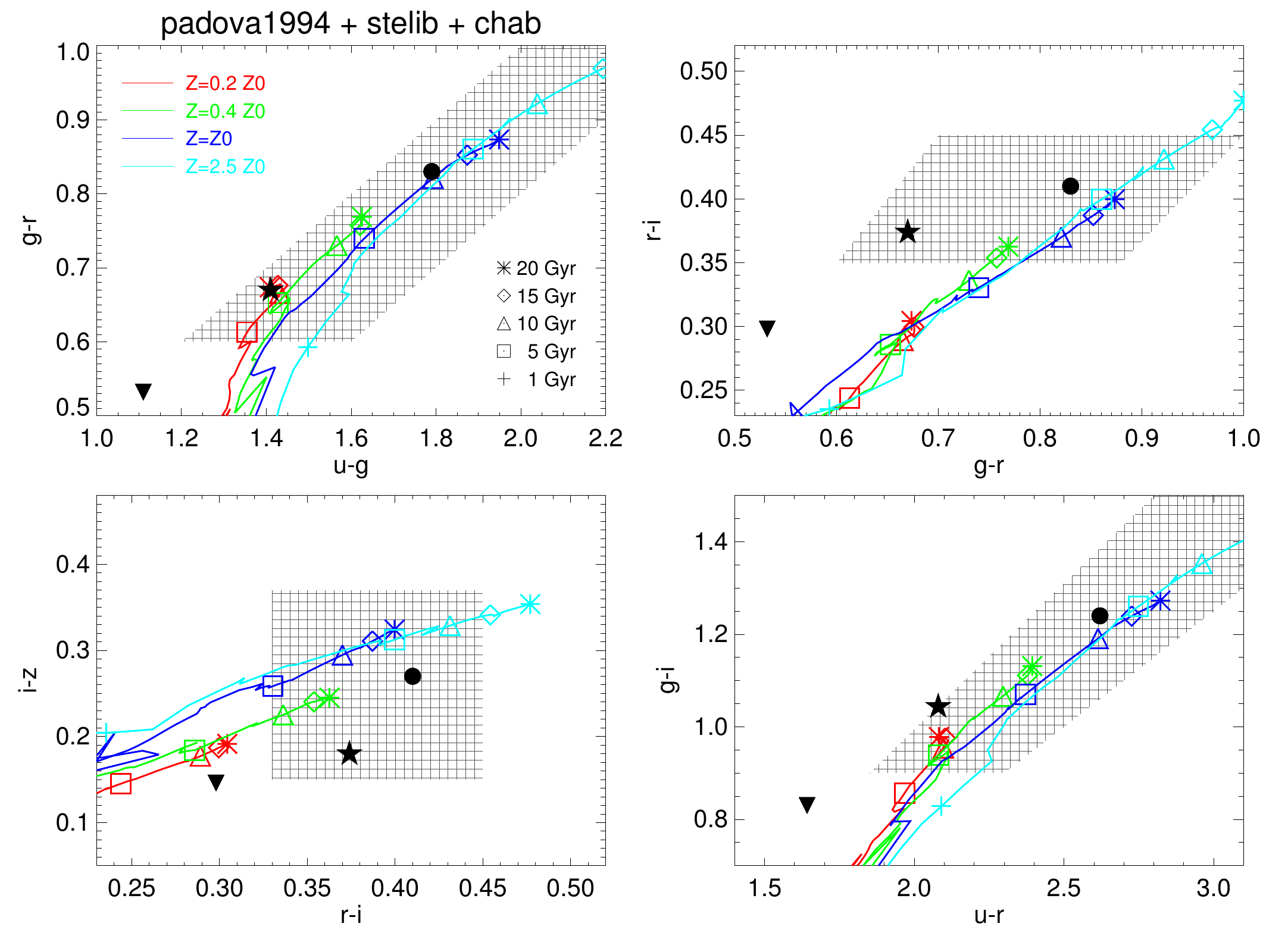}
 \caption{Stellar evolutionary models of \citet{bruzual2003}
   are calculated to four different metallicities, for the range of
   stellar ages of 1$-$20 Gyr.  The models use their ``Padova 1994''
   evolutionary tracks, the \citet{chabrier2003} Initial Mass Function
   (IMF), and STELIB spectral library \citep{leborgne2003}. The tracks are calculated for the 
same color indices as the observations shown in Fig. 6, and the black symbols
represent the same galaxy samples as before.
The ranges spanned by our observations are marked with the hatched regions.
}
\label{bl_bruzual}
\end{centering}
\end{figure*}

\vspace{0.2cm}

\noindent {\it (g-r)} vs. {\it (u-g)}: 

\noindent The bar components are mostly distributed to a narrow range
of {\it (g-r)} colors of $\sim$ 0.7$-$0.8, indicating that they are dominated
by K-giant stars. 
Overall the colors of the two bar components
(barlens and thin bar) fall into the same region in this
diagram. Therefore, taking into account that this diagram is also the
most sensitive to metallicity means that the metallicities of the two
bar components must be similar.  On the other hand, the central
regions of the galaxies (i.e., nuc1) are shifted to {\it
(g-r)} $>$ 0.8 and {\it (u-g)} $>$ 1.8.
For these high values of {\it (g-r)} and {\it (u-g)} colors the
age-metallicity degeneracy prevents of making any conclusions of their
possible higher metallicities or older ages. Simply high extinction in
the central regions could explain the observed colors.  Also a large
majority of the disks have similar colors as bars in this diagram, but
there are also some galaxies in which the disks are shifted to
bluer colors: however even these colors ({\it g-r} $>$ 0.6) are
representative of stellar populations dominated by K-giant stars.

\vspace{0.2cm}
\noindent {\it (r-i)} vs. {\it (g-r)}:

\noindent Also in this diagram the two bar components have very
similar colors.  However, the colors of the central peaks and the
disks deviate from those of bars: for a given {\it (r-i)} the {\it (g-r)} color is
redder for the central regions and bluer for the disks.  As the
reddening vector goes nearly perpendicular to the obtained shifts
between the different structure components, the shifts cannot be
explained by internal galactic extinction. Based on the simple models
in Fig. 7 the shifts are not due to metallicity either. It is possible
that the model tracks use too simple model for star formation and therefore
cannot account the observations well.

\vspace{0.2cm}
\noindent {\it (i-z)}  vs. {\it (r-i)}: 

\noindent Again, the two bar components have very similar colors,
whereas the central peaks and disks are shifted so that for a given
{\it (r-i)} color the central peaks have redder {\it (i-z)} color, and the disks
have bluer color.  According to \citet{shimasaku2001} there is 0.1$-$0.2
magnitude uncertainty in the {\it z}-band calibration of the SDSS images,
but that is not critical in this study because we are looking at the
relative colors between the structure components.  Again, these color
differences between the structure components cannot be explained
neither by internal galactic extinction, nor by metallicity, in
particular if solar or super-solar metallicities are concerned.
Obviously, the central peaks are more dominated by the infrared flux
observed in the {\it z}-band, whereas the bars and disks are less dominated
by the infrared flux.

\vspace{0.2cm}
\noindent {\it (g-i)} vs. {\it (u-r)}: 

\noindent The last panel uses {\it (u-r)} which is sensitive to the presence
of massive, short-lived stars that dominate the blue part of the
spectrum \citep{casado2015}. The color index {\it (u-r)} traces stellar
populations ages not much older than OB stars, traced by 
EW(H$\alpha$). Based on the studies of
a large galaxy sample and using the SDSS images, it has been shown
that the division line between active and passive galaxies appears at
{\it (u-r)} = 2.3 \citep{casado2015,mcintosh2014}. This also marks
the well known red and blue sequence among the galaxies in the Hubble
sequence. Looking at our Fig. 6 (lower right panel) it appears that
all the structure components studied by us are mostly passive. The
central peaks are even redder than the bars or disks, having
{\it (u-r)} = 2.5$-$3.0. However, among all the structures there are some
individual cases in which manifestations of active star formation
appears ({\it u-r} $<$ 2.3).
\vspace{0.15cm}

In conclusion: thin bars and barlenses have similar colors.  The
main body of them have colors best traced with the stellar
evolutionary tracks with stellar ages of 5$-$10 Gyr and solar
  metallicity (Z$_0$), mainly in the {\it (g-r)} vs. {\it (u-g)} and
  {\it (g-i)} vs. {\it (u-r)} color-color diagrams. However,
  the colors are not so well accounted when {\it (r-i)} appears in the
  diagram, which leaves uncertain the interpretation of the
  color-color plots using {\it (r-i)}.  In any case, taking into
  account the reddening vectors it is clear that extinction alone
  cannot explain the redder central colors. Using nuc2 instead of nuc1
  would reduce the difference between the central regions and
  barlenses, because possible central star forming rings are then
  covered. However, even the central regions appear redder than
  barlenses.  


\begin{figure*}
\begin{centering}
\includegraphics[scale=.9]{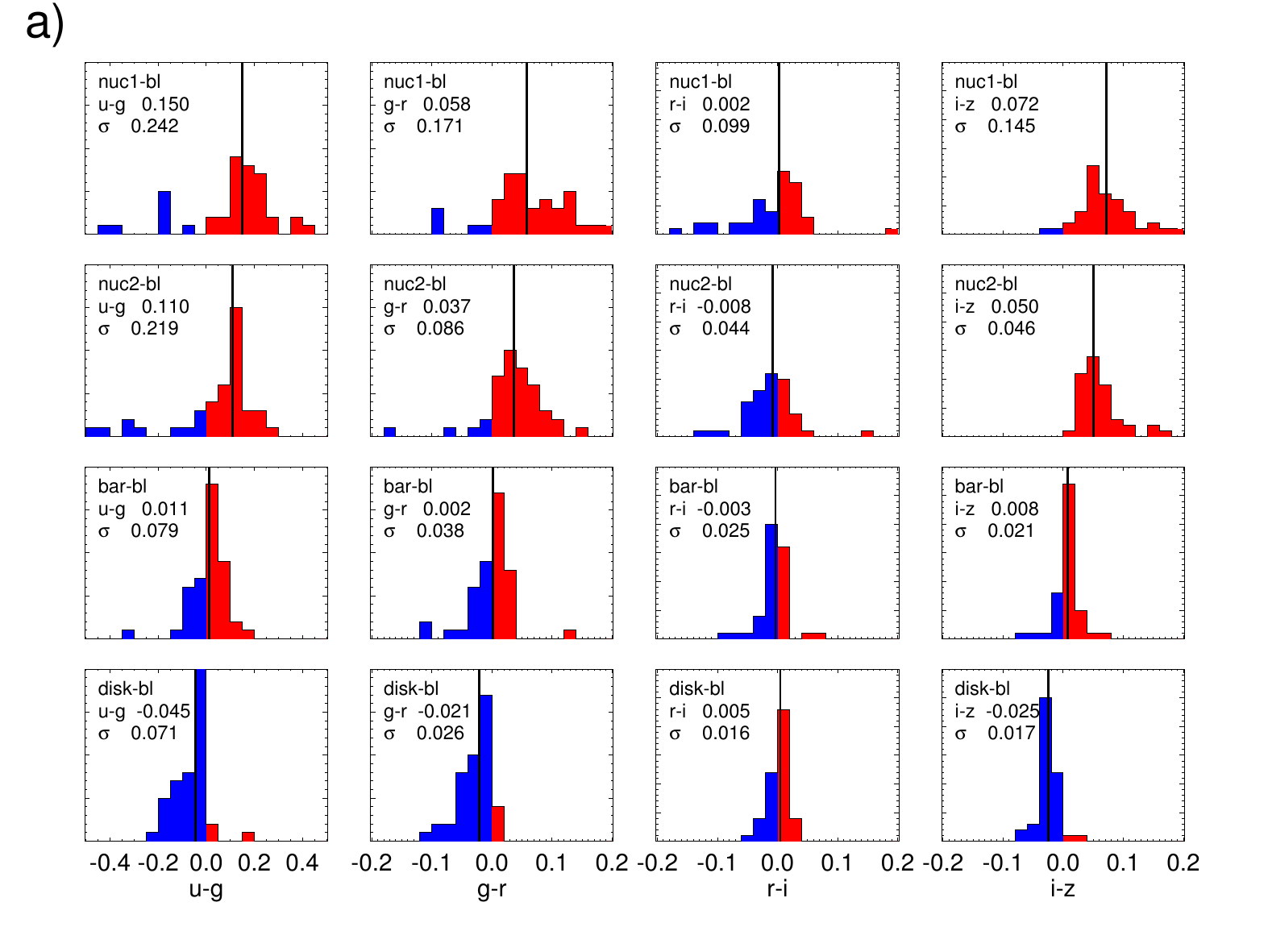}
\includegraphics[scale=.9]{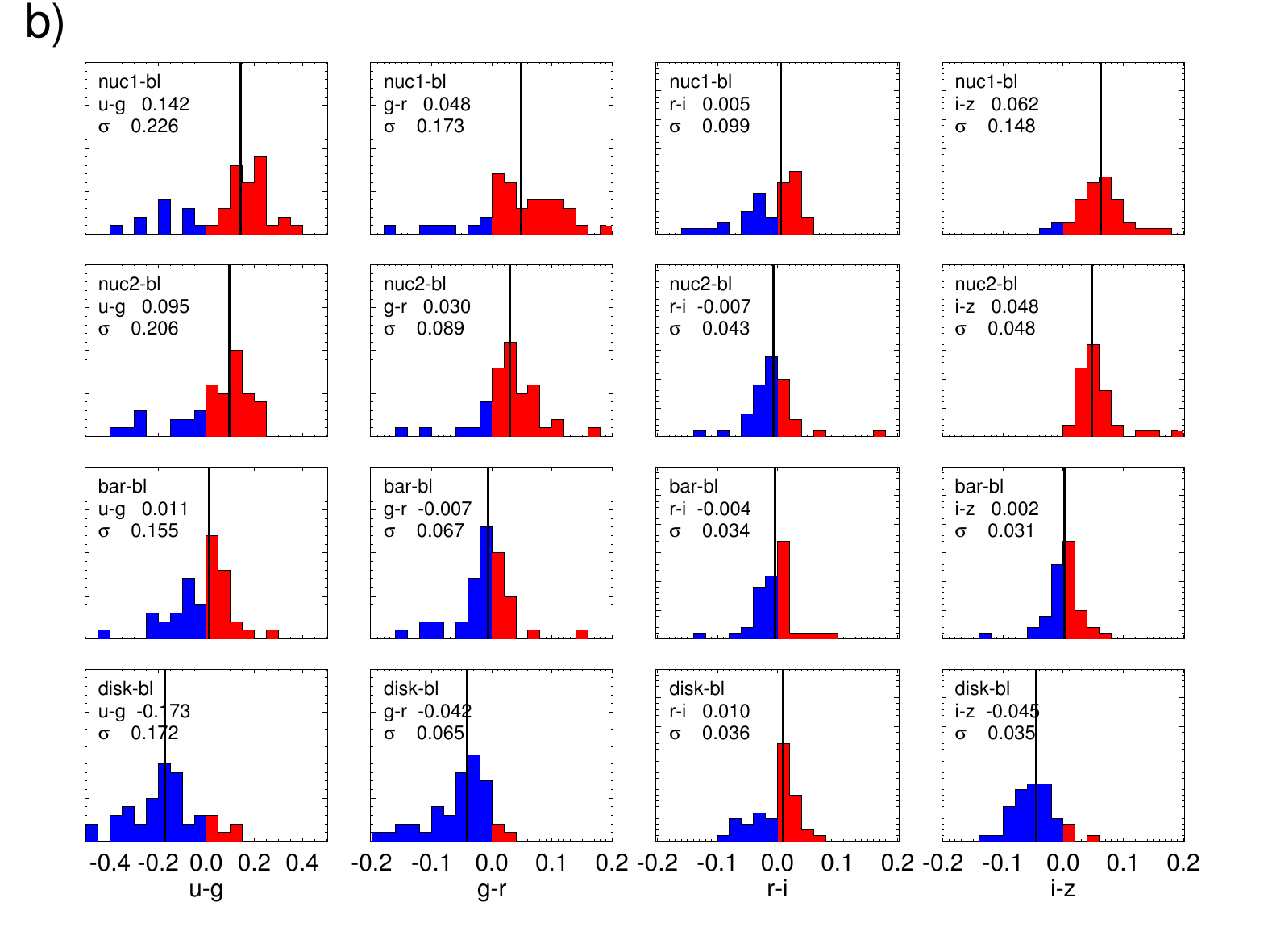}
\caption{Color differences between the structure components calculated
  separately for each galaxy. Shown are the histograms of the
  deviations from barlens color (blc), for the color
    subsample. The deviations are shown in four color indices. The
  red and blue colors indicate whether the deviations are toward
  redder or bluer color, when compared to the color of the
  barlens. The numbers are the median values of the deviations, 
    indicated also by the vertical lines, and the standard deviations
    ($\sigma$). Panel (a) uses the measurements from approach 1 
    and panel (b) those from approach 2 (see Section 4.2).
    The estimated photometric accuracy of the
  colors of individual galaxies is 0.03 magnitudes.
}
\label{bl_example}
\end{centering}
\end{figure*}

\subsection{Comparison of the colors of barlenses with respect to those of the other
structure components}

In the following the colors of the structure components are derived
galaxy by galaxy. In each color we calculated how much the barlens
color deviates from the color of another component, that is, from the
central peak, the thin bar, or the disk. These color differences are
shown in Fig. 8 using the same color indices as above. In the
histograms positive and negative deviations are shown with red and
blue colors, respectively. The median
values of the color differences are also indicated in the plots. 
The measurements using the two approaches explained in Sect. 4.2
are shown in panels (a) and (b).  
 
This analysis also shows that the colors of barlenses are very similar
to the colors of the thin bars of the same galaxies, manifested as
almost zero deviations in their average color differences. Small
deviations to both directions appear particularly in {\it (u-g)} and {\it (g-r)},
which can be associated to metallicity, star formation, or small
differences in the dominant stellar populations. This is not
unexpected having in mind that some of the barlenses have rich
structures manifested in their color maps, as will be shown in the
next section.

The central peaks (nuc1) are systematically redder than barlenses in
all the color indices except in {\it (r-i)}.  Using the larger apertures
(nuc2) gives somewhat bluer relative colors, because in some of the
galaxies the central color is contaminated by a star forming nuclear
ring.  Anyway, the obtained tendencies for the colors of the central
regions is independent of the aperture size used. However, there
are also some barlenses in which the central region is clearly bluer
than the barlens, which is manifested particularly in {\it (u-g)} and {\it (g-r)},
and a hint of that appears also in {\it (r-i)}. In {\it (r-i)} this might be related to
prominent emission lines in star forming regions in {\it r}-band.
On the other hand, the colors of the disks within the bar radii are
systematically bluer than those of barlenses. Again, this is the case
in all the other colors except in {\it (r-i)}. 

\subsection{Color profiles of individual galaxies}

Color profiles in {\it (g-r)} and {\it (i-z)} for the barlens
  galaxies in our color subsample are shown in Fig. 9 and Appendix B. The color
  profiles were obtained from the corresponding color map, along the
  bar major axis, within the stripes shown in Fig. 10 (region
  between two parallel dashed lines).  The stripes were defined to have
  a width of 0.2 $\times$ r$_{\rm bl}$ centered on the galaxy center.
What is actually shown in the profiles
are the color deviations in respect of the median colors
along the bar major-axis. The radial coordinates are normalized
to r$_{\rm bar}$.


The color profiles were divided into groups based on how flat the
profiles are. F indicates a completely flat color profile 
in which any deviation from flatness is not larger than the random 
color variations of the profile. RP indicates that the galaxy has a red
central peak which is clearly larger than the random variations. 
RP+nr includes galaxies that have a red central peak
and also a nuclear ring clearly identifiable as two minima in the 
{\it (g-r)} color profile, one at each side of the peak.
nb contains galaxies with nuclear bars  classified by
\citet{buta2015}. B contains
barlenses with bluer nuclear features due to a nuclear ring or a blue
central peak, here the {\it (g-r)} profile shows a minimum in the central region.
D means that the color images show plenty of dust 
features inside the barlens manifested in the profiles as variable color. 
According to this division of the barlens galaxies to different groups,
red central peaks appear in 21 (45$\%$) of the galaxies. This number
includes also the five (11$\%$) red nuclear bars, and the five (11$\%$)
galaxies with star forming
nuclear rings surrounding the red central peaks (in Appendix B: RP, RP+nr,
nb).  In ten (21$\%$) of the galaxies the profiles are completely flat, and seven 
(15$\%$) galaxies have blue central peaks. 
In nine (19$\%$) of the galaxies the whole
barlens regions are full of dust and star forming regions.
The galaxies with star forming nuclear rings or
red nuclear bars are indicated in Table 1.
Negative
color gradients in barlenses appear at least in IC 1067 and NGC 2968,
and variable colors in some other galaxies, associated to complex
dusty structures within the barlenses. 
NGC 6014 represents a particular case in which a nuclear ring is manifested
as two minima in the {\it (g-r)} profile, however it shows signs of a bluer central
region. We have included this galaxy in the B group (Appendix B) but considered its 
nuclear ring in Table 1.

In conclusion: although the
small central regions of barlenses have on average redder colors than
the rest of the barlens, the color profiles along the bar major axis
hint to clearly red central peaks only in $\sim$ 23$\%$ of the
barlenses (those that belong to the RP group).

In order to characterize the extension of the red central peaks, 
we looked at those galaxies for which the FWHM of the 
images is clearly smaller than the radial extension of the peak (see Appendix B).
In such cases one can be sure that the observed structure is resolved and 
the size can be estimated in a reliable manner.
Visual inspection of the color profiles
revealed that the peak
extension is typically $\sim$ 10$\%$ that of the bar. However, it should be kept in
mind that extinction by dust can also affect this estimation of the peak extension.

\begin{figure}
\begin{centering}
\includegraphics[trim = 15mm 0mm 0mm 0mm, clip, scale=.7]{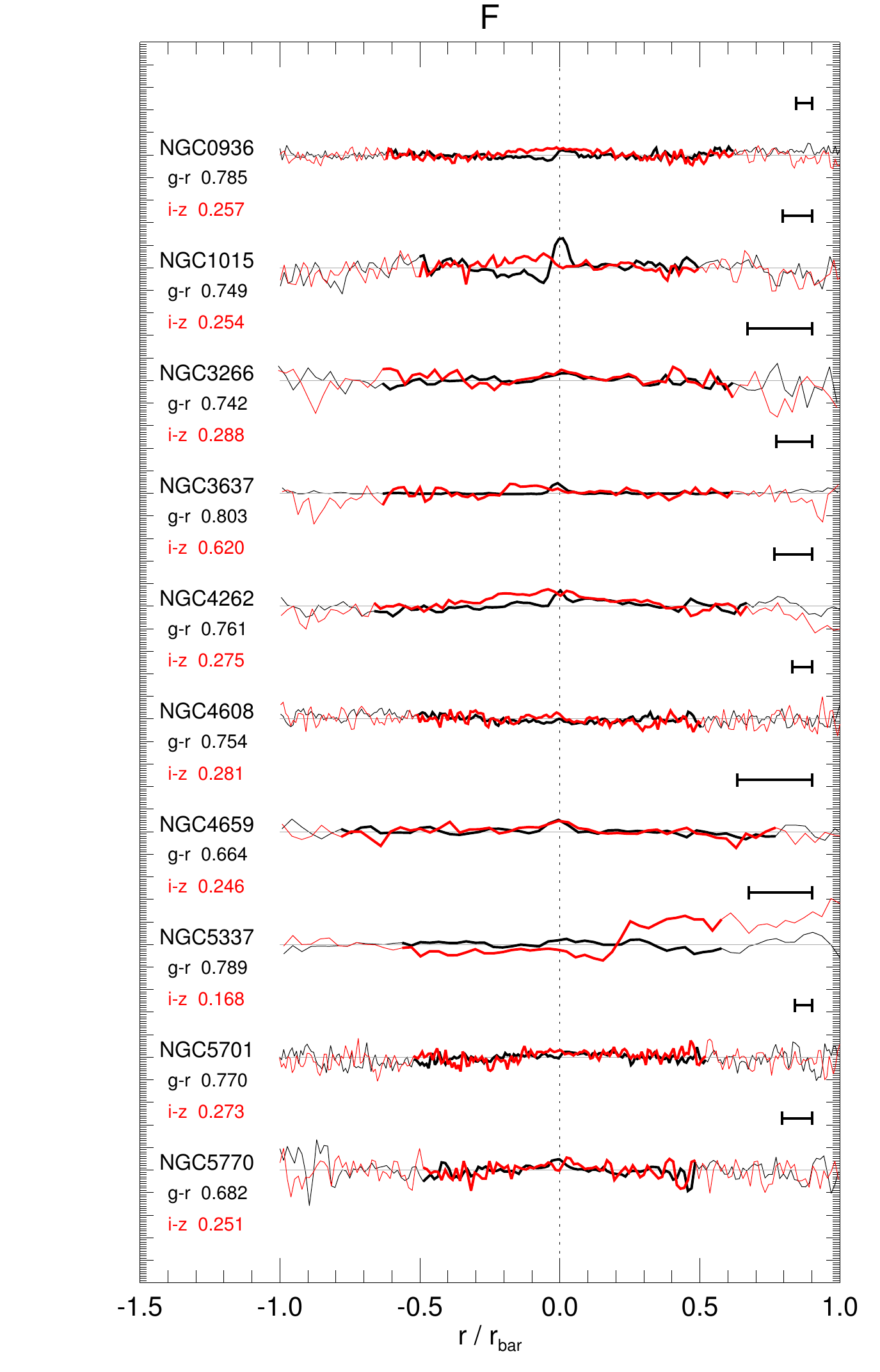}
\caption{Bar major axis color profiles in {\it (g-r)} (black) and {\it (i-z)}
   (red) for the barlens galaxies in the F group.  
   The radial coordinate is normalized to the radius of the bar
   (r$_{\rm bar}$), and the
   thick lines indicate the region inside the radius of the barlens 
   (r$_{\rm bl}$). The horizontal lines
   indicate the median color along the bar, and the individual
   profiles are shifted by 0.5 mags. The dashed vertical line indicates
the galaxy center. The labels in the left give
   the median {\it (g-r)} and {\it (i-z)} colors, and the bar on the right represents
   the FWHM of the {\it g}-band image for each galaxy.
   Similar plots for the RP, RP+nr, nb, B and D groups are shown in Appendix B.
   }
\label{bl_example_a}
\end{centering}
\end{figure}

Compared to the Atlas of Images of NUclear Rings \citep[AINUR, ][]{comeron2010}
we have also detected blue nuclear
rings in NGC 3185, NGC 3351, NGC 4245, and NGC 4314, which belong to our 
RP+nr group. On the other hand, nuclear
rings are reported in AINUR for the galaxies NGC 936 and NGC 4262,
classified by us as having flat color profiles. In fact, for NGC 4262
a nuclear ring might appear also in our color profile, but the signature is
not very strong. In total, we have 24 galaxies in common with AINUR but only
12 of them have the full set of optical images used here (NGC 4593 has only {\it g}-band imaging).
Apart from the two F and four RP+nr cases mentioned above, from AINUR we have classified one Rp (NGC 4371), 
two nb (NGC 2859 and NGC 4340), one B (NGC 3945), and two D (NGC 4448 and NGC 4579)
as can be seen in Fig. 9 and Appendix B. 

About 20$\%$ (15) of the barlens galaxies in our sample are shown to have some
type of nuclear activity in  HYPERLEDA \citep{makarov2014}, being mainly 
Seyfert 2 nuclei or LINERs. Only NGC 4639 has Seyfert 1 type
nucleus in our color subsample.
However, the galaxies with nuclear activity do not show
any peculiarities in their central colors. Eight of these galaxies belong
to our color subsample. We have compared their nuc1 median colors 
with those of the rest of galaxies without active galactic nuclei (AGN) classification.
The differences are $\Delta${\it (u-g)} = 0.024, $\Delta${\it (g-r)} = 0.016, 
$\Delta${\it (r-i)} = 0.034, 
and $\Delta${\it (i-z)} = 0.013 magnitudes. These differences are smaller than, for example, those among 
the disks and the bars of galaxies, except in the {\it (r-i)} index (see Section 5.1.2). This 
suggests that the central parts of barlens galaxies with and without AGN are rather similar.}

\subsection{Color maps of individual galaxies}

Color maps of the representative examples of barlens galaxies are
shown in Fig. 10. The {\it g}-band images, {\it (g-r)} and {\it (i-z)} 
color maps, as well as the color profiles along the major
and minor axes of the bar in both colors are shown. In order to facilitate the
interpretation of the profiles, the images were rotated so that the bar
major axis appears always horizontally.  In the color profiles, the
red and blue
vertical lines indicate the sizes of barlenses and the thin
bars, respectively. The selected galaxies are examples of the different categories
shown in Fig. 9 and Appendix B, marked in parenthesis below.  The colors of bars are
generally fairly smooth, although some of the bars have a lot of
fine-structure, related to dust extinction and recent star formation,
or separate structure components like nuclear bars or star forming
nuclear rings appear.
\vspace{0.15cm}

\noindent {\it NGC 5701 (F):} is an S0/a galaxy, at an inclination of
inc = 15.2$^\circ$.  The color profiles both along the bar major and
minor axes are completely flat, and the barlens shows no structure in
the optical colors.  The median colors are {\it (g-r)} $\sim$ 0.77 and
{\it (i-z)} $\sim$ 0.27, which are also
normal colors of elliptical galaxies according to \citet{shimasaku2001}.  
\vspace{0.15cm}

\noindent {\it NGC 4596 (RP):} 
is an S0/a galaxy with inclination of i = 35.5$^\circ$.
This galaxy has a red central peak at r $\sim$ 5\arcsec, which has colors of
{\it (g-r)} = 0.91 and {\it (i-z)} = 0.39. At {\it (g-r)} this is a typical color of an
elliptical galaxy (compared with 0.83 $\pm$ 0.14 in \citeauthor{shimasaku2001}), and at
{\it (i-z)} slightly redder than the mean value (compared with 0.27 $\pm$ 0.06 in
\citeauthor{shimasaku2001}). The barlens is featureless even at {\it (g-r)} color
showing very little intermediate aged star formation or features of
dust.  This is an example of galaxies in which small classical bulges
might be present in the central regions.
\vspace{0.15cm}

\noindent {\it NGC 4314 (RP+nr):} has a Hubble stage Sa, and has an
inclination inc = 20.4$^\circ$. This is an example of barlens galaxies
which have a red central peak in the color profile along the bar major
axis (in {\it g-r} and {\it i-z}). The central peak appears within r = 5\arcsec, and has
colors of {\it (g-r)} $\sim$ 0.92 and {\it (i-z)} $\sim$ 0.35, which are redder than
the mean values in normal elliptical galaxies.  The red peak is
surrounded by a star forming nuclear ring visible in {\it (g-r)}.  Outside
the nuclear ring spiral arms appear, which are red in {\it (g-r)} probably
being a manifestation of dust. The nuclear ring and the nuclear bar of
this galaxy have been previously discussed in the optical by \citet[][see references there]{erwin2003},
and in the infrared by \citet{lauri2011}. In \citet[][see their figure 1]{lauri2014} 
the observed surface brightness profiles along the bar major and minor
axis were shown and compared with those predicted by the
simulation model by \citet{atha2013b}. 
They concluded that 
these profiles are similar to those predicted by the simulation models.
\vspace{0.15cm}

\noindent {\it NGC 4143 (nb):} this is an S0 galaxy with an inclination of 
inc = 43.8$^\circ$. It hosts a nuclear bar with a radius of r$_{\rm nb}$ = 2\arcsec
\citep{lauri2011}. Looking at its {\it (i-z)} color profiles, it seems likely that
the observed central peak is related to this feature. This is suggested
by the peak radial extension, which roughly corresponds to that of the nuclear bar.
This can also be seen 
in Appendix B, given that r$_{\rm nb}$,
normalized to the size of the main bar (r$_{\rm bar}$) is $\sim$ 0.1, which is 
roughly the same extension of the central peak.
However, in this case the central peak is unresolved because the FWHM is larger
than the peak extension.
\vspace{0.15cm}

\noindent {\it NGC 3380 (B):} this galaxy has Hubble stage S0/a and
galaxy inclination of inc = 20.8$^\circ$. Belonging to category B, the
barlens has a blue central region within a few arcseconds inside a
redder barlens. The blue nucleus within a few arcseconds appears in both
colors, but it is particularly clear in {\it (g-r)}, indicating that it
consists of younger stellar populations than the barlens. The bar is
blue in its outskirts, and in the zone where it crosses the
surrounding ring-like structure (rs).

 \vspace{0.15cm}
\noindent {\it NGC 4579 (D):} this is an Sa galaxy at
inc = 41.6$^\circ$.  This is an example of a dusty barlens galaxies.
Particularly in the {\it (g-r)} color map a lot of fine structure
appears inside the barlens. The star forming regions inside the
barlens are oriented along the disk, but they are much more elongated than
the outer disk, meaning that they are clearly associated to the barlens
morphology.  In the {\it (i-z)} color profile there is a small
red central peak.
The minor-axis profile further illustrates that the red
color is associated to the whole barlens, which color steadily turns
bluer from the galaxy center toward the edge of the barlens.

\begin{figure}
\begin{centering}
\includegraphics[scale=0.45,angle=0]{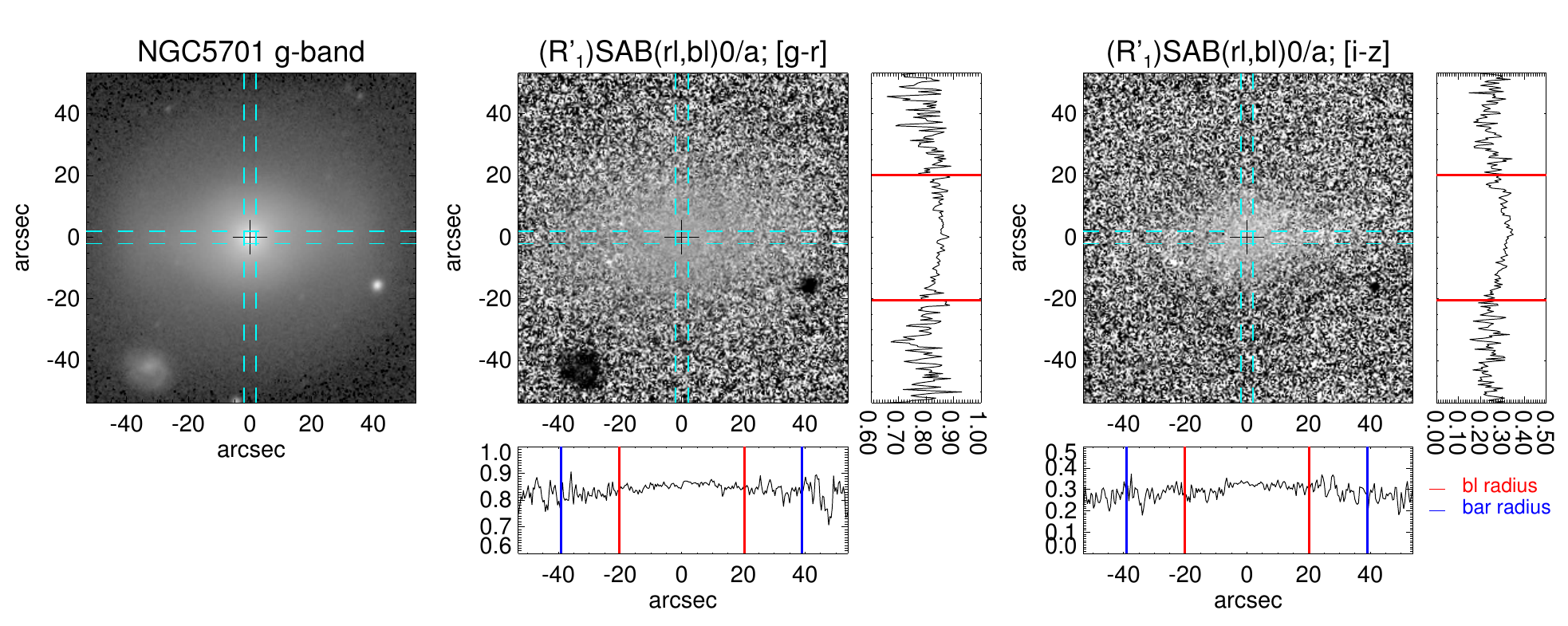}

\includegraphics[scale=0.45,angle=0]{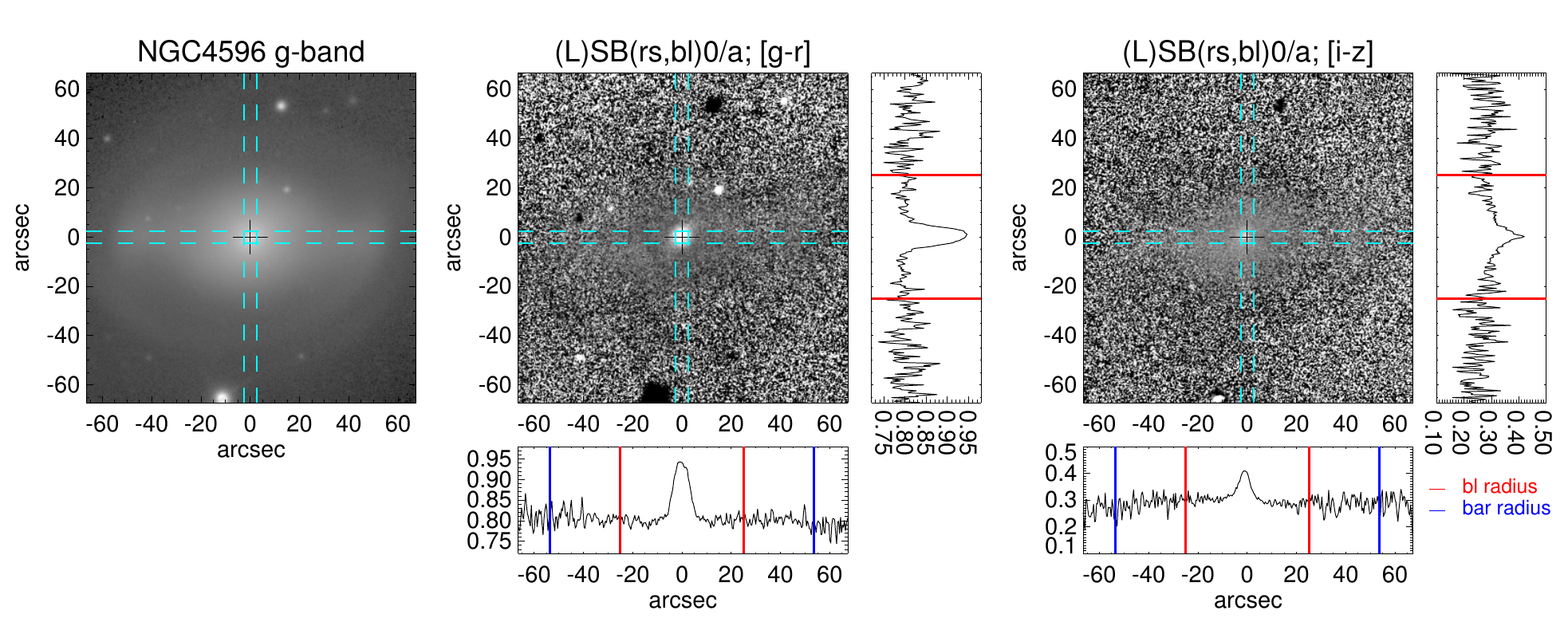}

\includegraphics[scale=0.45,angle=0]{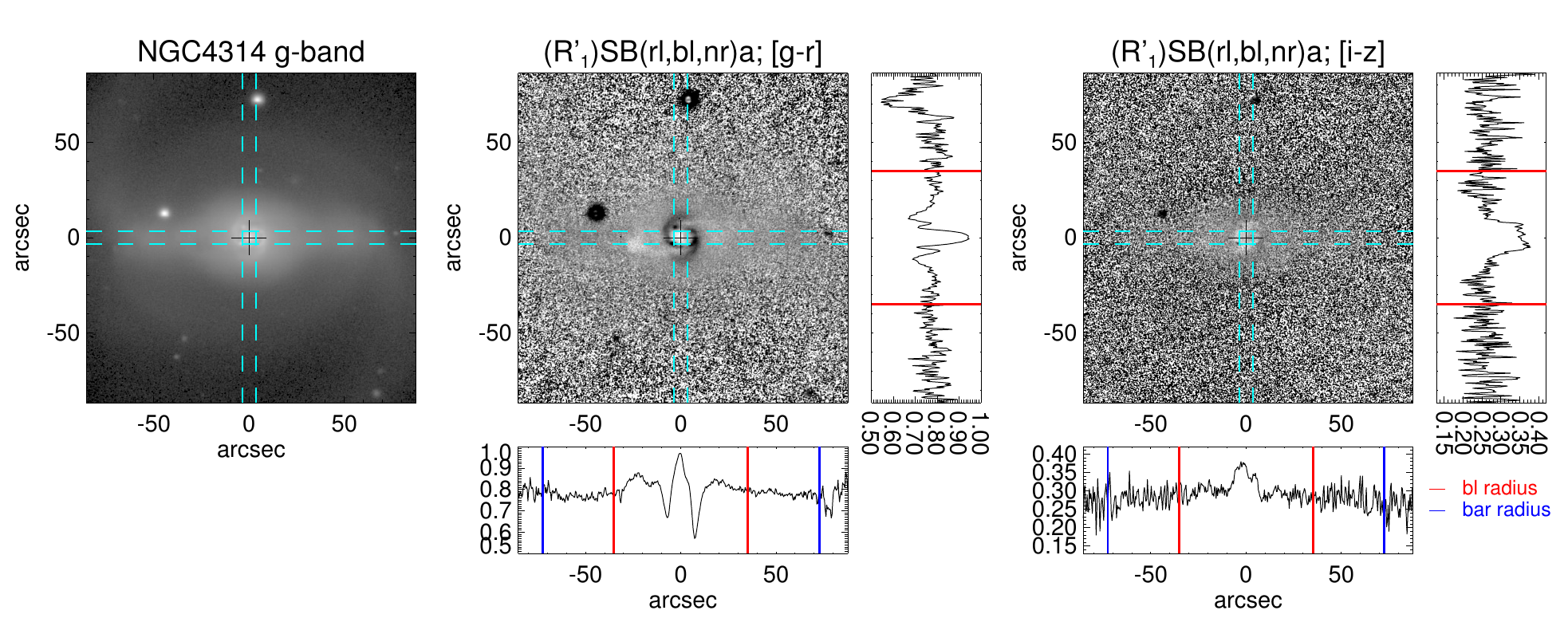}

\includegraphics[scale=0.45]{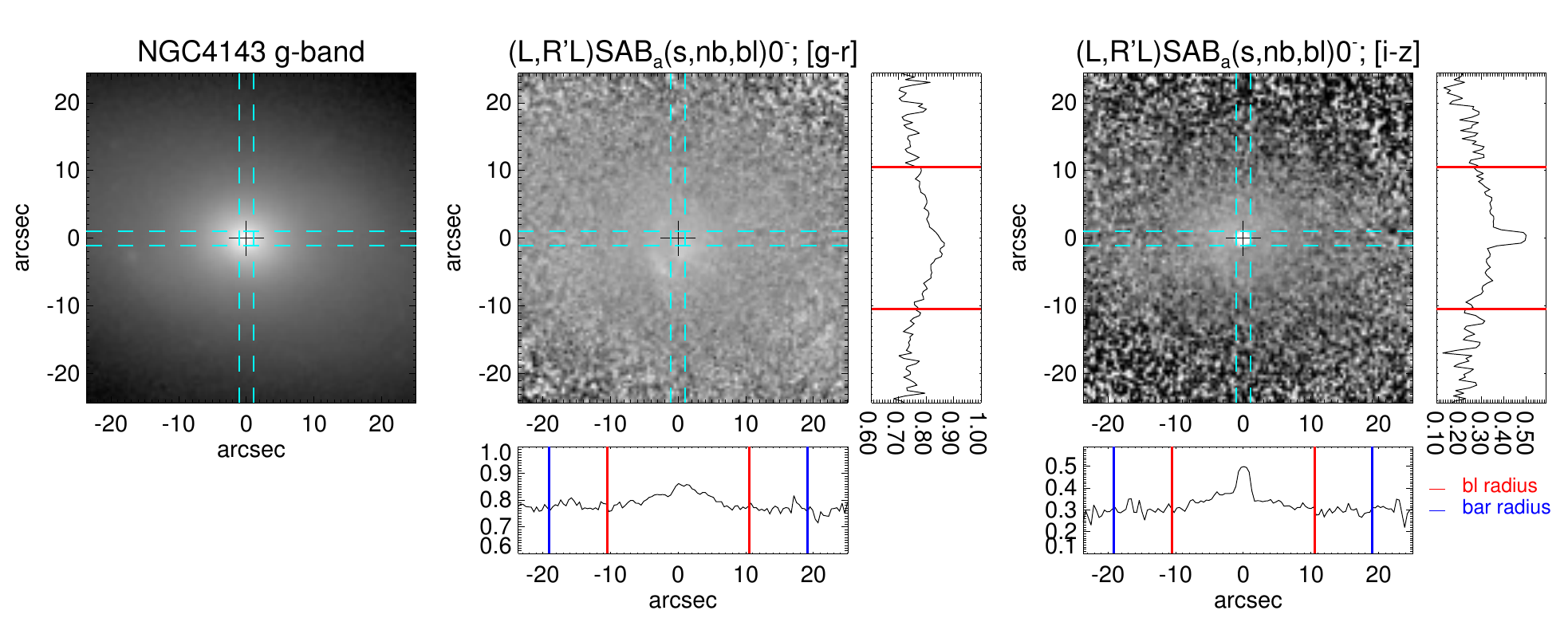}
%

\includegraphics[scale=0.45,angle=0]{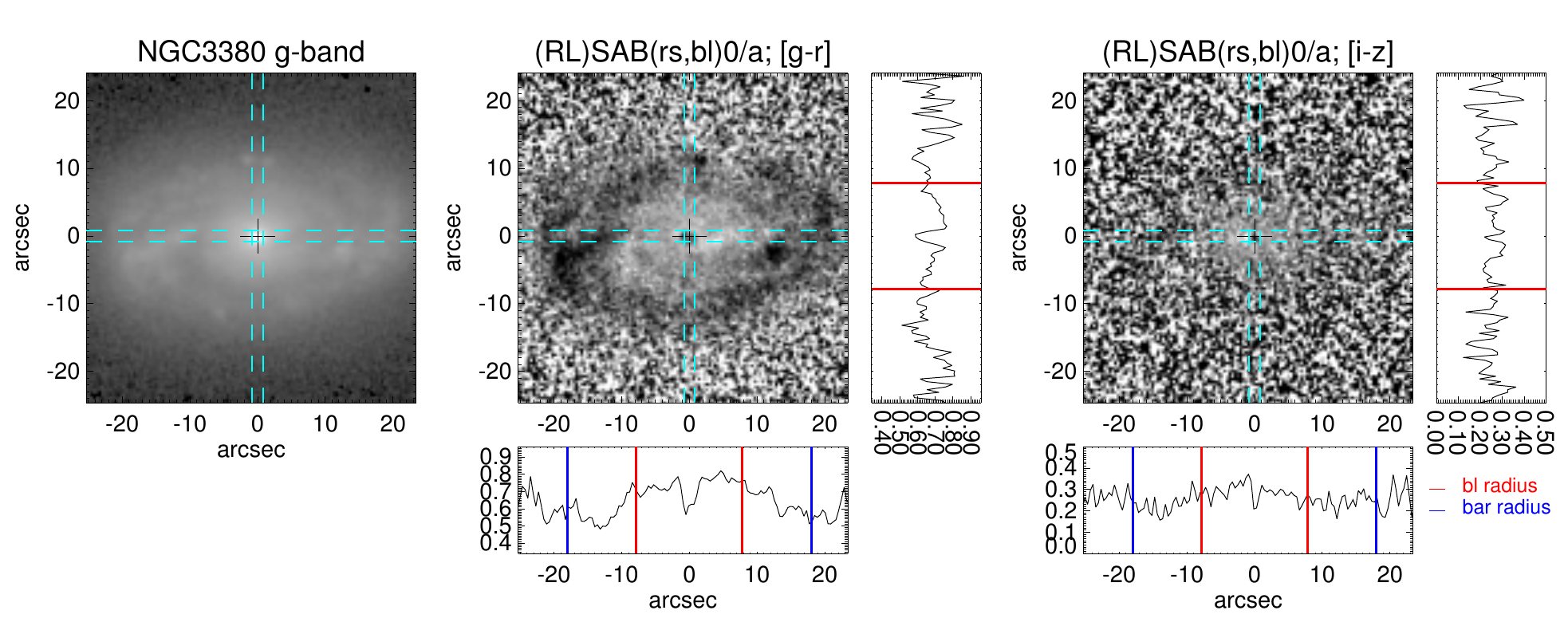}

\includegraphics[scale=0.45,angle=0]{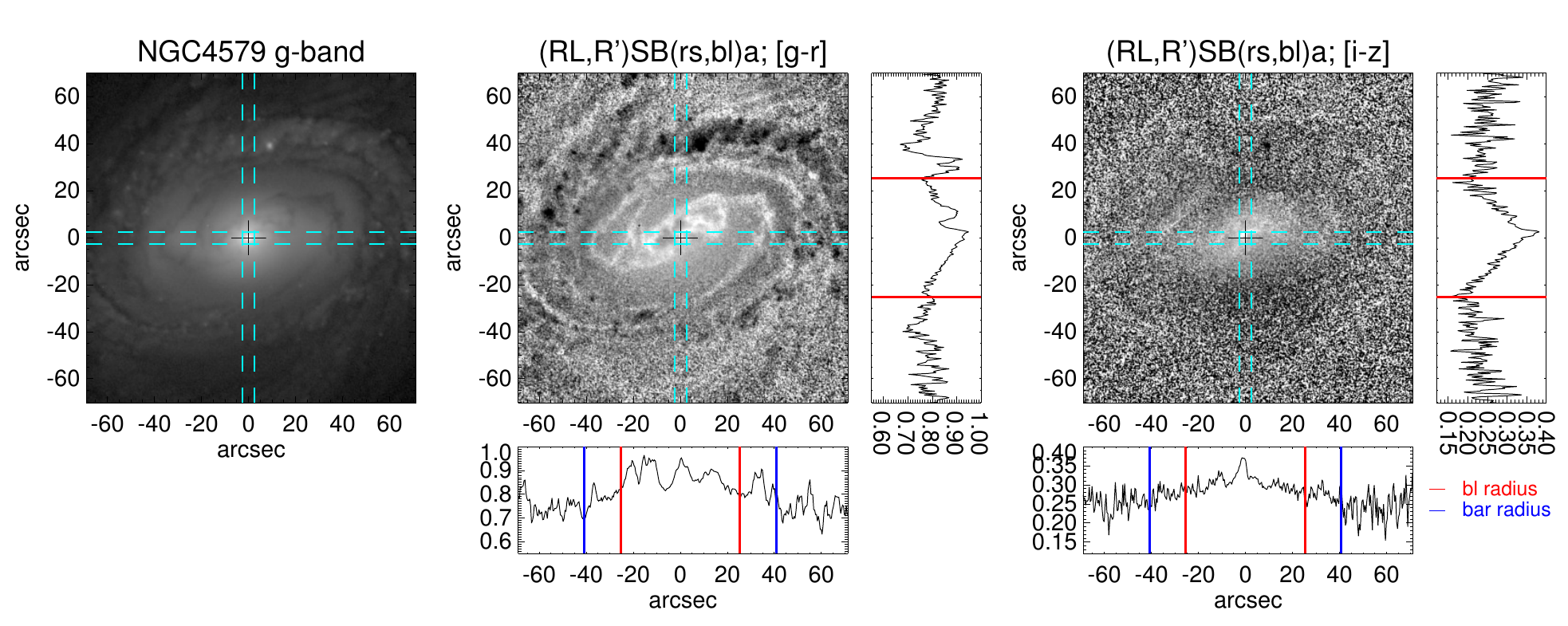}
 \caption{Examples of the different types of barlenses discussed in the
   text.  For each galaxy shown are: {\it g}-band image (left), {\it (g-r)} color
   map (middle), and {\it (i-z)} color map (right). For the color maps shown
   are also the color profiles along the bar major and minor axes,
   measured within the regions marked with dashed lines. The vertical
   lines in the profiles indicate the bar (blue) and the
   barlens (red) radii.  All the dimensions are in arcseconds,
   and the images are rotated so that the bar (measured from the 3.6
   or 2.2 $\mu$m image) is horizontally oriented. Similar plots for the rest of
   galaxies in the color subsample are shown in Appendix A.}
\label{bl_example_1}
\end{centering}
\end{figure}

\section{Discussion}

\subsection{Barlenses form part of the bar}

One of the main motivations of this study was to answer the question
whether the optical broadband colors support the idea that barlenses
form part of the bar, in a similar manner as the boxy/peanut bulges
do.  We have shown that the mean colors of the two bar components,
barlenses and the thin bar, are identical. The colors of the two bar
components also fall into the same regions in the color-color diagrams
studied by us, which means that they must have fairly similar
metallicities, {dust content, and dominant stellar
populations. However, one must keep in mind that the straight comparison
of measured colors with those predicted by models provides no clear, unambiguous
age information. If we look at the individual galaxies and the obtained
color differences between the two bar components we end up with the same
conclusion: no systematic deviations either to redder or bluer
colors appear in any of the color indices. The range of deviations is
extremely narrow in {\it (r-i)} and {\it (i-z)}, and somewhat broader in {\it (u-g)} and
{\it (g-r)}.  Obviously some of the barlenses are dusty and have a variable
contribution of young and intermediate aged stars, which naturally
explains the broadening. Our method for obtaining the colors is
robust, and does not depend on whether the region of the thin bar
within the barlens radius is included to the color measurement or
not. So the answer is that the colors obtained for the two bar
components in this study indeed are consistent with the idea that
barlenses form part of the bar. The main body of the stars in
barlenses were formed at the same time with the stars of the thin
bars. 
Whether the barlens structure also formed at the same time
with the rest of the bar is a matter of making detailed dynamical
models for these galaxies.

The colors do not answer the question whether barlenses are also thick
in the vertical direction. However, support for the vertical
thickening of barlenses comes from our analysis of their apparent
isophotal orientations with respect to the thin bars and the disks
line-of-nodes.  A comparison with the predictions of the theoretical
models also provide some insight to this question.  Formation of
barlenses in the simulation models have been discussed by \citet{atha2015}.
Their simulations use GADGET-3, combining stellar
N-body with SPH methods, which makes possible to model star formation
during the formation and evolution of the bar. In these models the
vertically thick inner bar components are protruded from the disk
material soon after the bar forms.  In the edge-on view they are
boxy/peanuts or X-shapes structures, and in face-on view have the
appearance of barlenses.  No bulges were added to these models while
starting the simulations, while a central concentration could grow
during the simulation via bar induced gas inflow and star
formation. More details of these models are explained by \citet{atha2013b}.
For several galaxies in our sample (NGC 1022, NGC 1512,
NGC 4314, NGC 4608, and NGC 5101) \citet{atha2015} suggested that
the K$_{\rm s}$-band or 3.6 $\mu$m profiles along the bar major axis can be
explained by their models without invoking any assumption of a pre-existing
classical bulge. Direct comparisons with the model profiles were done
for NGC 5101 in \citet[][their Fig. 3]{atha2015} and for NGC 4314
in \citet[][their Fig. 1]{lauri2014}. 
They observed that the surface brightness profile along the bar 
minor axis is exponential. Along the bar major axis the profile is less steep 
and has an exponential subsection associated to the barlens.
Similar bar profiles were obtained in the models by
\citeauthor{atha2015} although they were not specific models to those galaxies.

Although no predictions of specific colors of the structure components
are given in these models, they do predict that the oldest stars in
the vertically thin and thick bar components should have similar
ages. When bars evolve in their models they are rejuvenated by later
gas accretion and star formation which can be manifested throughout
the bar. This is in a qualitative agreement with the similar colors
obtained for the two bar components in this study, although the colors
still leave open the question what are actually the ages of the oldest
stars in these structures. Many of the barlenses in our sample also
have rejuvenated stellar populations and strong features of dust,
indicating that they have accreted gas also after the bar formation,
of which gas only some fraction has ended up to the central regions of
these galaxies.


\subsection{Classical bulges embedded in barlenses?}

{Bulges in S0s are generally thought to be massive, although
multi-component structural decompositions have reduced the mass
estimates \citep{lauri2005,lauri2010,lauri2014,erwin2015}.  However,
we would still expect to see the most massive classical bulges in the
S0s.  Having this in mind the different groups of the barlens color
profiles studied in this work are interesting. A typical profile type
is flat along the bar major axis, showing no evidence of a massive
(classical) bulge redder than the bar itself. In one of the barlens
groups the barlens is even covered by irregular dusty features, of
which NGC 2968 and NGC 1022 are good examples. Such dust features
might be manifestations of ongoing gas accretion to the barlenses.  In
the rest of the barlens groups the characteristic features appear in
the central regions of the galaxies, either associated directly with recent star
formation, or manifesting themselves in the form of small red central peaks.  In one the
groups the red central peaks are due to red nuclear bars, but
there are also barlenses in which the red central peaks are rather
unresolved dusty nuclei, or small central concentrations of the old
stellar populations.

Our study does not support the idea that massive classical bulges
appear in S0s, but it does not rule out either the possibility that small
bulges other than barlenses appear in these galaxies.  As colors alone
are not sufficient to distinguish stellar population ages, detailed
stellar population analysis for individual galaxies, combined with
simulation models, is needed for retrieving the star formation histories
of barlenses.}

\section{Summary and conclusions}

We have analyzed a sample of 79 barlens galaxies that appear in the
classification by \citet{buta2015} in the {\it Spitzer} Survey of Stellar
Structure in Galaxies \citep[S$^4$G;][]{sheth2010}, and in the Near-IR
S0 galaxy Survey by \citet[][NIRS0S]{lauri2011}. The infrared
images of these surveys were used to measure the sizes and orientations
of barlenses. The conclusions made from the study of these parameters
are the following:

\vspace{0.15cm}
\noindent {\it Barlens orientations} were compared with the
orientations of the thin bars and the lines-of-nodes.  We showed that
the region where barlens galaxies appear in the $|$PA$_{\rm
  bar}$ - PA$_{\rm disk}|$ vs. $|$PA$_{\rm bl}$ - PA$_{\rm disk}|$ plane is
as expected if they were vertically thick components embedded in a
flat disk. In galaxies seen in nearly face-on view
($inc <$ 40$^\circ$) barlenses are often aligned with the
thin bar indicating small intrinsic elongation along the thin bar.
\vspace{0.15cm}

\noindent {\it The size and orientation measurements of barlenses}
obtained in this study are consistent with those previously made by
\citet{herrera2015} for the same galaxies.
\vspace{0.2cm}

For a subset of 47 galaxies optical colors using the
Sloan Digital Sky Survey (SDSS) images at {\it u, g, r, i}, and {\it z}-bands 
from \citet{knapen2014}, were studied.
The
colors of the different structure components, including the two bar
components (barlens and thin bar), the disks, and the central
regions of the galaxies were measured. Color maps and color-color diagrams were also constructed.
Based on the color profiles along the bar major axis, the galaxies
were divided into different groups.  
The following conclusions
were made:
\vspace{0.2cm}

\noindent {\it Barlenses vs. thin bars:} the two bar components were
found to have very similar colors in all the color index images
studied here.  It means that their stellar populations,
metallicities, and the values of internal galactic extinction must on
average be fairly similar. 
Thus, the very similar 
colors that we found for the barlenses and the thin bars, are a manifestation that
barlenses indeed form part of bars, in a similar manner as boxy/peanut
bulges. 
The mean colors at {\it (u-g), (g-r), (r-i)} and
{\it (i-z)} are also similar to the mean colors of normal elliptical
galaxies. Galaxy by galaxy studies further showed that
galaxies which have irregular features related to rejuvenated stellar
populations in barlenses do exist.
\vspace{0.15cm}

\noindent {\it Barlenses vs. central peaks:} 
except for the {\it (r-i)} color index, the central peaks are
slightly shifted toward redder colors in all of our color-color
diagrams.
This can be due to internal galactic extinction or higher
metallicities, but based on the stellar population grids used by us no
unambiguous explanations for these redder colors could be
obtained. Many of the barlenses 
(11, accounting for $\sim$ 23 $\%$ of the color subsample)
also have central star forming nuclear
rings or redder nuclear bars.
\vspace{0.15cm}

\noindent {\it Barlenses vs. the underlying disks:} disks are
systematically bluer than bars (or barlenses) in all the color indices
studied here, except in {\it (r-i)}. The difference is best visible in 
the color-color diagrams {\it (r-i)} vs. {\it (g-r)} 
and {\it (i-z)} vs. {\it (r-i)}.
\vspace{0.15cm}

\begin{acknowledgements}

The authors thank the anonymous referee for comments that improved this paper.
The authors acknowledge financial support from to the DAGAL network
   from the People Programme (Marie Curie Actions) of the European
   Union's Seventh Framework Programme FP7/2007-2013 under REA grant
   agreement number PITN-GA-2011-289313. Special acknowledgment for
   the S4G-team (PI Kartik Sheth) for making this database available
   for us.  We also acknowledge NTT at ESO, as well as NOT and WHT in
   La Palma, where the K$_{\rm s}$-band images used in this study were originally
   obtained.  EL and HS acknowledge financial support from the Academy
   of Finland. JHK acknowledges financial support from the Spanish Ministry of
Economy and Competitiveness (MINECO) under grant number
AYA2013-41243-P, and thanks the Astrophysics Research Institute of
Liverpool John Moores University for their hospitality, and the
Spanish Ministry of Education, Culture and Sports for financial
support of his visit there, through grant number PR2015-00512.
MHE also acknowledges Santiago Erroz Ferrer,
S\'ebastien Comer\'on and Sim\'on D\'iaz Garc\'ia 
for useful discussions.

\end{acknowledgements}

\bibliographystyle{aa} 
\bibliography{references2} 

\begin{thebibliography}{56}
\expandafter\ifx\csname natexlab\endcsname\relax\def\natexlab#1{#1}\fi

\bibitem[{{Athanassoula}(2005)}]{atha2005}
{Athanassoula}, E. 2005, \mnras, 358, 1477

\bibitem[{{Athanassoula} \& {Beaton}(2006)}]{atha2006}
{Athanassoula}, E. \& {Beaton}, R.~L. 2006, \mnras, 370, 1499

\bibitem[{{Athanassoula} {et~al.}(2015){Athanassoula}, {Laurikainen}, {Salo},
  \& {Bosma}}]{atha2015}
{Athanassoula}, E., {Laurikainen}, E., {Salo}, H., \& {Bosma}, A. 2015, \mnras,
  454, 3843

\bibitem[{{Athanassoula} {et~al.}(2013){Athanassoula}, {Machado}, \&
  {Rodionov}}]{atha2013b}
{Athanassoula}, E., {Machado}, R.~E.~G., \& {Rodionov}, S.~A. 2013, \mnras,
  429, 1949

\bibitem[{{Balcells} \& {Peletier}(1994)}]{balcells1994}
{Balcells}, M. \& {Peletier}, R.~F. 1994, \aj, 107, 135

\bibitem[{{Bergvall} {et~al.}(2010){Bergvall}, {Zackrisson}, \&
  {Caldwell}}]{bergvall2010}
{Bergvall}, N., {Zackrisson}, E., \& {Caldwell}, B. 2010, \mnras, 405, 2697

\bibitem[{{Bettoni} \& {Galletta}(1994)}]{bettoni1994}
{Bettoni}, D. \& {Galletta}, G. 1994, \aap, 281, 1

\bibitem[{{Bruzual} \& {Charlot}(2003)}]{bruzual2003}
{Bruzual}, G. \& {Charlot}, S. 2003, \mnras, 344, 1000

\bibitem[{{Bureau} {et~al.}(2006){Bureau}, {Aronica}, {Athanassoula},
  {Dettmar}, {Bosma}, \& {Freeman}}]{bureau2006}
{Bureau}, M., {Aronica}, G., {Athanassoula}, E., {et~al.} 2006, \mnras, 370,
  753

\bibitem[{{Buta} {et~al.}(2015){Buta}, {Sheth}, {Athanassoula}, {Bosma},
  {Knapen}, {Laurikainen}, {Salo}, {Elmegreen}, {Ho}, {Zaritsky}, {Courtois},
  {Hinz}, {Mu{\~n}oz-Mateos}, {Kim}, {Regan}, {Gadotti}, {Gil de Paz}, {Laine},
  {Men{\'e}ndez-Delmestre}, {Comer{\'o}n}, {Erroz Ferrer}, {Seibert},
  {Mizusawa}, {Holwerda}, \& {Madore}}]{buta2015}
{Buta}, R.~J., {Sheth}, K., {Athanassoula}, E., {et~al.} 2015, \apjs, 217, 32

\bibitem[{{Casado} {et~al.}(2015){Casado}, {Ascasibar}, {Gavil{\'a}n},
  {Terlevich}, {Terlevich}, {Hoyos}, \& {D{\'{\i}}az}}]{casado2015}
{Casado}, J., {Ascasibar}, Y., {Gavil{\'a}n}, M., {et~al.} 2015, \mnras, 451,
  888

\bibitem[{{Chabrier}(2003)}]{chabrier2003}
{Chabrier}, G. 2003, \pasp, 115, 763

\bibitem[{{Combes} {et~al.}(1990){Combes}, {Debbasch}, {Friedli}, \&
  {Pfenniger}}]{combes1990}
{Combes}, F., {Debbasch}, F., {Friedli}, D., \& {Pfenniger}, D. 1990, \aap,
  233, 82

\bibitem[{{Comer{\'o}n} {et~al.}(2010){Comer{\'o}n}, {Knapen}, {Beckman},
  {Laurikainen}, {Salo}, {Mart{\'{\i}}nez-Valpuesta}, \& {Buta}}]{comeron2010}
{Comer{\'o}n}, S., {Knapen}, J.~H., {Beckman}, J.~E., {et~al.} 2010, \mnras,
  402, 2462

\bibitem[{{Elmegreen} \& {Elmegreen}(1985)}]{elmegreen1985}
{Elmegreen}, B.~G. \& {Elmegreen}, D.~M. 1985, \apj, 288, 438

\bibitem[{{Erwin} \& {Debattista}(2013)}]{erwin2013}
{Erwin}, P. \& {Debattista}, V.~P. 2013, \mnras, 431, 3060

\bibitem[{{Erwin} {et~al.}(2015){Erwin}, {Saglia}, {Fabricius}, {Thomas},
  {Nowak}, {Rusli}, {Bender}, {Vega Beltr{\'a}n}, \& {Beckman}}]{erwin2015}
{Erwin}, P., {Saglia}, R.~P., {Fabricius}, M., {et~al.} 2015, \mnras, 446, 4039

\bibitem[{{Erwin} \& {Sparke}(2003)}]{erwin2003}
{Erwin}, P. \& {Sparke}, L.~S. 2003, \apjs, 146, 299

\bibitem[{{Fazio} {et~al.}(2004){Fazio}, {Hora}, {Allen}, {Ashby}, {Barmby},
  {Deutsch}, {Huang}, {Kleiner}, {Marengo}, {Megeath}, {Melnick}, {Pahre},
  {Patten}, {Polizotti}, {Smith}, {Taylor}, {Wang}, {Willner}, {Hoffmann},
  {Pipher}, {Forrest}, {McMurty}, {McCreight}, {McKelvey}, {McMurray}, {Koch},
  {Moseley}, {Arendt}, {Mentzell}, {Marx}, {Losch}, {Mayman}, {Eichhorn},
  {Krebs}, {Jhabvala}, {Gezari}, {Fixsen}, {Flores}, {Shakoorzadeh}, {Jungo},
  {Hakun}, {Workman}, {Karpati}, {Kichak}, {Whitley}, {Mann}, {Tollestrup},
  {Eisenhardt}, {Stern}, {Gorjian}, {Bhattacharya}, {Carey}, {Nelson},
  {Glaccum}, {Lacy}, {Lowrance}, {Laine}, {Reach}, {Stauffer}, {Surace},
  {Wilson}, {Wright}, {Hoffman}, {Domingo}, \& {Cohen}}]{fazio2004}
{Fazio}, G.~G., {Hora}, J.~L., {Allen}, L.~E., {et~al.} 2004, \apjs, 154, 10

\bibitem[{{Fisher} \& {Drory}(2016)}]{fisher2016}
{Fisher}, D.~B. \& {Drory}, N. 2016, in Astrophysics and Space Science Library,
  Vol. 418, Galactic Bulges, ed. {E. Laurikainen, R. Peletier, and D. Gadotti
  (Springer International Publishing)}, 41

\bibitem[{{Fukugita} {et~al.}(1996){Fukugita}, {Ichikawa}, {Gunn}, {Doi},
  {Shimasaku}, \& {Schneider}}]{fukugita1996}
{Fukugita}, M., {Ichikawa}, T., {Gunn}, J.~E., {et~al.} 1996, \aj, 111, 1748

\bibitem[{{Gadotti} \& {de Souza}(2006)}]{gadotti2006}
{Gadotti}, D.~A. \& {de Souza}, R.~E. 2006, \apjs, 163, 270

\bibitem[{{Gunn} {et~al.}(1998){Gunn}, {Carr}, {Rockosi}, {Sekiguchi}, {Berry},
  {Elms}, {de Haas}, {Ivezi{\'c}}, {Knapp}, {Lupton}, {Pauls}, {Simcoe},
  {Hirsch}, {Sanford}, {Wang}, {York}, {Harris}, {Annis}, {Bartozek},
  {Boroski}, {Bakken}, {Haldeman}, {Kent}, {Holm}, {Holmgren}, {Petravick},
  {Prosapio}, {Rechenmacher}, {Doi}, {Fukugita}, {Shimasaku}, {Okada}, {Hull},
  {Siegmund}, {Mannery}, {Blouke}, {Heidtman}, {Schneider}, {Lucinio}, \&
  {Brinkman}}]{gunn1998}
{Gunn}, J.~E., {Carr}, M., {Rockosi}, C., {et~al.} 1998, \aj, 116, 3040

\bibitem[{{Herrera-Endoqui} {et~al.}(2015){Herrera-Endoqui},
  {D{\'{\i}}az-Garc{\'{\i}}a}, {Laurikainen}, \& {Salo}}]{herrera2015}
{Herrera-Endoqui}, M., {D{\'{\i}}az-Garc{\'{\i}}a}, S., {Laurikainen}, E., \&
  {Salo}, H. 2015, \aap, 582, A86

\bibitem[{{Kennicutt}(1992)}]{kennicutt1992}
{Kennicutt}, Jr., R.~C. 1992, \apjs, 79, 255

\bibitem[{{Knapen} {et~al.}(2014){Knapen}, {Erroz-Ferrer}, {Roa}, {Bakos},
  {Cisternas}, {Leaman}, \& {Szymanek}}]{knapen2014}
{Knapen}, J.~H., {Erroz-Ferrer}, S., {Roa}, J., {et~al.} 2014, \aap, 569, A91

\bibitem[{{Kormendy}(1983)}]{kormendy1983}
{Kormendy}, J. 1983, \apj, 275, 529

\bibitem[{{Kormendy}(2016)}]{kormendy2016}
{Kormendy}, J. 2016, in Astrophysics and Space Science Library, Vol. 418,
  Galactic Bulges, ed. {E. Laurikainen, R. Peletier, and D. Gadotti (Springer
  International Publishing)}, 431

\bibitem[{{Kurucz}(1991)}]{kurucz1991}
{Kurucz}, R.~L. 1991, in Precision Photometry: Astrophysics of the Galaxy, ed.
  A.~G.~D. {Philip}, A.~R. {Upgren}, \& K.~A. {Janes}, 27

\bibitem[{{Laurikainen} \& {Salo}(2016)}]{laurisalo2016}
{Laurikainen}, E. \& {Salo}, H. 2016, in Astrophysics and Space Science
  Library, Vol. 418, Galactic Bulges, ed. {E. Laurikainen, R. Peletier, and D.
  Gadotti (Springer International Publishing)}, 77

\bibitem[{{Laurikainen} {et~al.}(2014){Laurikainen}, {Salo}, {Athanassoula},
  {Bosma}, \& {Herrera-Endoqui}}]{lauri2014}
{Laurikainen}, E., {Salo}, H., {Athanassoula}, E., {Bosma}, A., \&
  {Herrera-Endoqui}, M. 2014, \mnras, 444, L80

\bibitem[{{Laurikainen} {et~al.}(2005){Laurikainen}, {Salo}, \&
  {Buta}}]{lauri2005}
{Laurikainen}, E., {Salo}, H., \& {Buta}, R. 2005, \mnras, 362, 1319

\bibitem[{{Laurikainen} {et~al.}(2011){Laurikainen}, {Salo}, {Buta}, \&
  {Knapen}}]{lauri2011}
{Laurikainen}, E., {Salo}, H., {Buta}, R., \& {Knapen}, J.~H. 2011, \mnras,
  418, 1452

\bibitem[{{Laurikainen} {et~al.}(2010){Laurikainen}, {Salo}, {Buta}, {Knapen},
  \& {Comer{\'o}n}}]{lauri2010}
{Laurikainen}, E., {Salo}, H., {Buta}, R., {Knapen}, J.~H., \& {Comer{\'o}n},
  S. 2010, \mnras, 405, 1089

\bibitem[{{Le Borgne} {et~al.}(2003){Le Borgne}, {Bruzual}, {Pell{\'o}},
  {Lan{\c c}on}, {Rocca-Volmerange}, {Sanahuja}, {Schaerer}, {Soubiran}, \&
  {V{\'{\i}}lchez-G{\'o}mez}}]{leborgne2003}
{Le Borgne}, J.-F., {Bruzual}, G., {Pell{\'o}}, R., {et~al.} 2003, \aap, 402,
  433

\bibitem[{{Lenz} {et~al.}(1998){Lenz}, {Newberg}, {Rosner}, {Richards}, \&
  {Stoughton}}]{lenz1998}
{Lenz}, D.~D., {Newberg}, J., {Rosner}, R., {Richards}, G.~T., \& {Stoughton},
  C. 1998, \apjs, 119, 121

\bibitem[{{L{\"u}tticke} {et~al.}(2000){L{\"u}tticke}, {Dettmar}, \&
  {Pohlen}}]{lutticke2000a}
{L{\"u}tticke}, R., {Dettmar}, R.-J., \& {Pohlen}, M. 2000, \aaps, 145, 405

\bibitem[{{Makarov} {et~al.}(2014){Makarov}, {Prugniel}, {Terekhova},
  {Courtois}, \& {Vauglin}}]{makarov2014}
{Makarov}, D., {Prugniel}, P., {Terekhova}, N., {Courtois}, H., \& {Vauglin},
  I. 2014, \aap, 570, A13

\bibitem[{{McIntosh} {et~al.}(2014){McIntosh}, {Wagner}, {Cooper}, {Bell},
  {Kere{\v s}}, {Bosch}, {Gallazzi}, {Haines}, {Mann}, {Pasquali}, \&
  {Christian}}]{mcintosh2014}
{McIntosh}, D.~H., {Wagner}, C., {Cooper}, A., {et~al.} 2014, \mnras, 442, 533

\bibitem[{{M{\'e}ndez-Abreu} {et~al.}(2008){M{\'e}ndez-Abreu}, {Corsini},
  {Debattista}, {De Rijcke}, {Aguerri}, \& {Pizzella}}]{mendezabreu2008}
{M{\'e}ndez-Abreu}, J., {Corsini}, E.~M., {Debattista}, V.~P., {et~al.} 2008,
  \apjl, 679, L73

\bibitem[{{M{\'e}ndez-Abreu} {et~al.}(2014){M{\'e}ndez-Abreu}, {Debattista},
  {Corsini}, \& {Aguerri}}]{mendezabreu2014}
{M{\'e}ndez-Abreu}, J., {Debattista}, V.~P., {Corsini}, E.~M., \& {Aguerri},
  J.~A.~L. 2014, \aap, 572, A25

\bibitem[{{Mu{\~n}oz-Mateos} {et~al.}(2009){Mu{\~n}oz-Mateos}, {Gil de Paz},
  {Zamorano}, {Boissier}, {Dale}, {P{\'e}rez-Gonz{\'a}lez}, {Gallego},
  {Madore}, {Bendo}, {Boselli}, {Buat}, {Calzetti}, {Moustakas}, \&
  {Kennicutt}}]{munoz2009}
{Mu{\~n}oz-Mateos}, J.~C., {Gil de Paz}, A., {Zamorano}, J., {et~al.} 2009,
  \apj, 703, 1569

\bibitem[{{Mu{\~n}oz-Mateos} {et~al.}(2015){Mu{\~n}oz-Mateos}, {Sheth},
  {Regan}, {Kim}, {Laine}, {Erroz-Ferrer}, {Gil de Paz}, {Comeron}, {Hinz},
  {Laurikainen}, {Salo}, {Athanassoula}, {Bosma}, {Bouquin}, {Schinnerer},
  {Ho}, {Zaritsky}, {Gadotti}, {Madore}, {Holwerda}, {Men{\'e}ndez-Delmestre},
  {Knapen}, {Meidt}, {Querejeta}, {Mizusawa}, {Seibert}, {Laine}, \&
  {Courtois}}]{munoz2015}
{Mu{\~n}oz-Mateos}, J.~C., {Sheth}, K., {Regan}, M., {et~al.} 2015, \apjs, 219,
  3

\bibitem[{{Okamura} \& {Takase}(1976)}]{okamura1976}
{Okamura}, S. \& {Takase}, B. 1976, \apss, 41, 275

\bibitem[{{Pfenniger} \& {Friedli}(1991)}]{pfenniger1991}
{Pfenniger}, D. \& {Friedli}, D. 1991, \aap, 252, 75

\bibitem[{{Pohlen} \& {Trujillo}(2006)}]{pohlen2006}
{Pohlen}, M. \& {Trujillo}, I. 2006, \aap, 454, 759

\bibitem[{{Raha} {et~al.}(1991){Raha}, {Sellwood}, {James}, \&
  {Kahn}}]{raha1991}
{Raha}, N., {Sellwood}, J.~A., {James}, R.~A., \& {Kahn}, F.~D. 1991, \nat,
  352, 411

\bibitem[{{Salo} {et~al.}(2015){Salo}, {Laurikainen}, {Laine}, {Comer{\'o}n},
  {Gadotti}, {Buta}, {Sheth}, {Zaritsky}, {Ho}, {Knapen}, {Athanassoula},
  {Bosma}, {Laine}, {Cisternas}, {Kim}, {Mu{\~n}oz-Mateos}, {Regan}, {Hinz},
  {Gil de Paz}, {Menendez-Delmestre}, {Mizusawa}, {Erroz-Ferrer}, {Meidt}, \&
  {Querejeta}}]{salo2015}
{Salo}, H., {Laurikainen}, E., {Laine}, J., {et~al.} 2015, \apjs, 219, 4

\bibitem[{{Schlafly} \& {Finkbeiner}(2011)}]{schlafly2011}
{Schlafly}, E.~F. \& {Finkbeiner}, D.~P. 2011, \apj, 737, 103

\bibitem[{{Schlegel} {et~al.}(1998){Schlegel}, {Finkbeiner}, \&
  {Davis}}]{schlegel1998}
{Schlegel}, D.~J., {Finkbeiner}, D.~P., \& {Davis}, M. 1998, \apj, 500, 525

\bibitem[{{Sheth} {et~al.}(2010){Sheth}, {Regan}, {Hinz}, {Gil de Paz},
  {Men{\'e}ndez-Delmestre}, {Mu{\~n}oz-Mateos}, {Seibert}, {Kim},
  {Laurikainen}, {Salo}, {Gadotti}, {Laine}, {Mizusawa}, {Armus},
  {Athanassoula}, {Bosma}, {Buta}, {Capak}, {Jarrett}, {Elmegreen},
  {Elmegreen}, {Knapen}, {Koda}, {Helou}, {Ho}, {Madore}, {Masters},
  {Mobasher}, {Ogle}, {Peng}, {Schinnerer}, {Surace}, {Zaritsky},
  {Comer{\'o}n}, {de Swardt}, {Meidt}, {Kasliwal}, \& {Aravena}}]{sheth2010}
{Sheth}, K., {Regan}, M., {Hinz}, J.~L., {et~al.} 2010, \pasp, 122, 1397

\bibitem[{{Shimasaku} {et~al.}(2001){Shimasaku}, {Fukugita}, {Doi}, {Hamabe},
  {Ichikawa}, {Okamura}, {Sekiguchi}, {Yasuda}, {Brinkmann}, {Csabai},
  {Ichikawa}, {Ivezi{\'c}}, {Kunszt}, {Schneider}, {Szokoly}, {Watanabe}, \&
  {York}}]{shimasaku2001}
{Shimasaku}, K., {Fukugita}, M., {Doi}, M., {et~al.} 2001, \aj, 122, 1238

\bibitem[{{Terlevich} \& {Forbes}(2002)}]{terlevich2002}
{Terlevich}, A.~I. \& {Forbes}, D.~A. 2002, \mnras, 330, 547

\bibitem[{{Weiland} {et~al.}(1994){Weiland}, {Arendt}, {Berriman}, {Dwek},
  {Freudenreich}, {Hauser}, {Kelsall}, {Lisse}, {Mitra}, {Moseley}, {Odegard},
  {Silverberg}, {Sodroski}, {Spiesman}, \& {Stemwedel}}]{weiland1994}
{Weiland}, J.~L., {Arendt}, R.~G., {Berriman}, G.~B., {et~al.} 1994, \apjl,
  425, L81

\bibitem[{{Williams} {et~al.}(2012){Williams}, {Bureau}, \&
  {Kuntschner}}]{williams2012}
{Williams}, M.~J., {Bureau}, M., \& {Kuntschner}, H. 2012, \mnras, 427, L99

\bibitem[{{York} {et~al.}(2000){York}, {Adelman}, {Anderson}, {Anderson},
  {Annis}, {Bahcall}, {Bakken}, {Barkhouser}, {Bastian}, {Berman}, {Boroski},
  {Bracker}, {Briegel}, {Briggs}, {Brinkmann}, {Brunner}, {Burles}, {Carey},
  {Carr}, {Castander}, {Chen}, {Colestock}, {Connolly}, {Crocker}, {Csabai},
  {Czarapata}, {Davis}, {Doi}, {Dombeck}, {Eisenstein}, {Ellman}, {Elms},
  {Evans}, {Fan}, {Federwitz}, {Fiscelli}, {Friedman}, {Frieman}, {Fukugita},
  {Gillespie}, {Gunn}, {Gurbani}, {de Haas}, {Haldeman}, {Harris}, {Hayes},
  {Heckman}, {Hennessy}, {Hindsley}, {Holm}, {Holmgren}, {Huang}, {Hull},
  {Husby}, {Ichikawa}, {Ichikawa}, {Ivezi{\'c}}, {Kent}, {Kim}, {Kinney},
  {Klaene}, {Kleinman}, {Kleinman}, {Knapp}, {Korienek}, {Kron}, {Kunszt},
  {Lamb}, {Lee}, {Leger}, {Limmongkol}, {Lindenmeyer}, {Long}, {Loomis},
  {Loveday}, {Lucinio}, {Lupton}, {MacKinnon}, {Mannery}, {Mantsch}, {Margon},
  {McGehee}, {McKay}, {Meiksin}, {Merelli}, {Monet}, {Munn}, {Narayanan},
  {Nash}, {Neilsen}, {Neswold}, {Newberg}, {Nichol}, {Nicinski}, {Nonino},
  {Okada}, {Okamura}, {Ostriker}, {Owen}, {Pauls}, {Peoples}, {Peterson},
  {Petravick}, {Pier}, {Pope}, {Pordes}, {Prosapio}, {Rechenmacher}, {Quinn},
  {Richards}, {Richmond}, {Rivetta}, {Rockosi}, {Ruthmansdorfer}, {Sandford},
  {Schlegel}, {Schneider}, {Sekiguchi}, {Sergey}, {Shimasaku}, {Siegmund},
  {Smee}, {Smith}, {Snedden}, {Stone}, {Stoughton}, {Strauss}, {Stubbs},
  {SubbaRao}, {Szalay}, {Szapudi}, {Szokoly}, {Thakar}, {Tremonti}, {Tucker},
  {Uomoto}, {Vanden Berk}, {Vogeley}, {Waddell}, {Wang}, {Watanabe},
  {Weinberg}, {Yanny}, {Yasuda}, \& {SDSS Collaboration}}]{york2000}
{York}, D.~G., {Adelman}, J., {Anderson}, Jr., J.~E., {et~al.} 2000, \aj, 120,
  1579

\end{thebibliography}

\clearpage

\begin{appendix}
\section{Color maps and color profiles of the barlens galaxies in the color subsample}
In this appendix we include similar plots as those presented in Fig. 10 
for the complete barlens color subsample. For each galaxy we show the {\it g}-band image 
in the bar region (left panel), the {\it (g-r)} color map and color profiles both along
the major and minor axes of the bar (middle panels), and the corresponding {\it (i-z)} color
map and profiles (right panels).

\noindent
\includegraphics[scale=0.45]{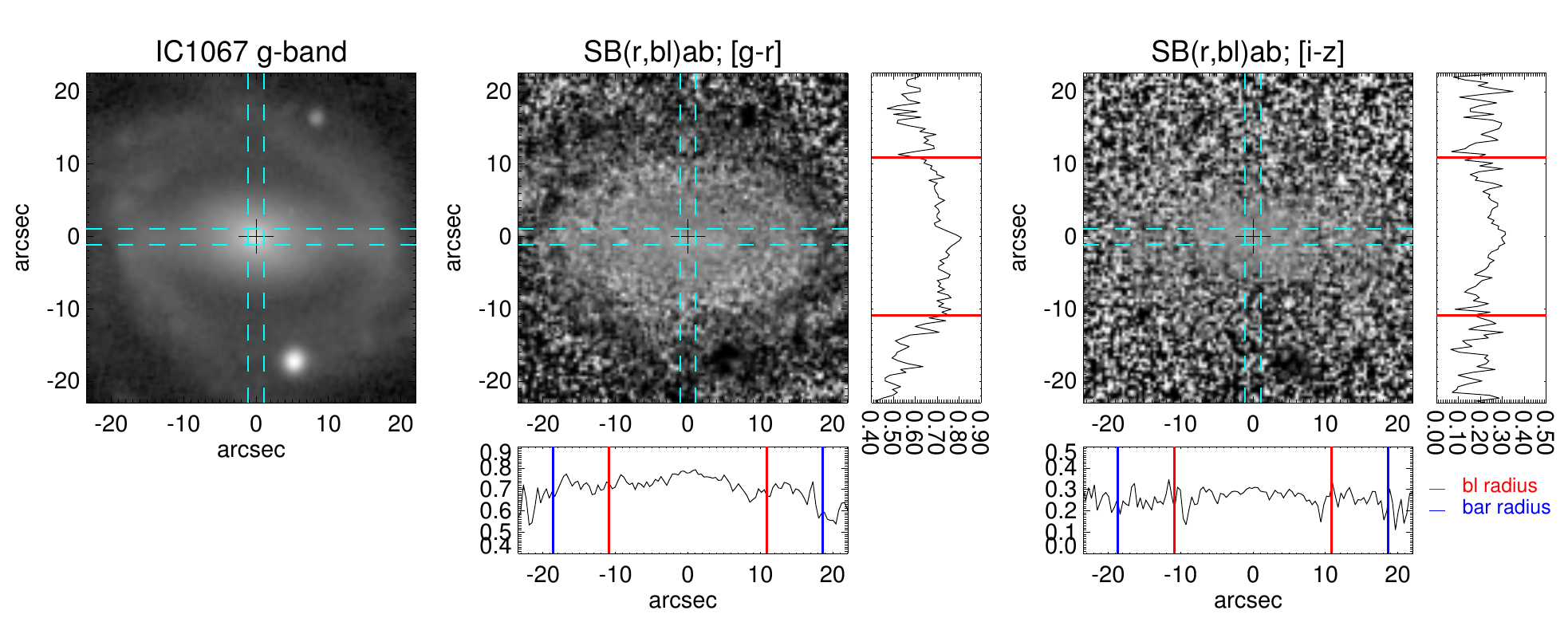}
\includegraphics[scale=0.45]{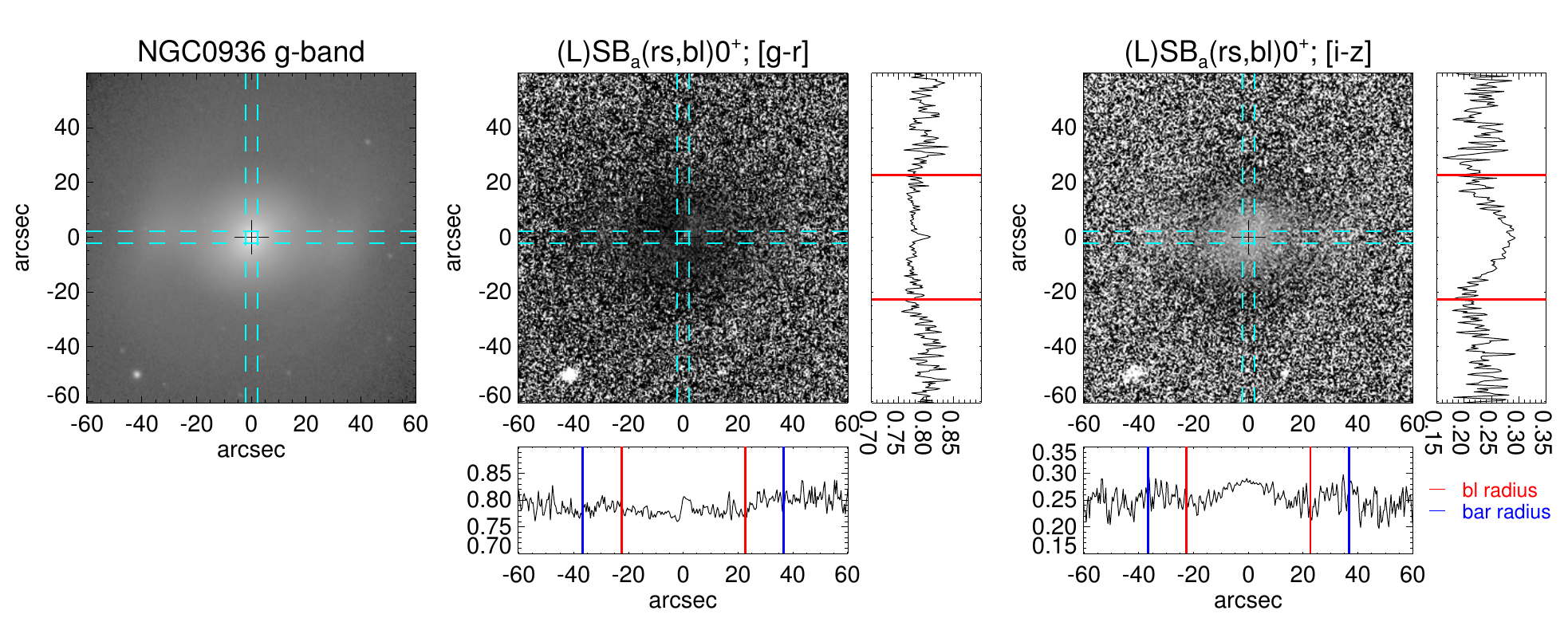}
\includegraphics[scale=0.45]{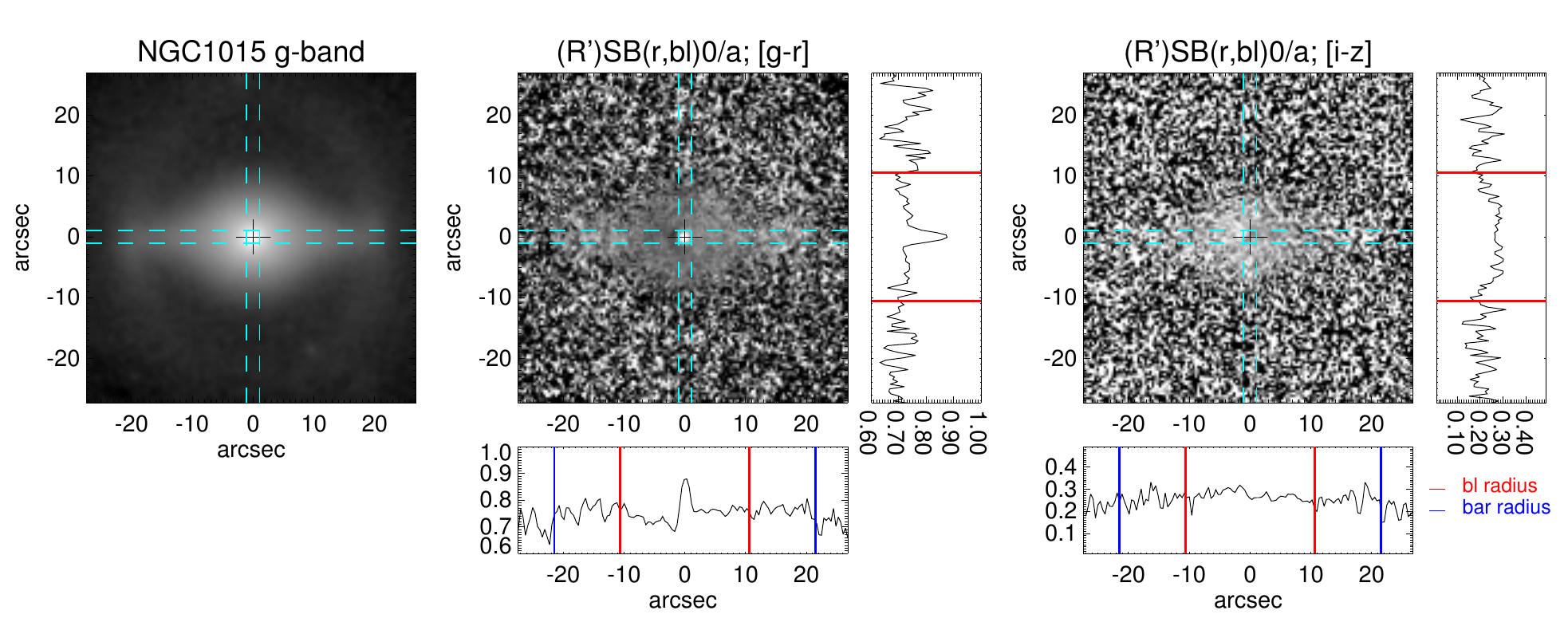}
\includegraphics[scale=0.45]{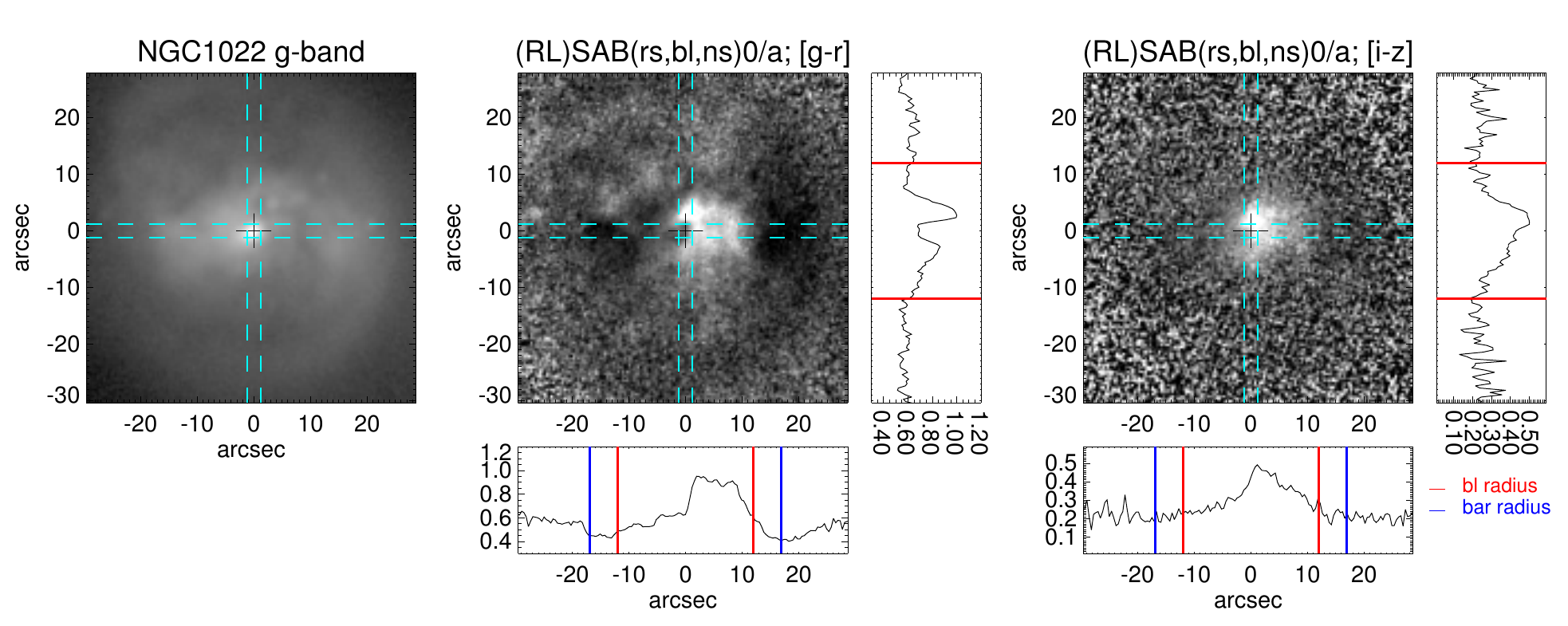}
\includegraphics[scale=0.45]{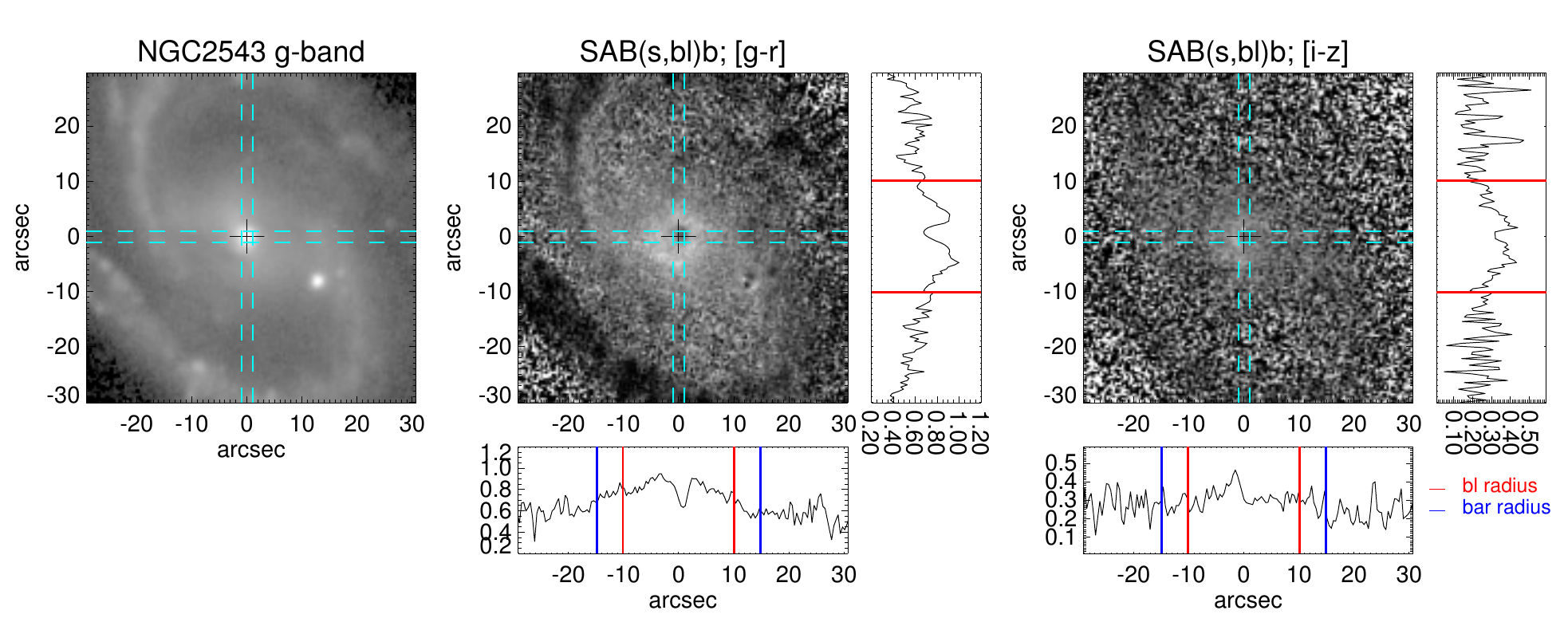}
\includegraphics[scale=0.45]{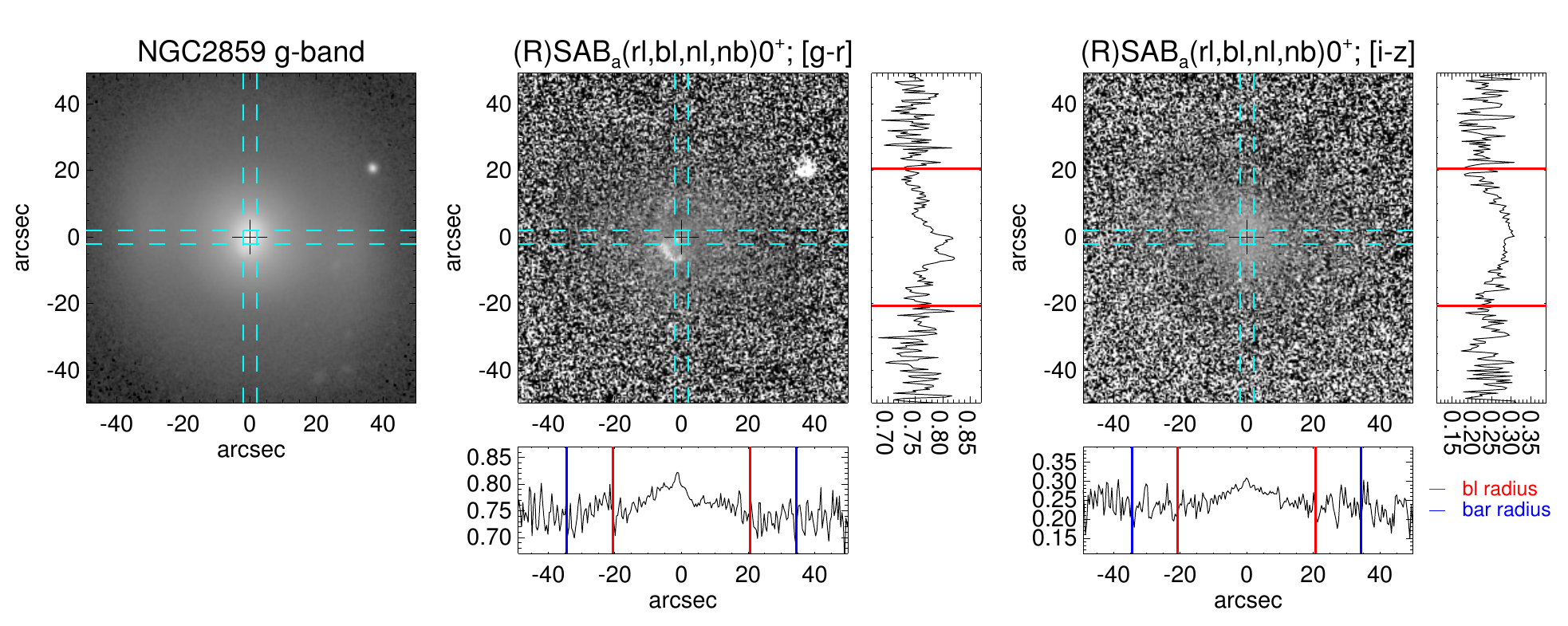}
\includegraphics[scale=0.45]{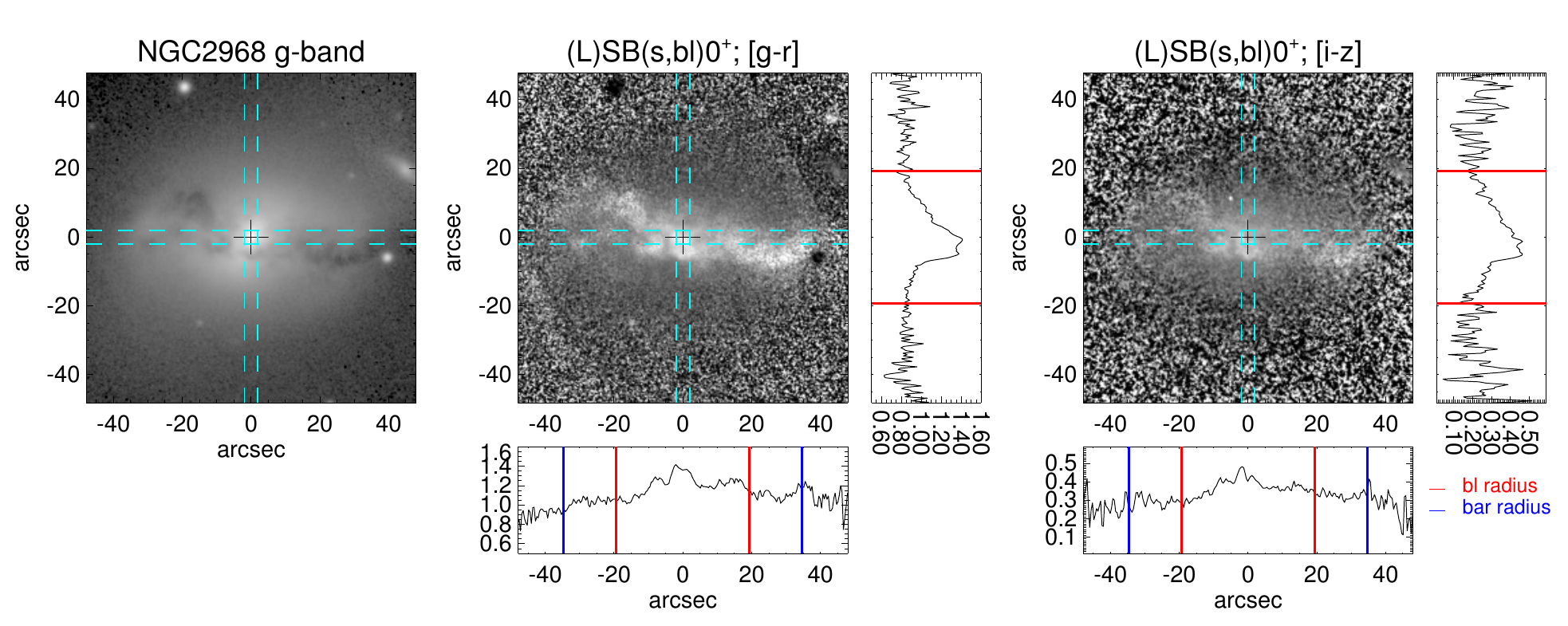}
\includegraphics[scale=0.45]{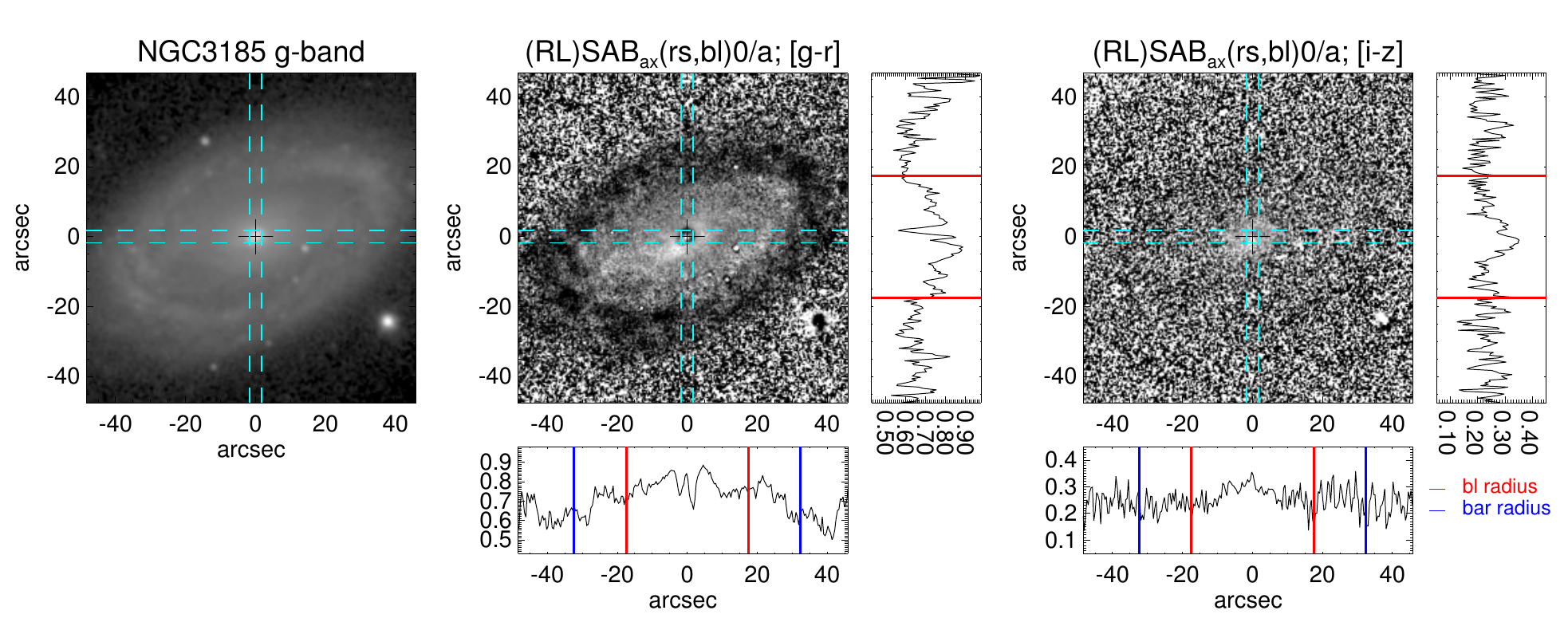}
\includegraphics[scale=0.45]{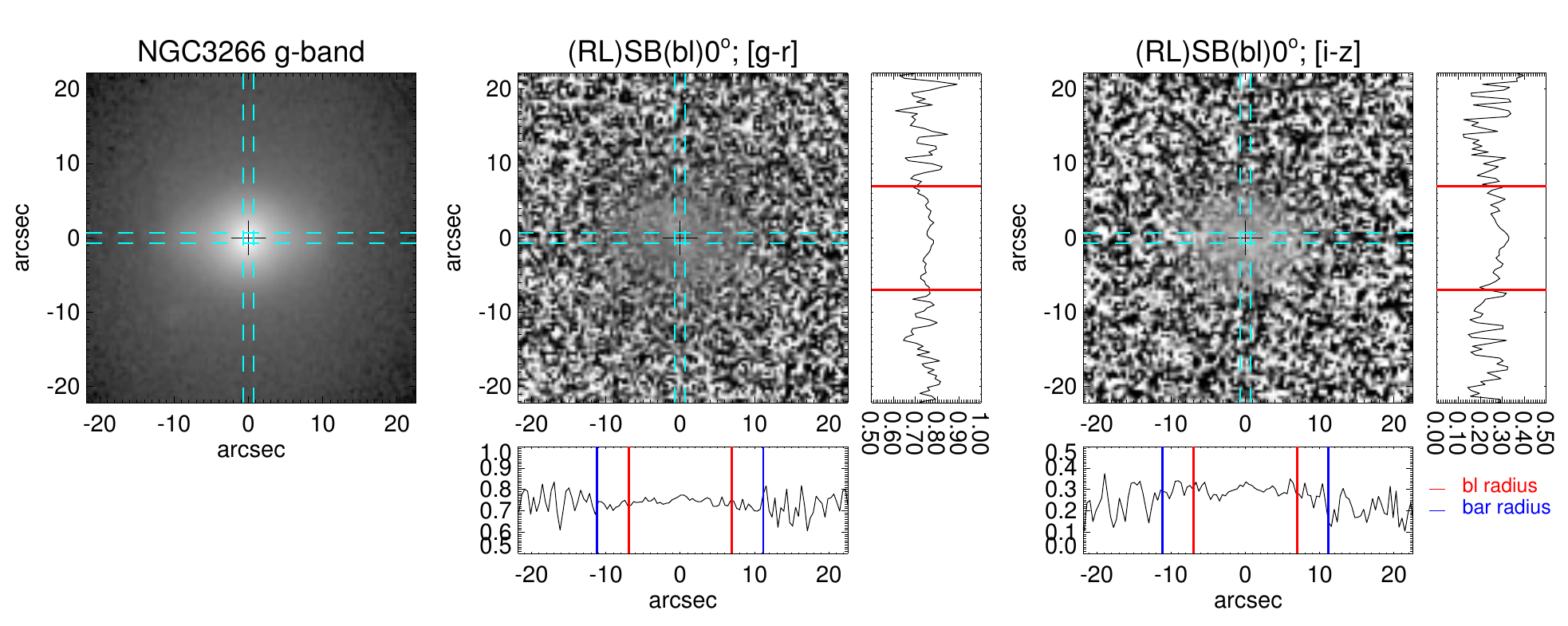}
\includegraphics[scale=0.45]{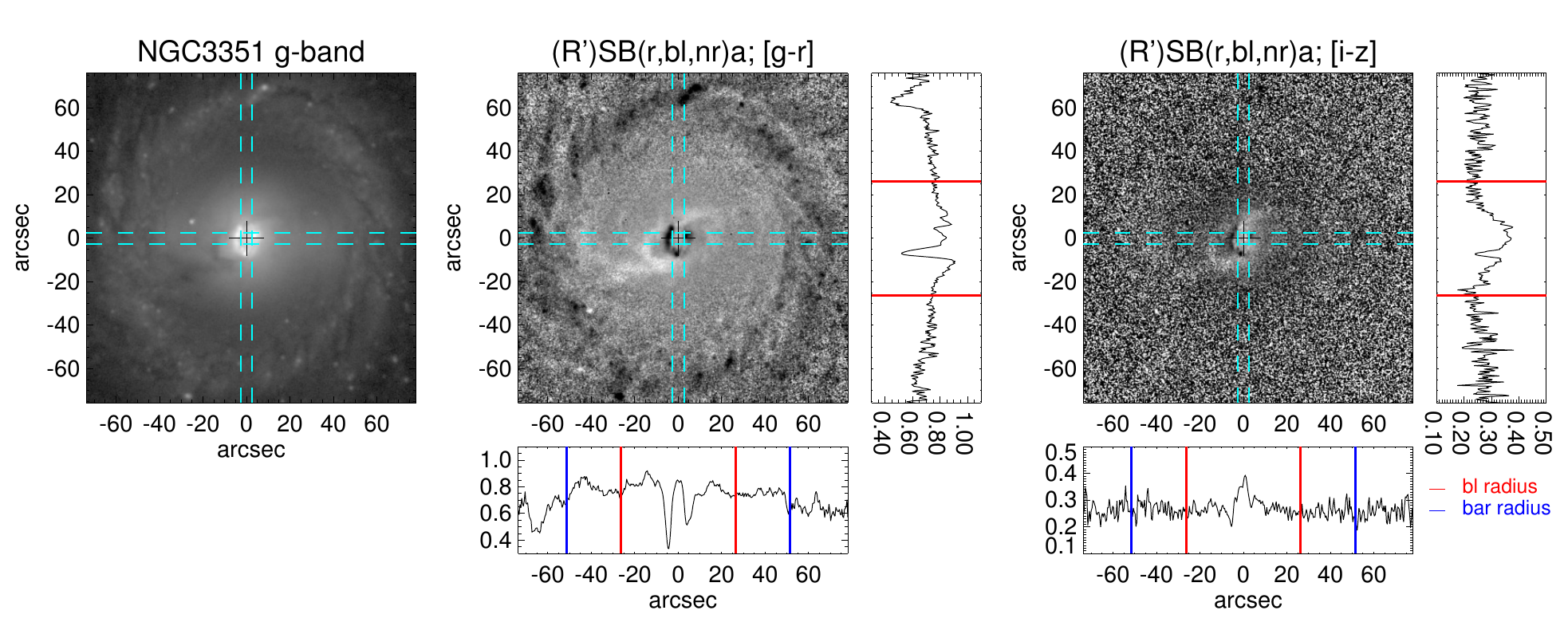}
\includegraphics[scale=0.45]{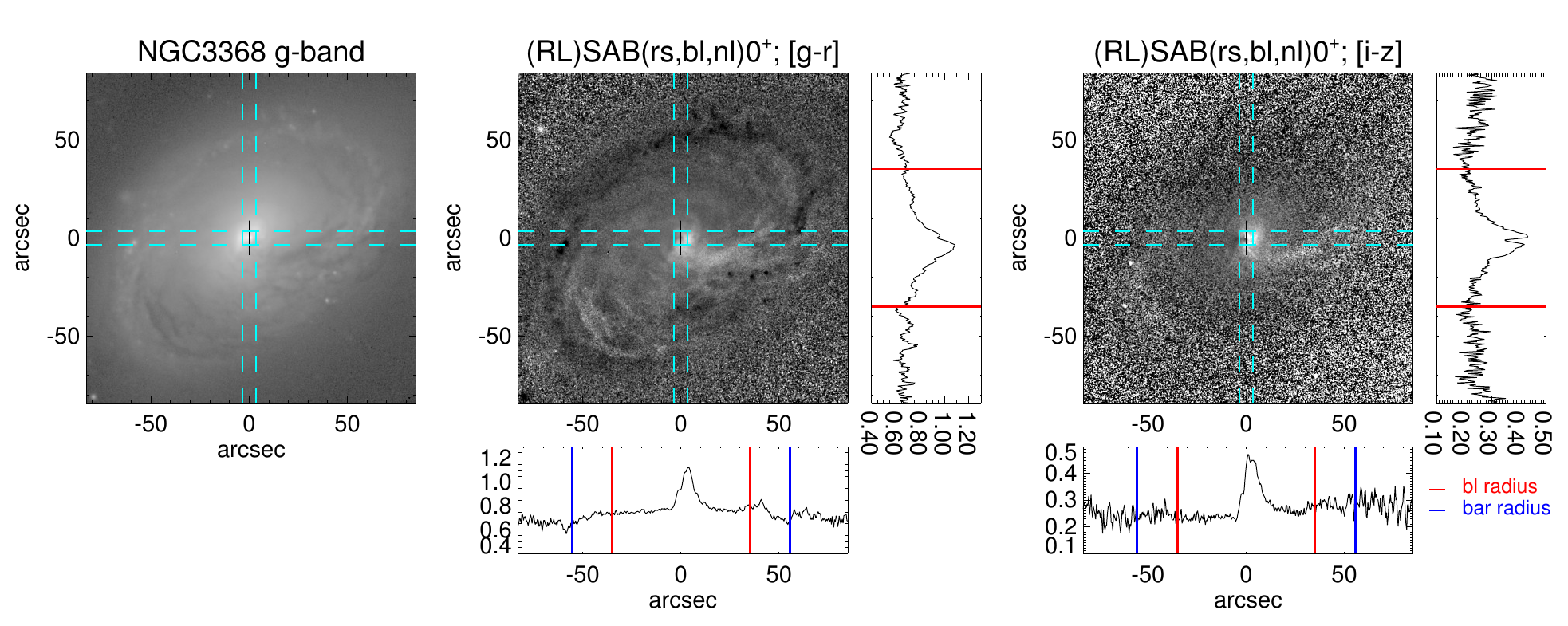}
\includegraphics[scale=0.45]{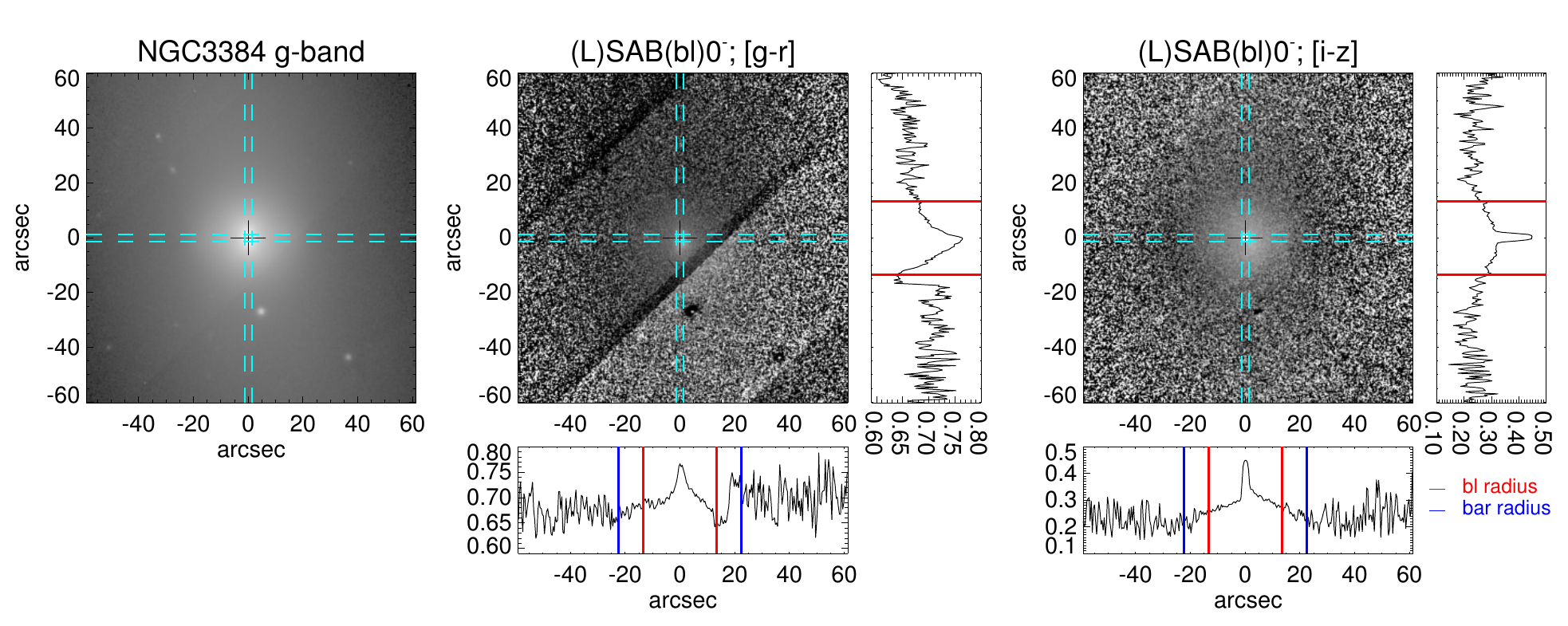}
\includegraphics[scale=0.45]{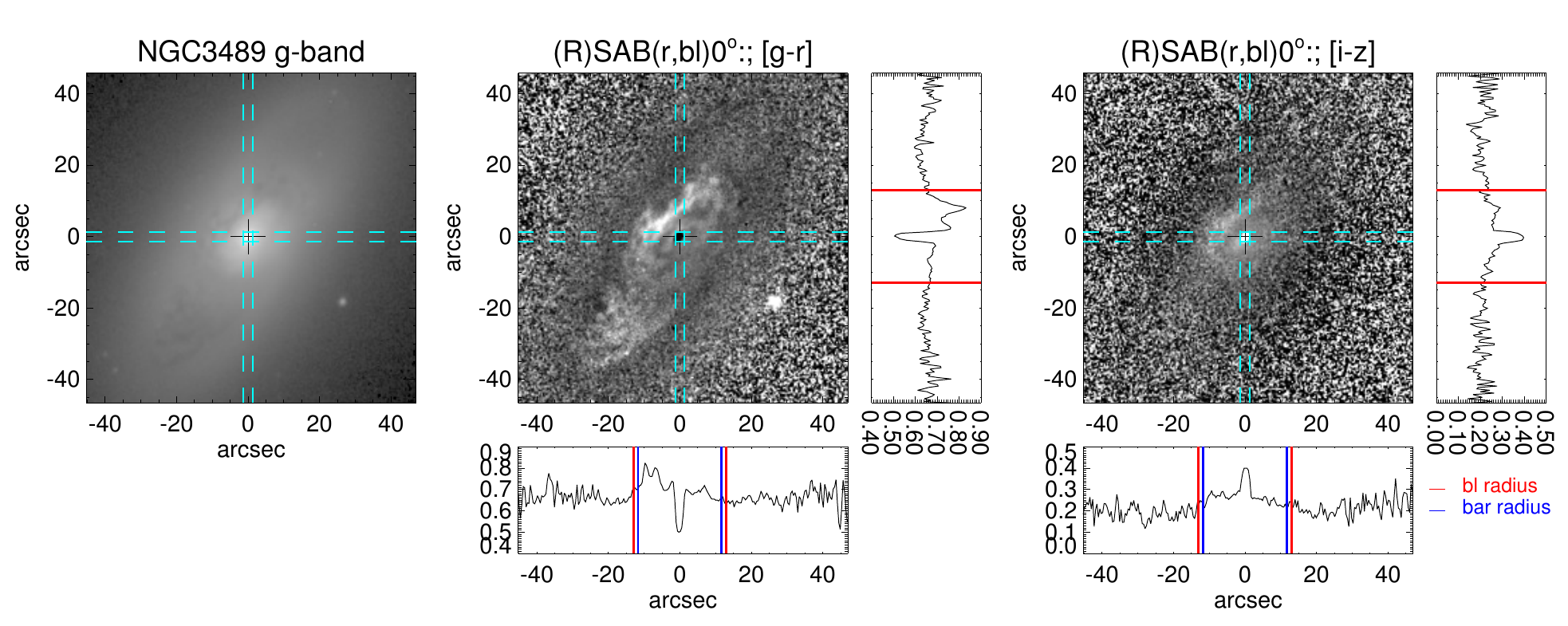}
\includegraphics[scale=0.45]{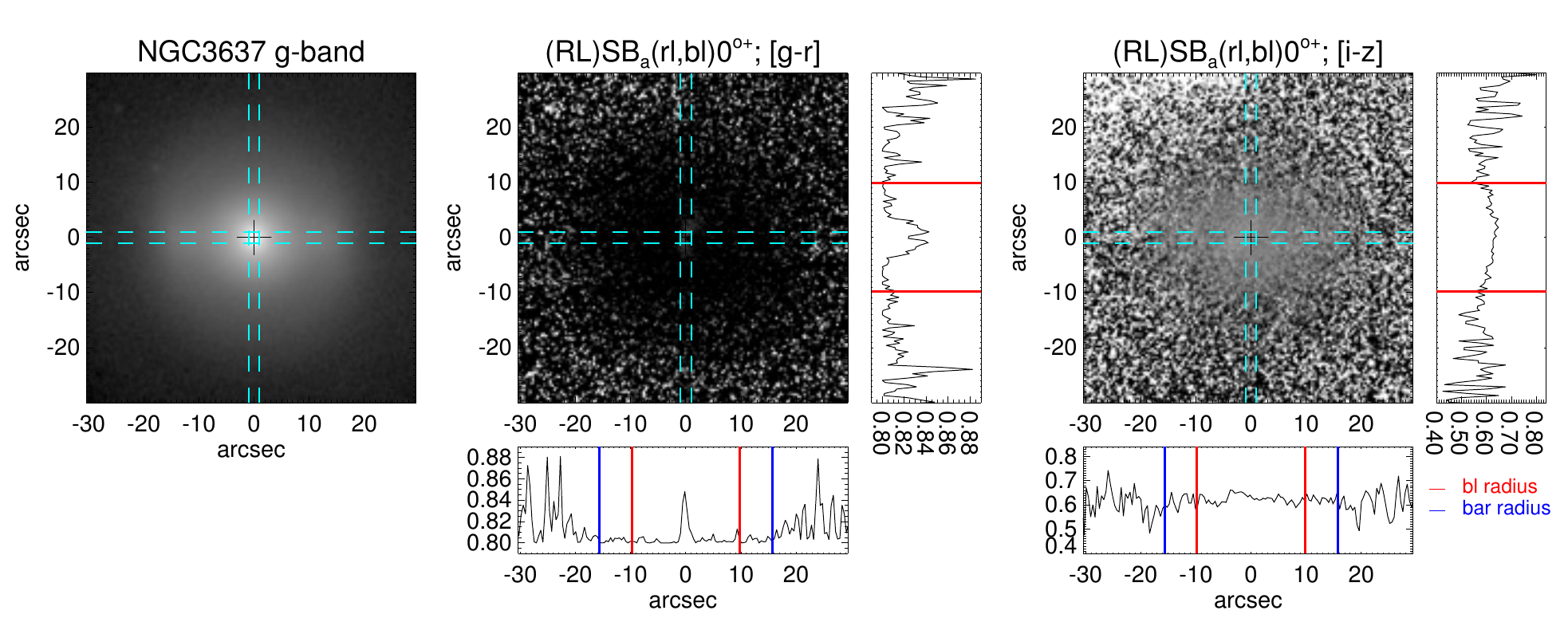}
\includegraphics[scale=0.45]{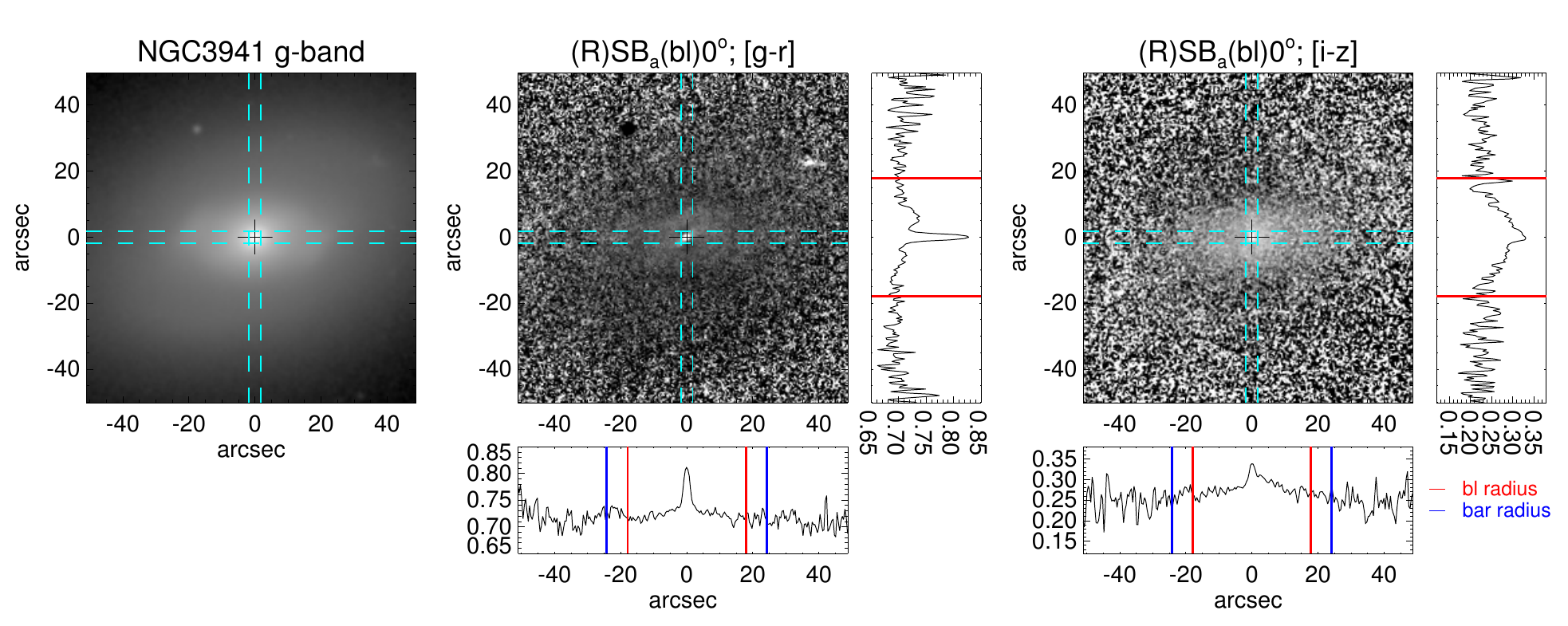}
\includegraphics[scale=0.45]{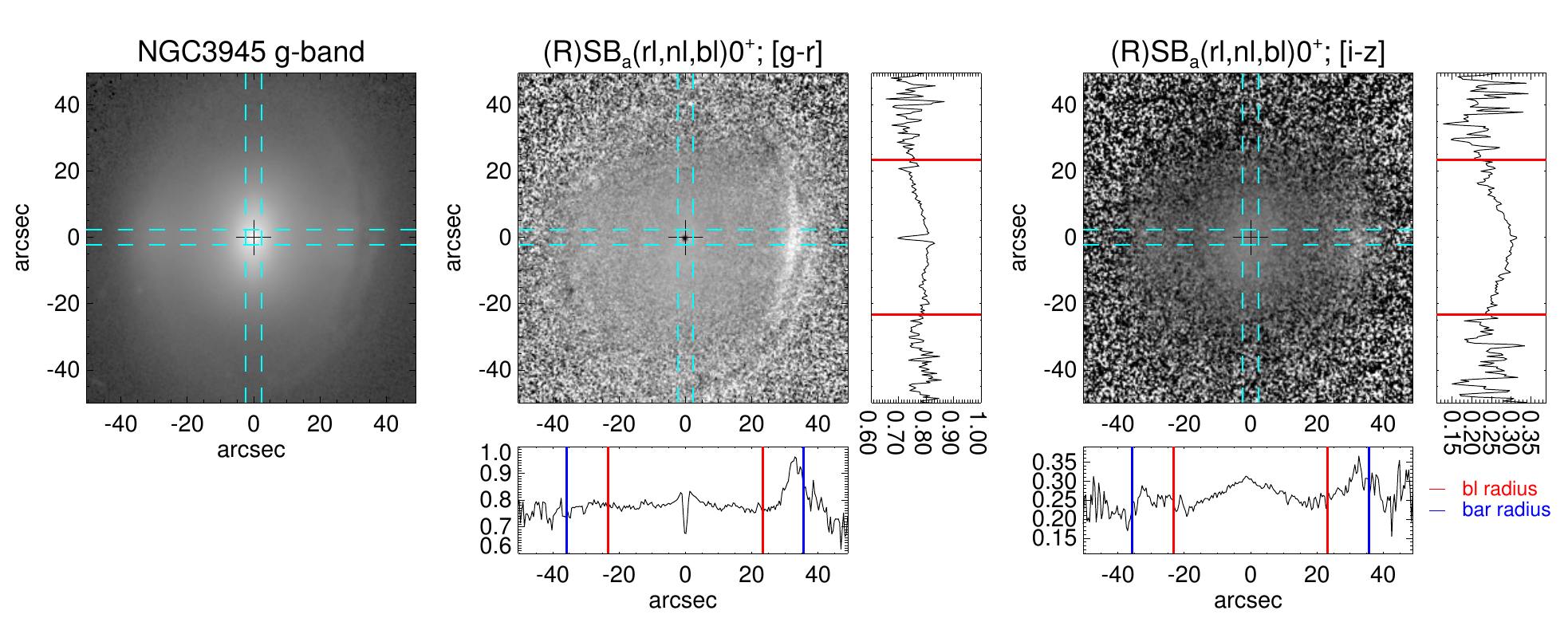}
\includegraphics[scale=0.45]{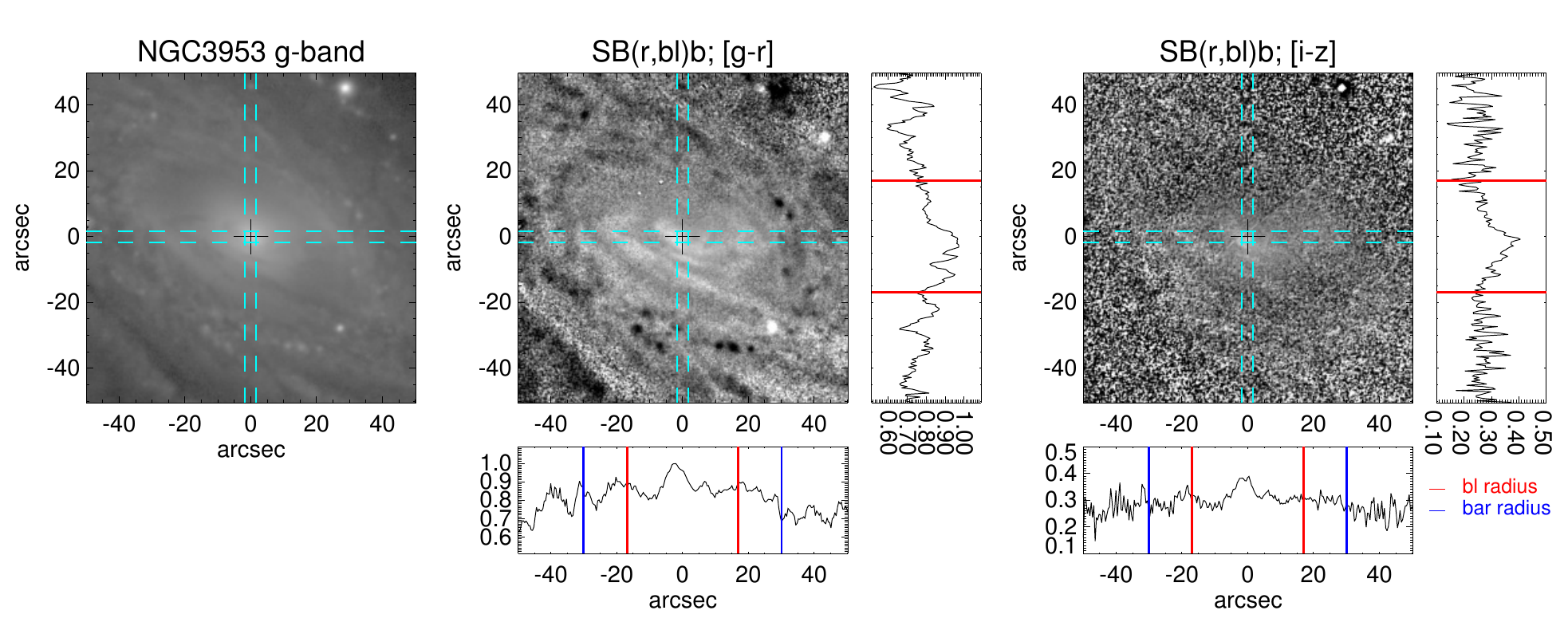}
\includegraphics[scale=0.45]{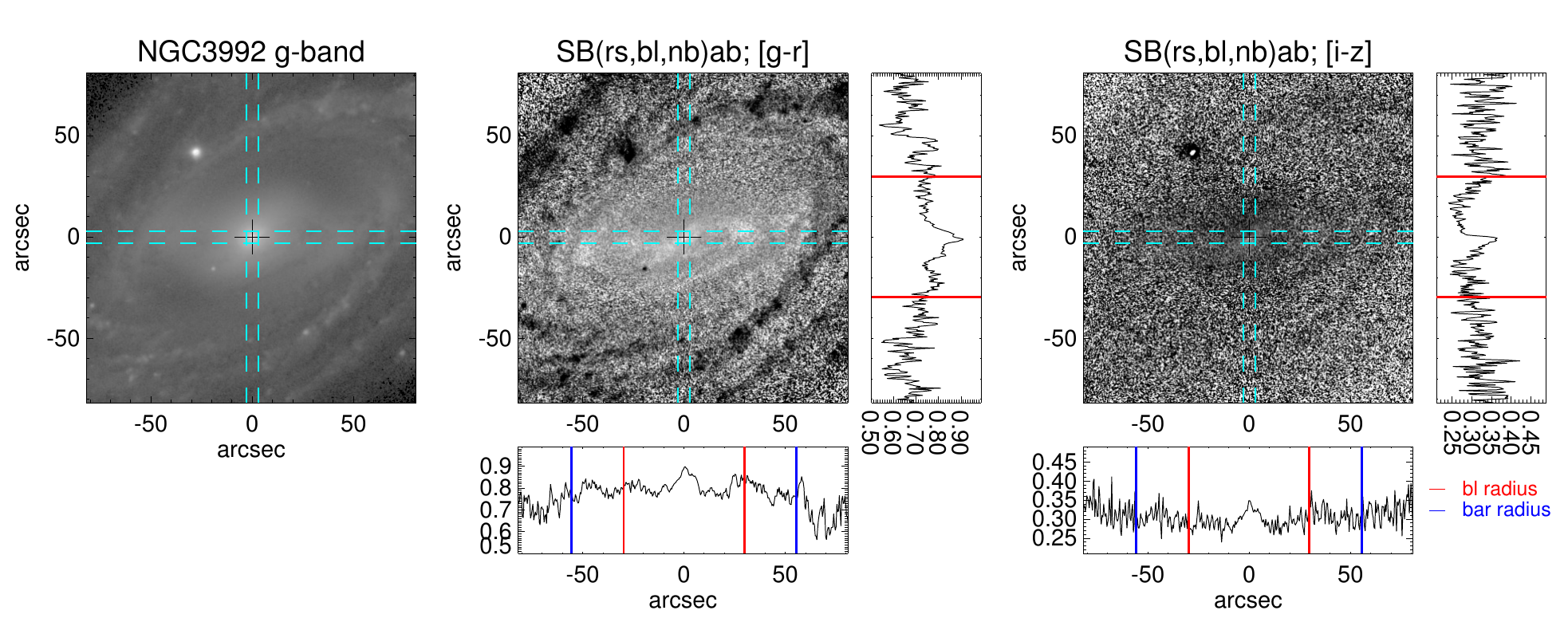}
\includegraphics[scale=0.45]{NGC4143_PGC0038654_colormap_v9_second_g-r-i-z_new.pdf}
\includegraphics[scale=0.45]{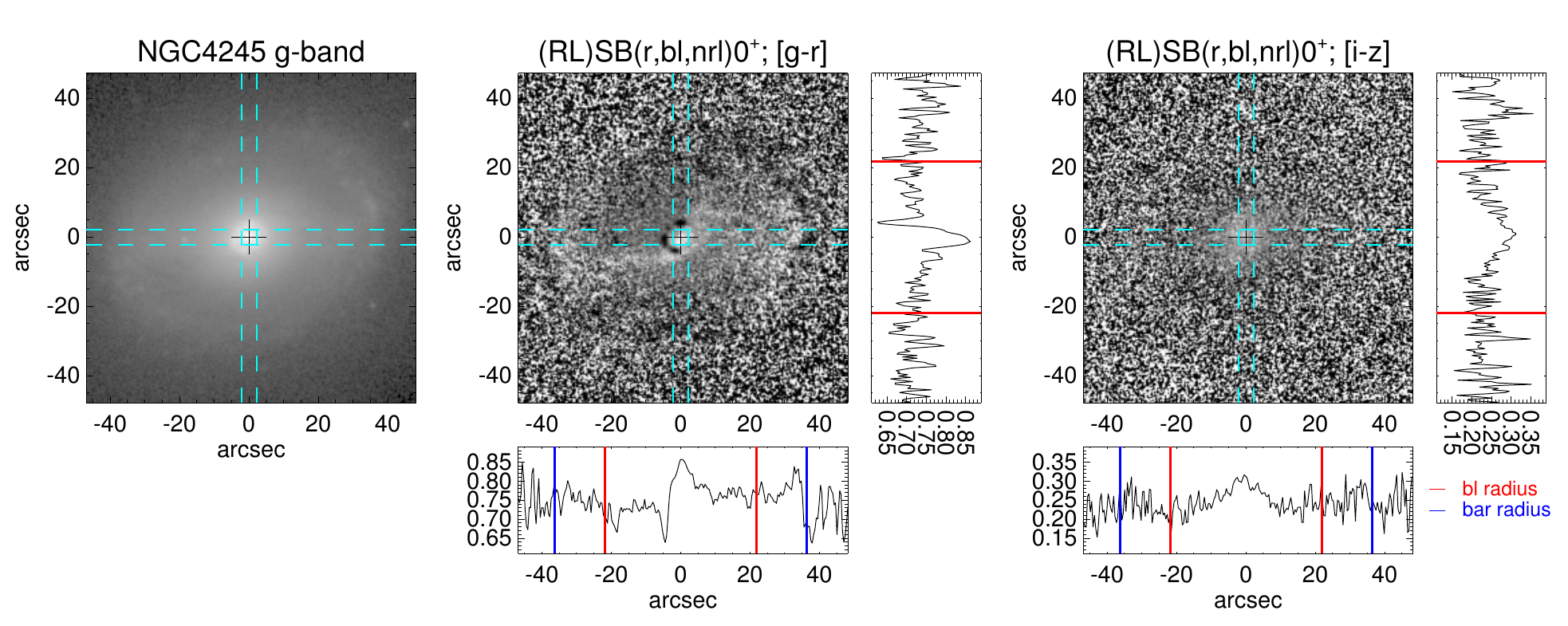}
\includegraphics[scale=0.45]{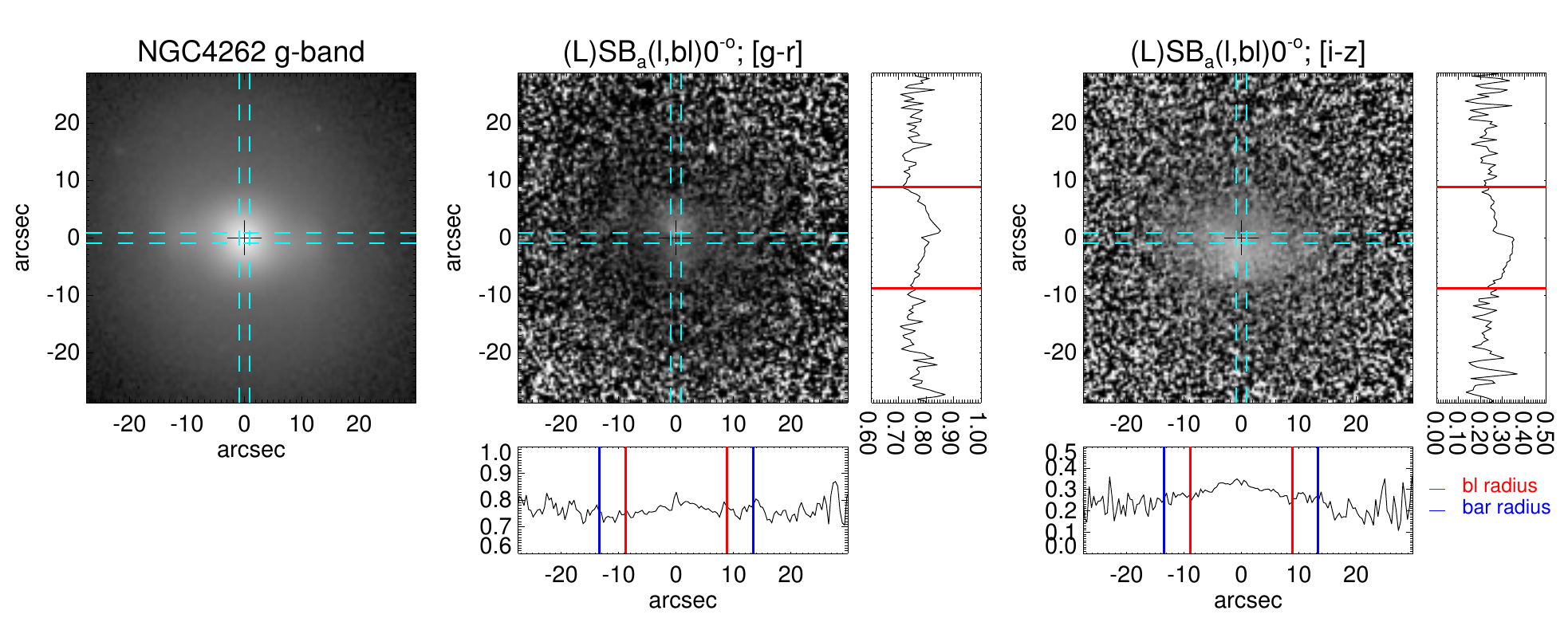}
\includegraphics[scale=0.45]{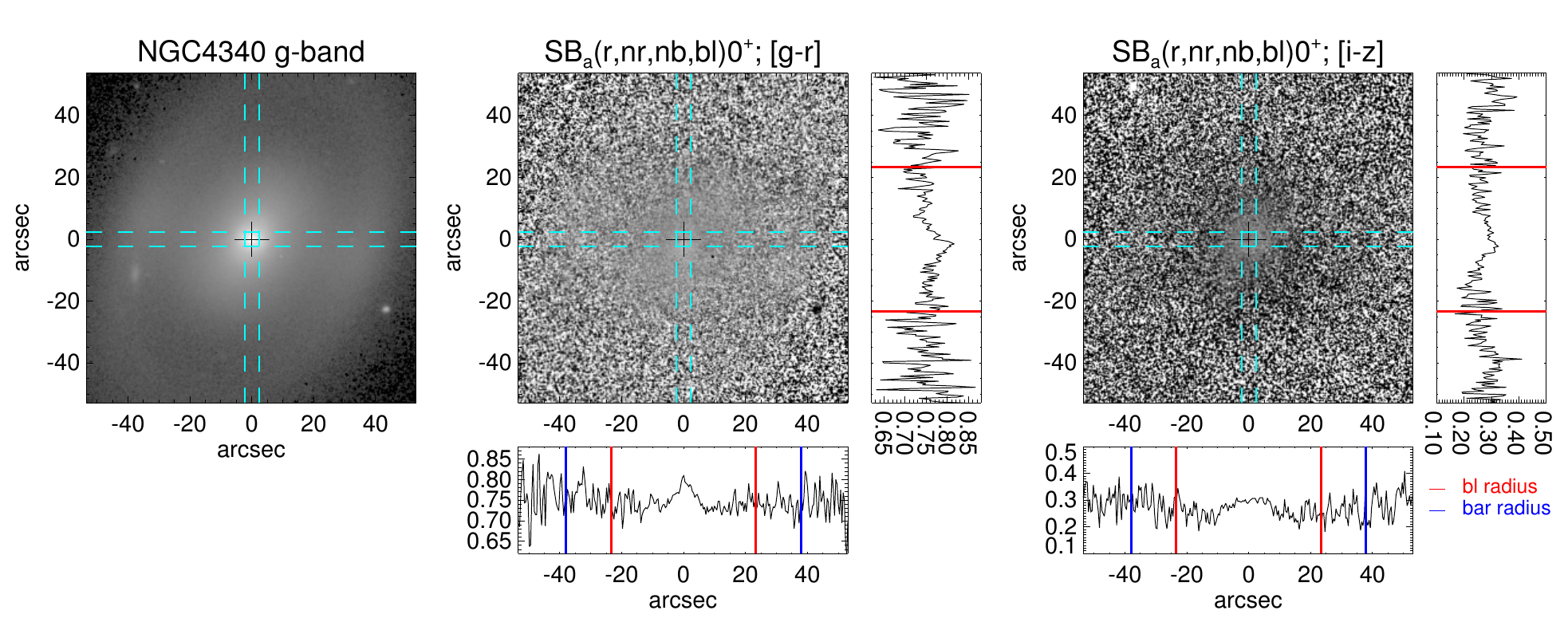}
\includegraphics[scale=0.45]{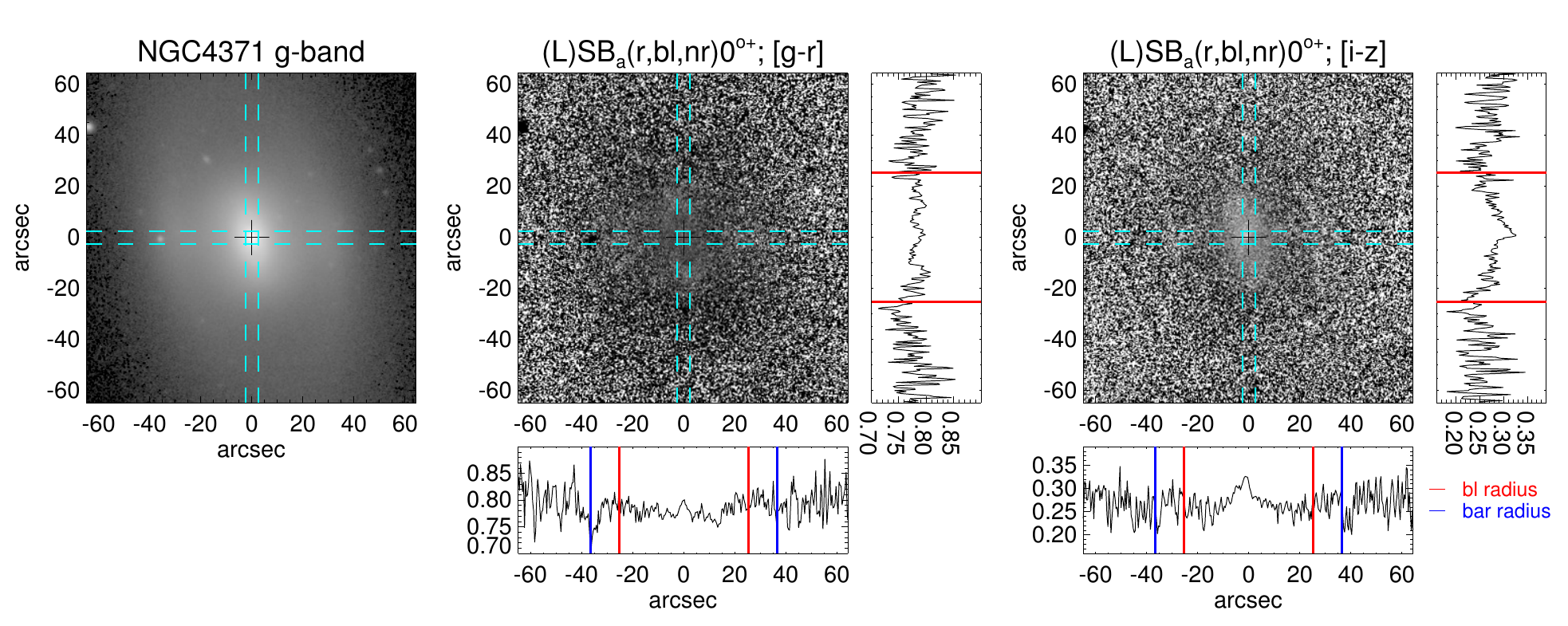}
\includegraphics[scale=0.45]{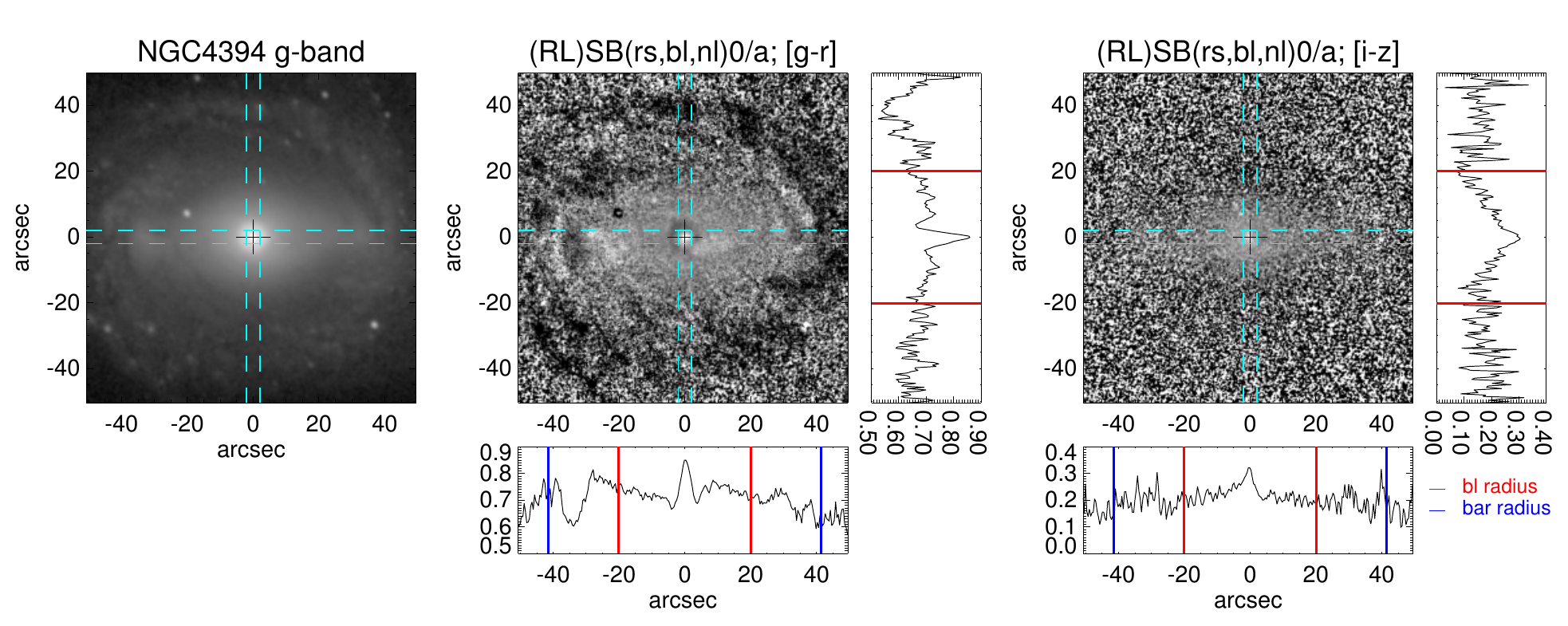}
\includegraphics[scale=0.45]{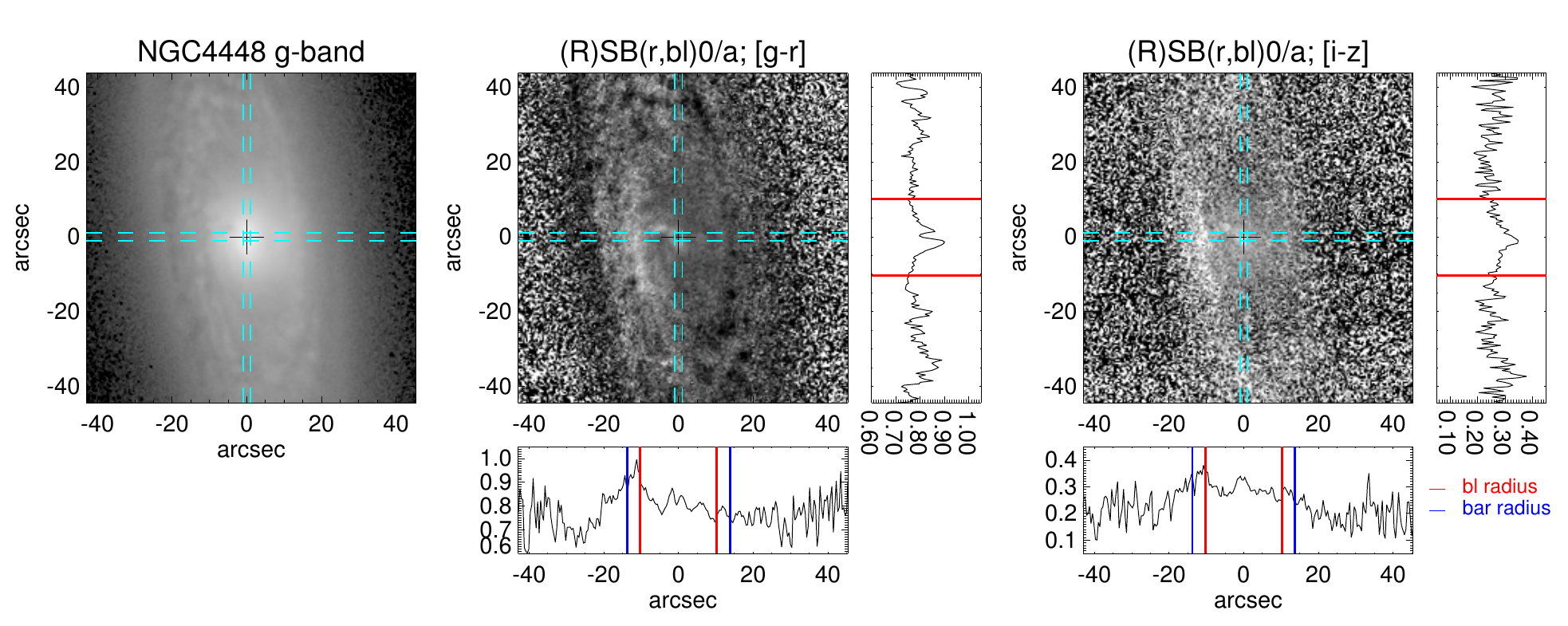}
\includegraphics[scale=0.45]{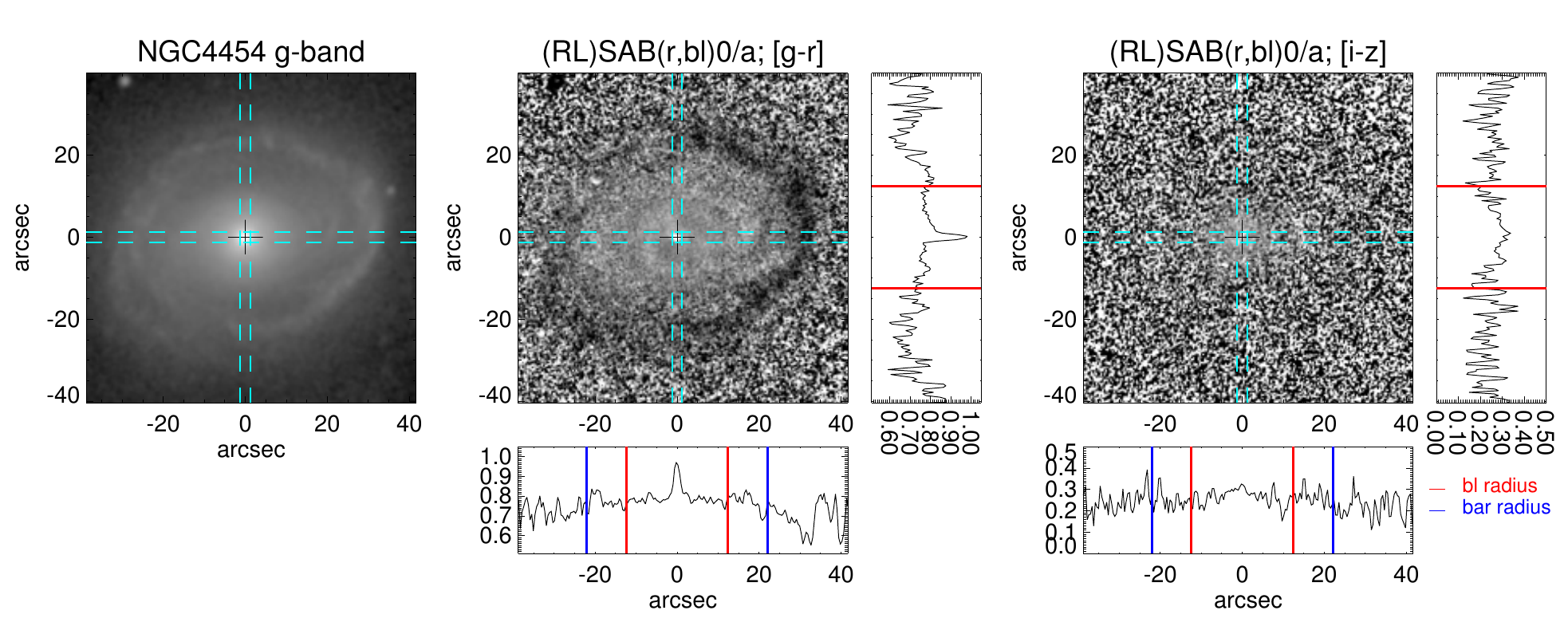}
\includegraphics[scale=0.45]{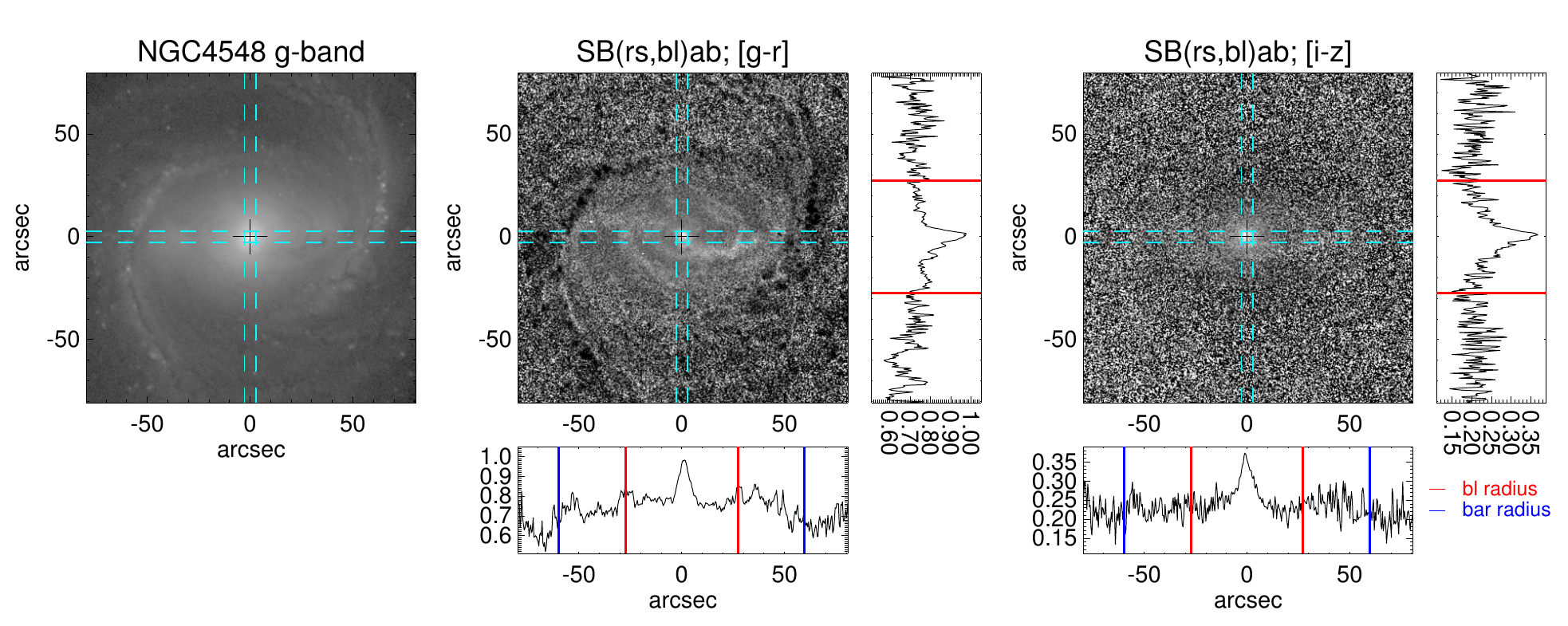}
\includegraphics[scale=0.45]{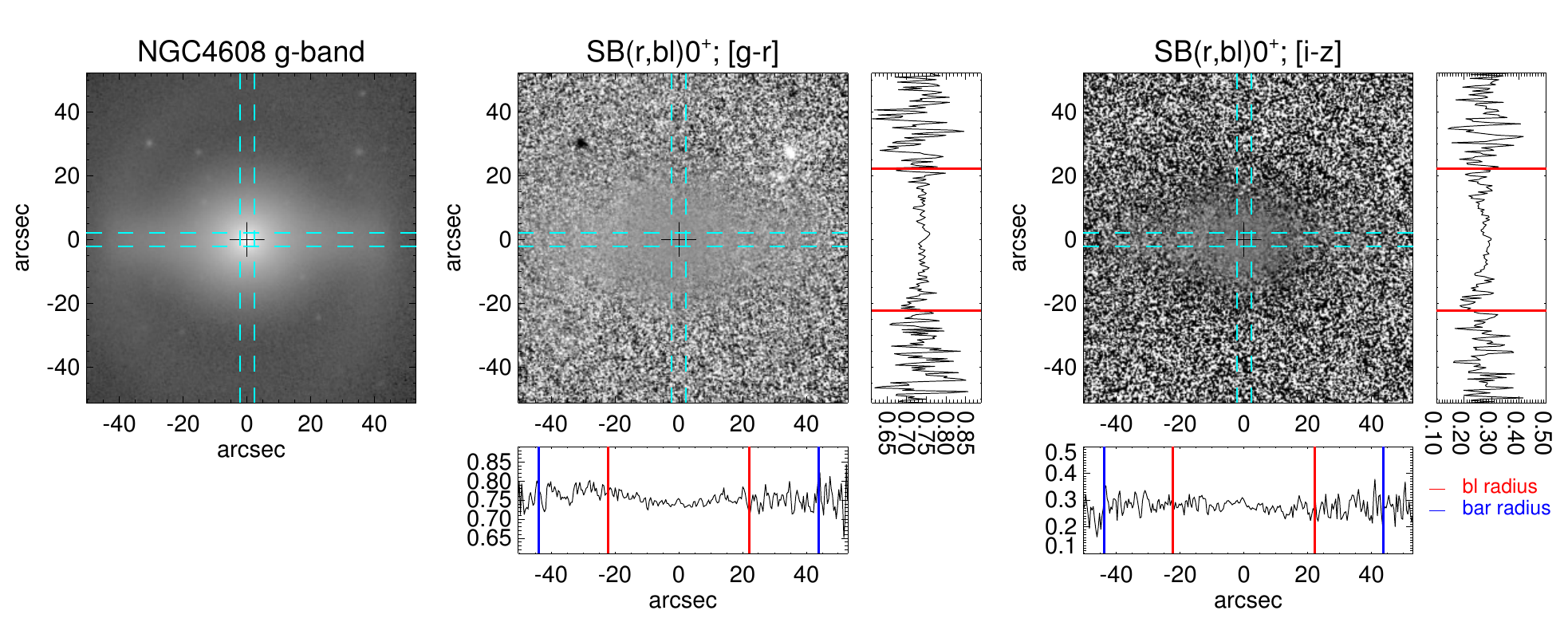}
\includegraphics[scale=0.45]{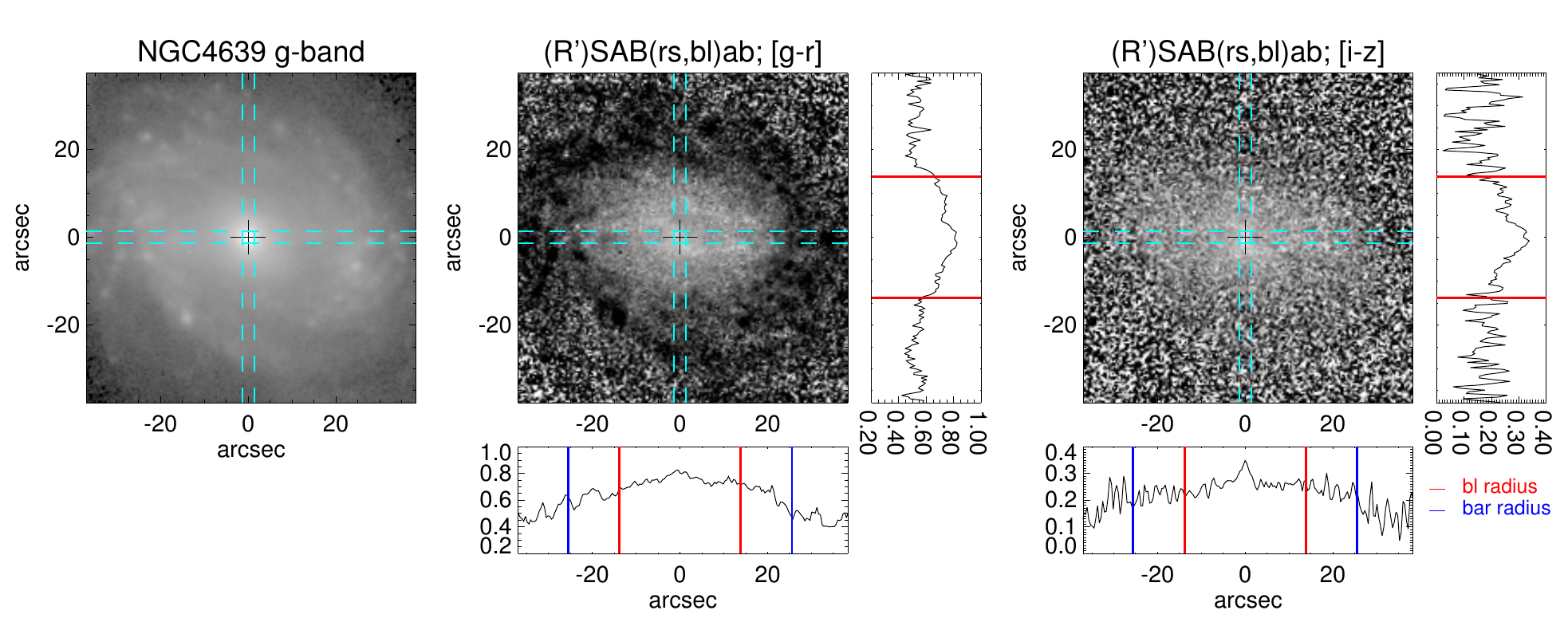}
\includegraphics[scale=0.45]{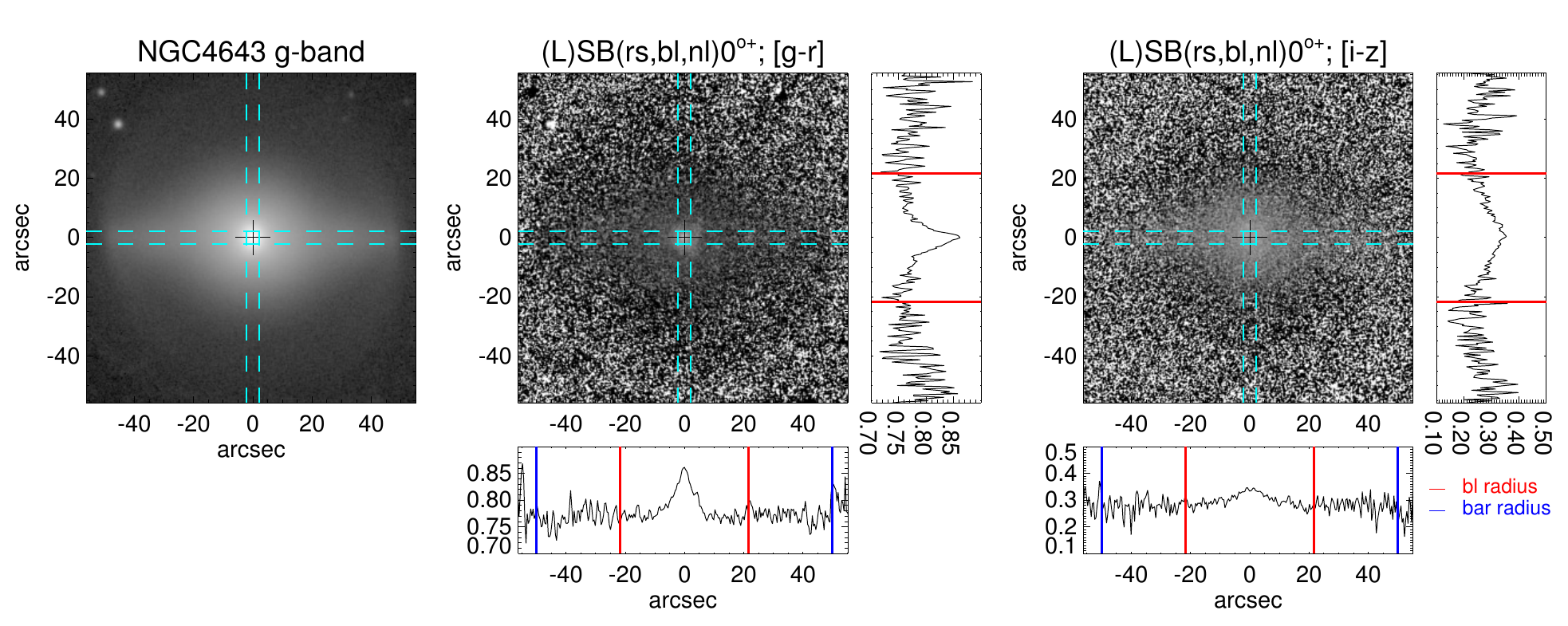}
\includegraphics[scale=0.45]{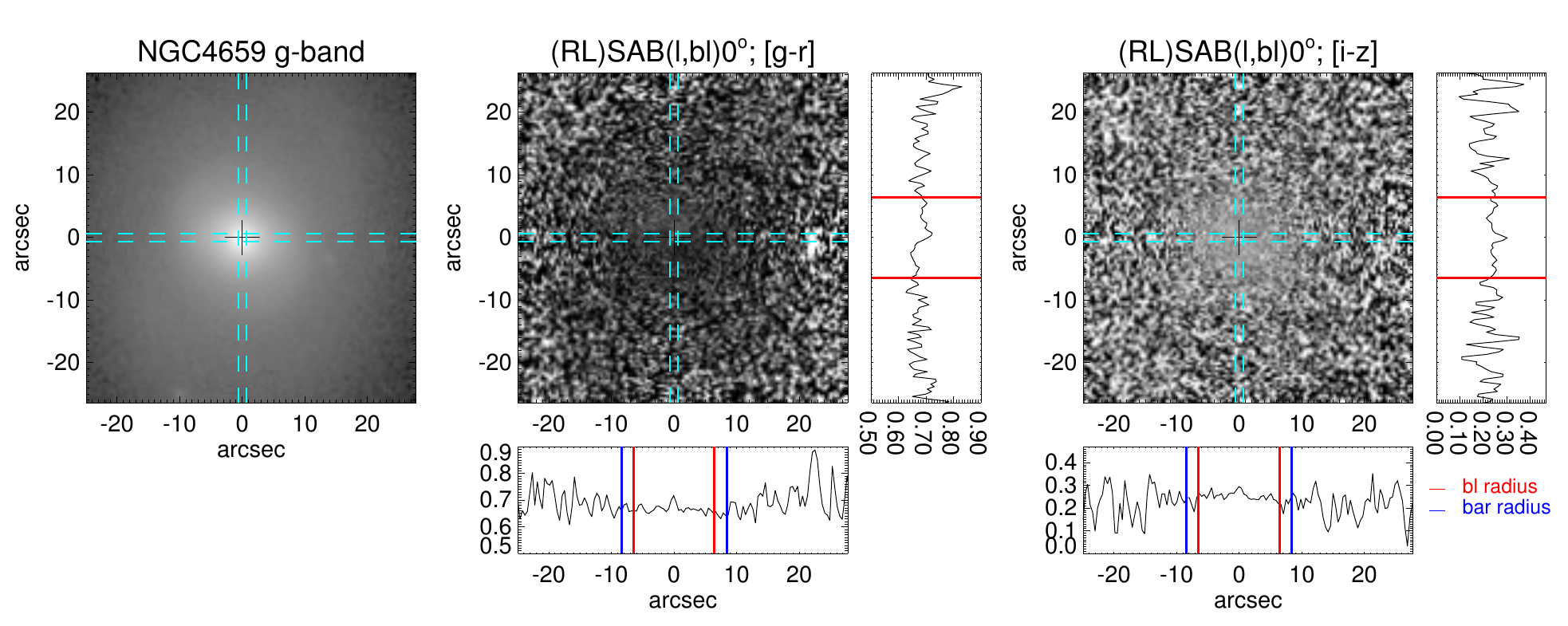}
\includegraphics[scale=0.45]{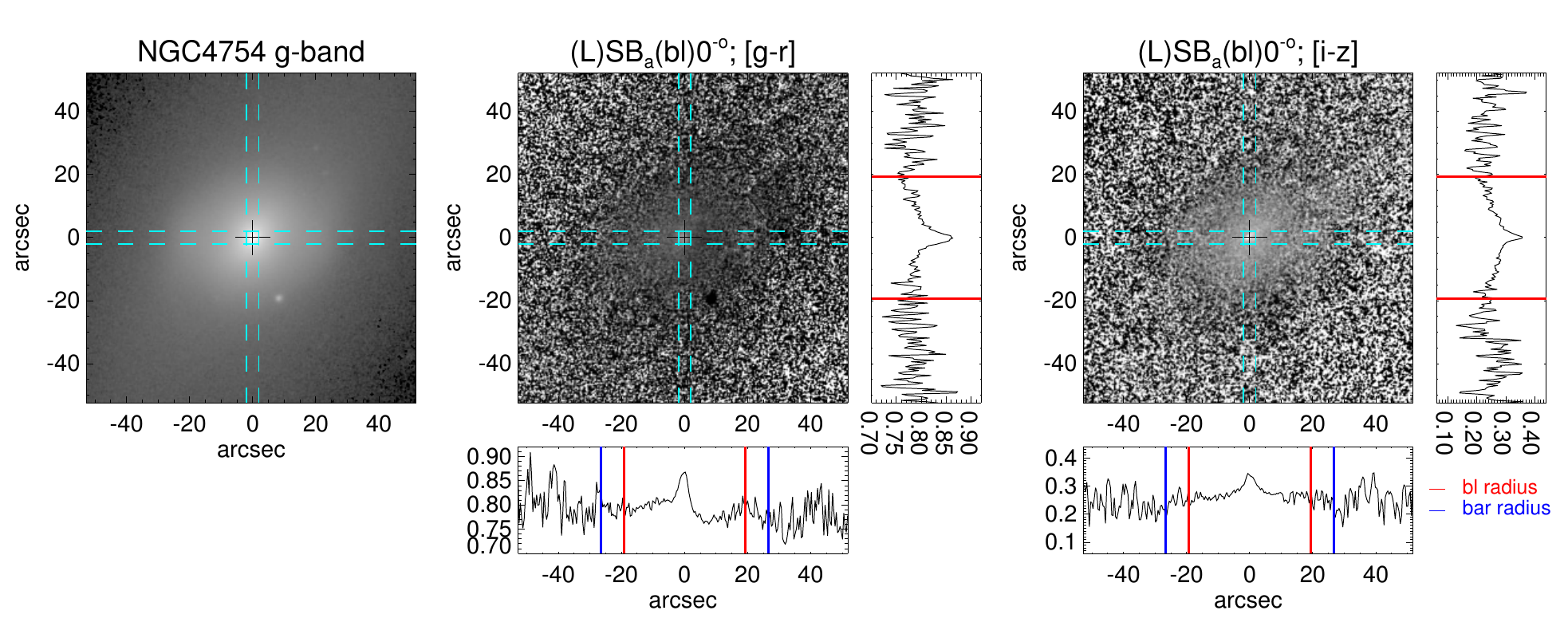}
\includegraphics[scale=0.45]{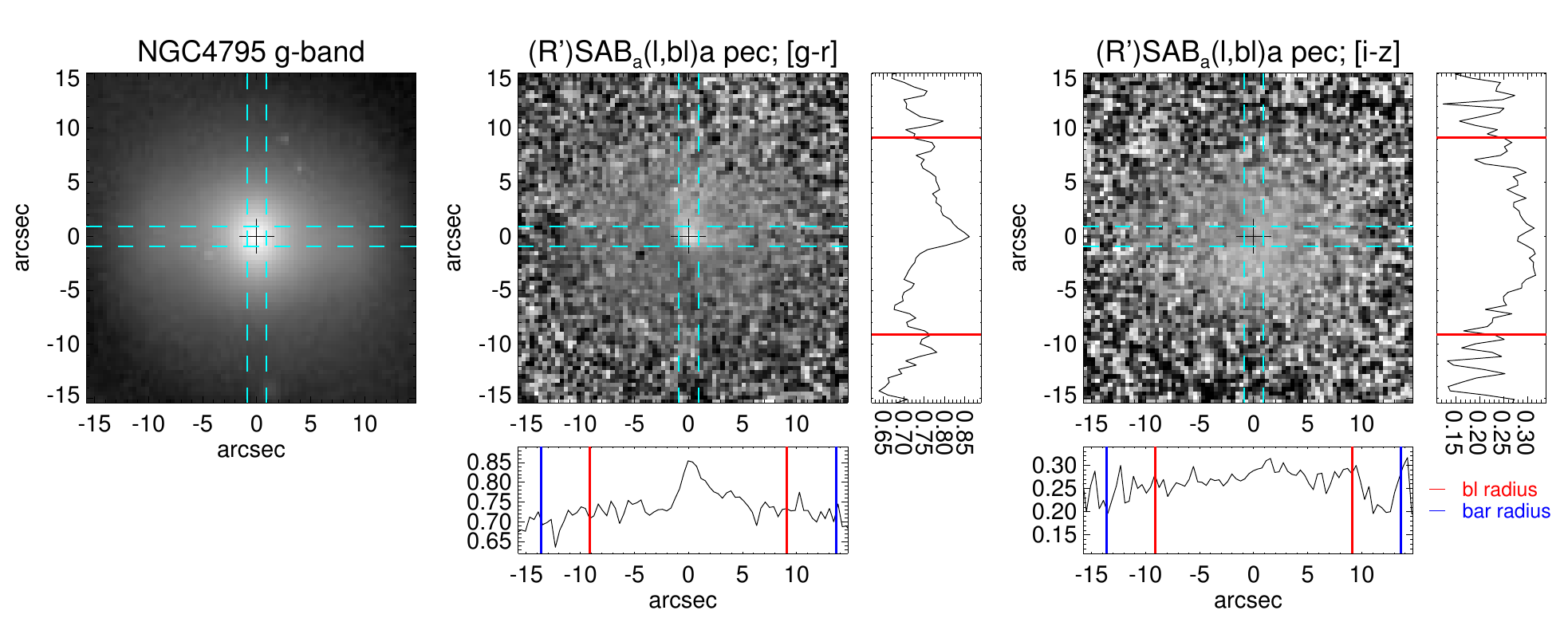}
\includegraphics[scale=0.45]{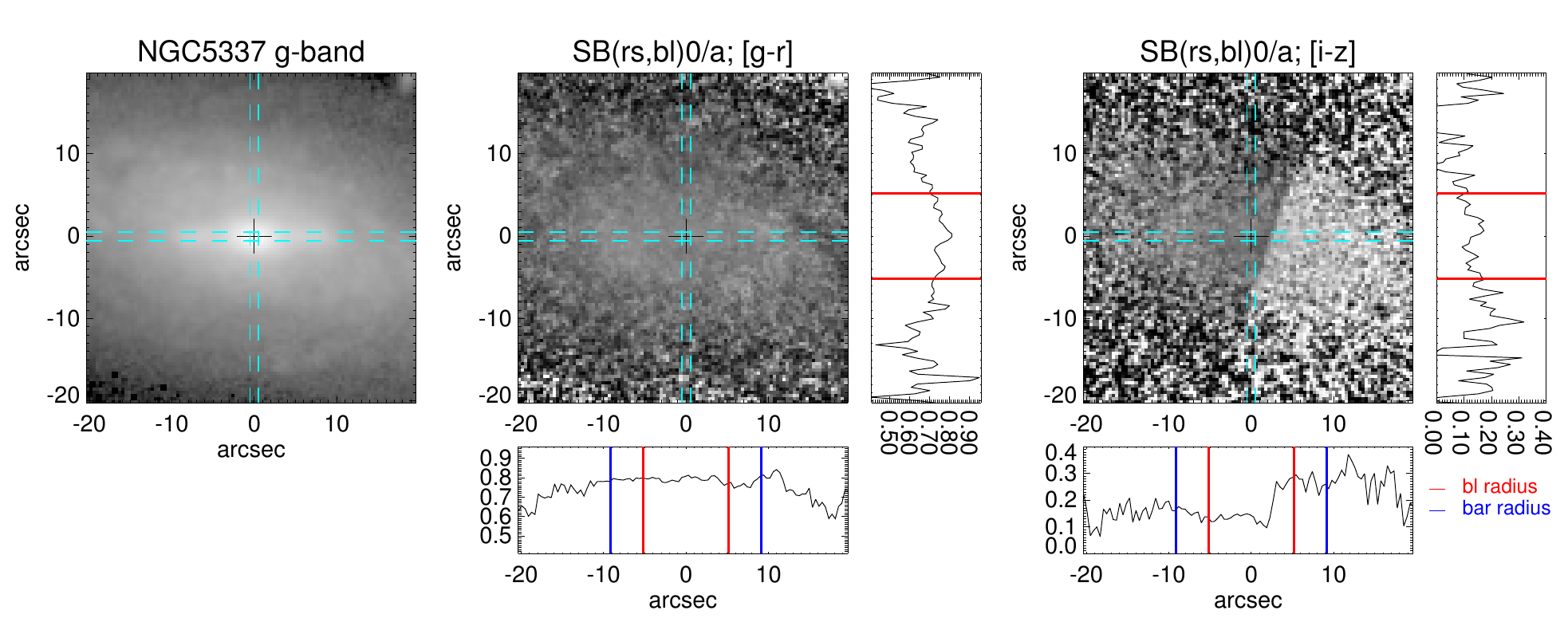}
\includegraphics[scale=0.45]{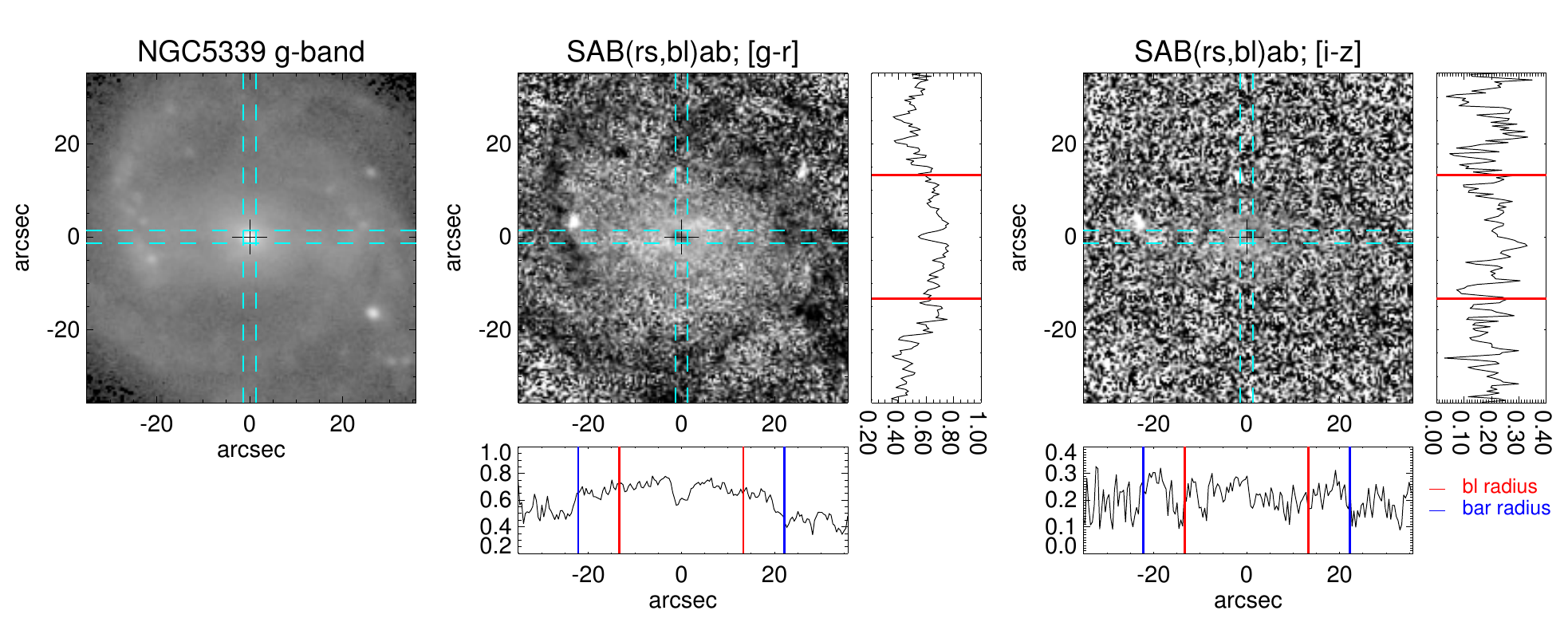}
\includegraphics[scale=0.45]{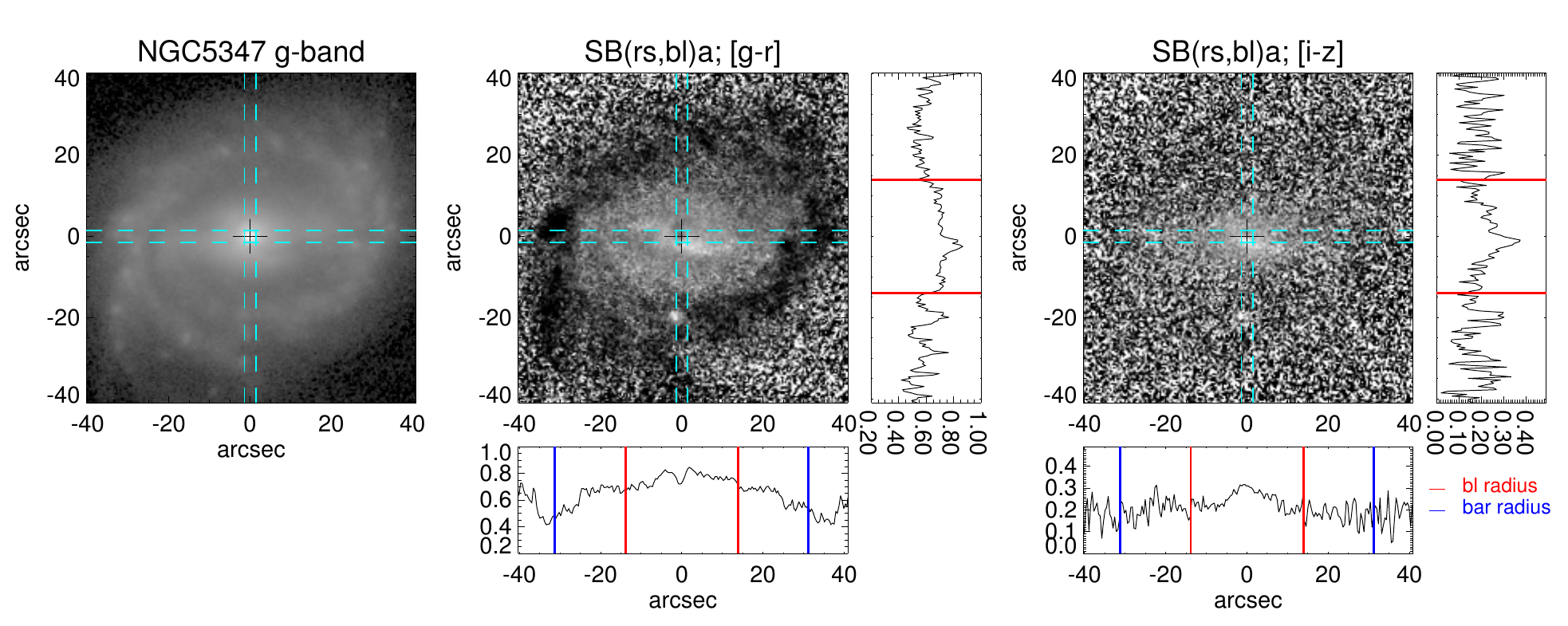}
\includegraphics[scale=0.45]{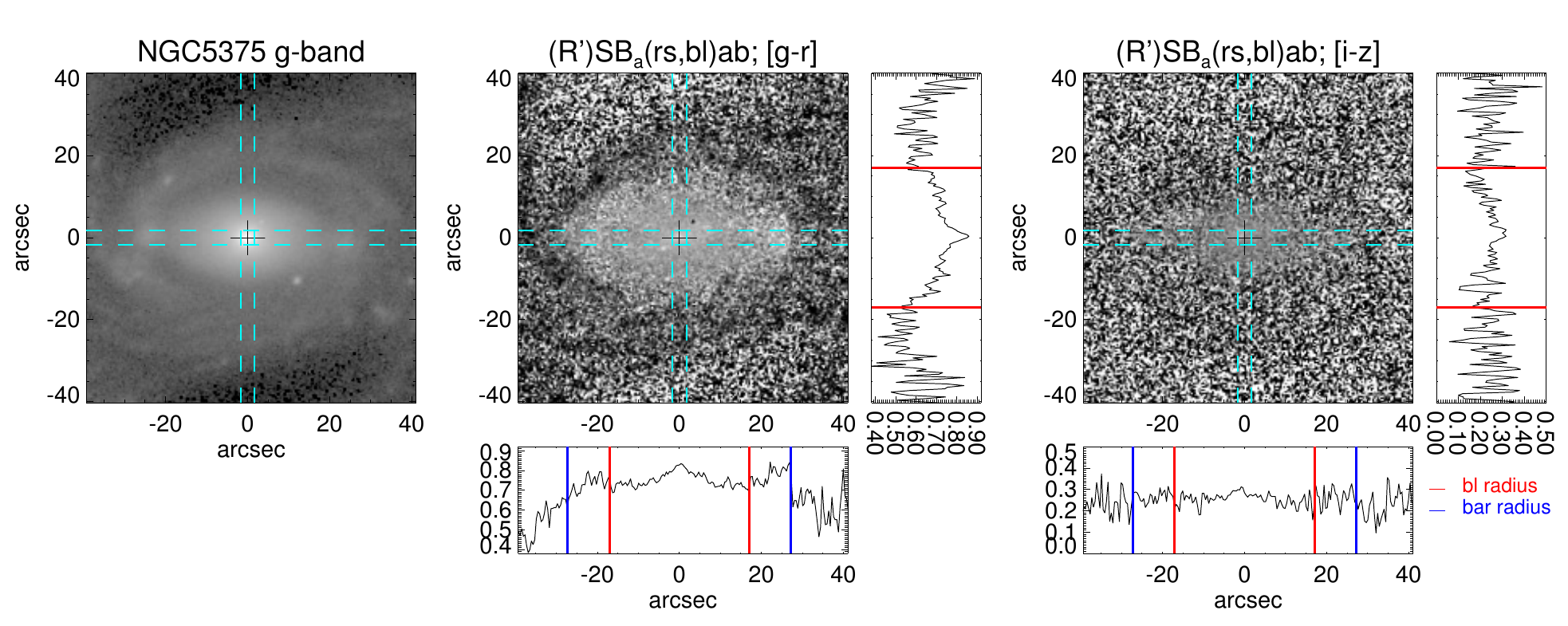}
\includegraphics[scale=0.45]{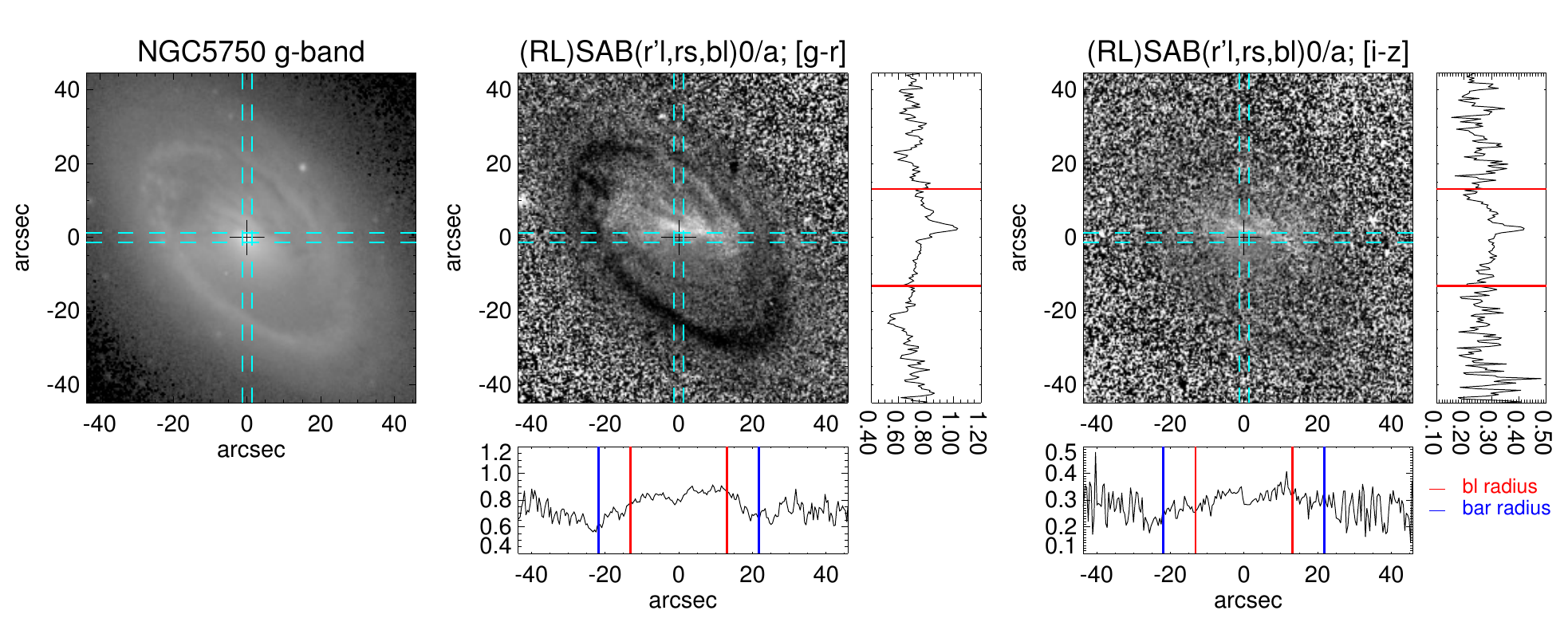}
\includegraphics[scale=0.45]{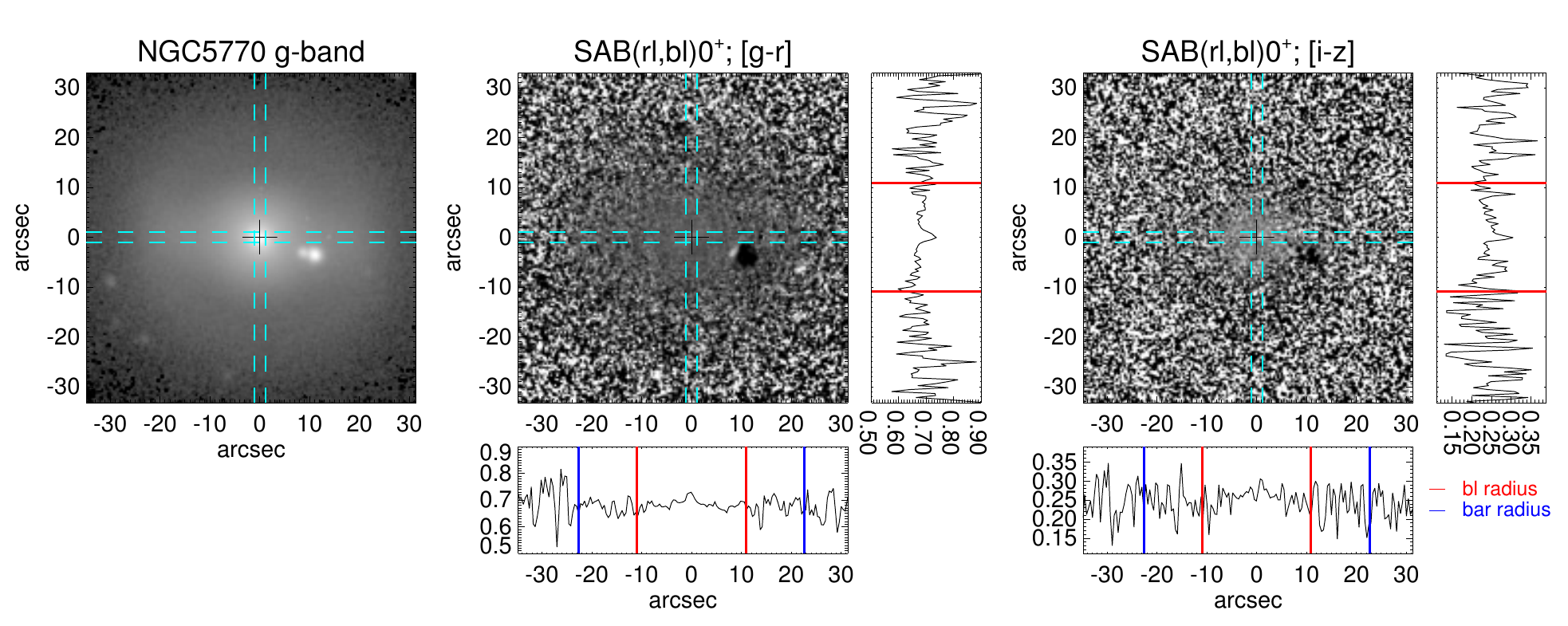}
\includegraphics[scale=0.45]{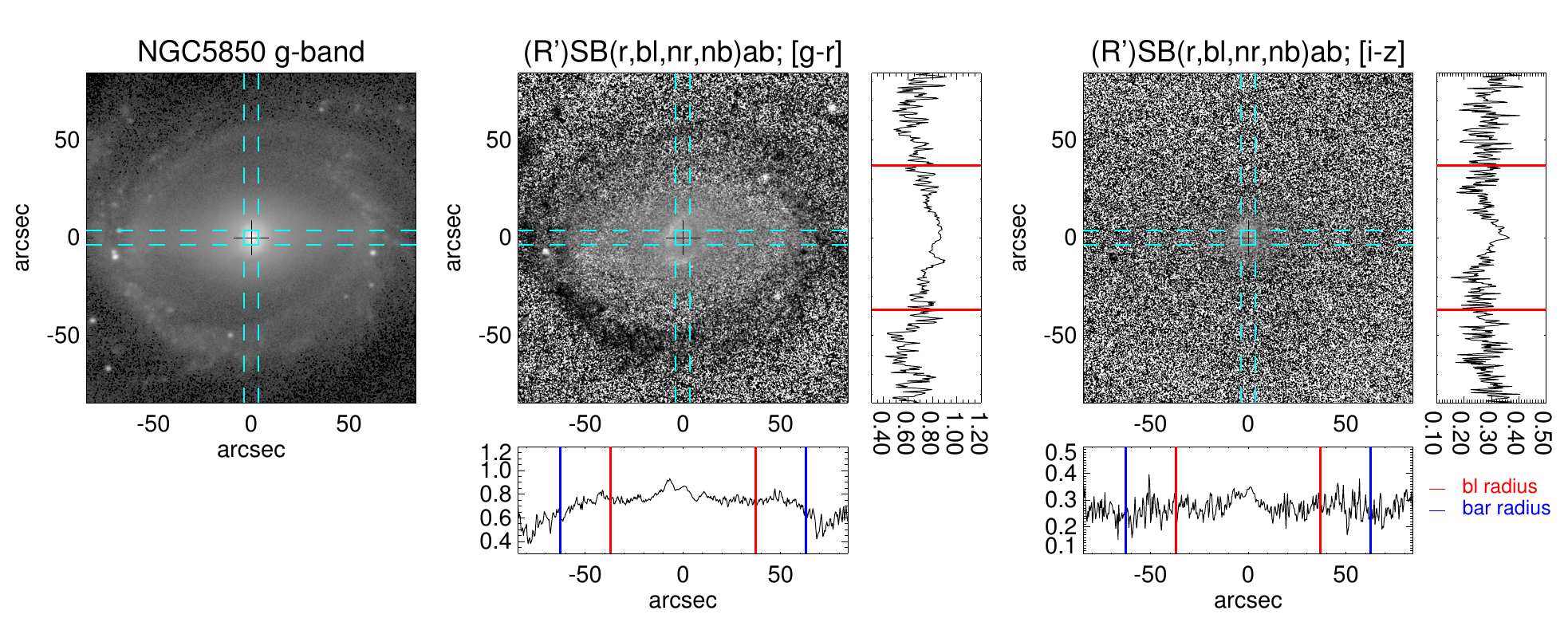}
\includegraphics[scale=0.45]{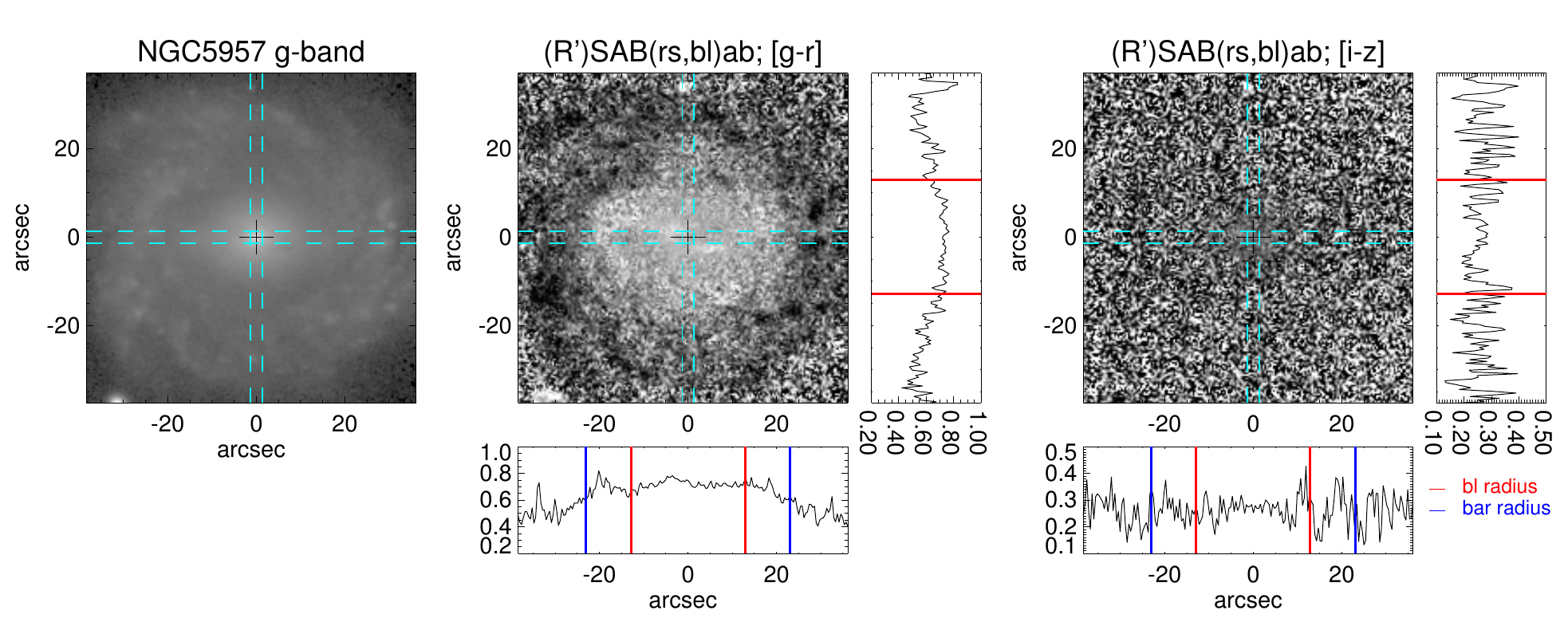}
\includegraphics[scale=0.45]{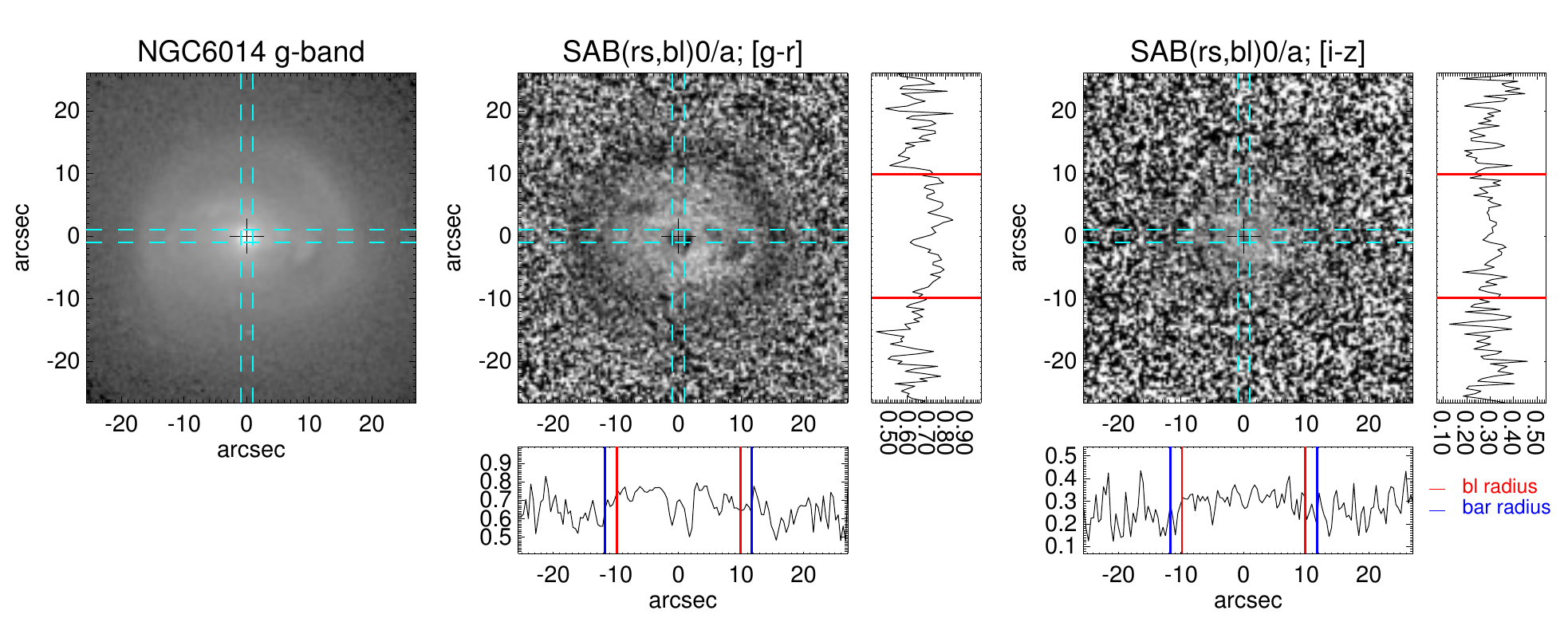}

\clearpage


\section{Color profiles of the barlens galaxies in the color subsample}
In this appendix we show the same information as in Fig. 9 for the RP, RP+nr,
nb, B and D profile groups (see text for details).

\noindent
\includegraphics[scale=.62]{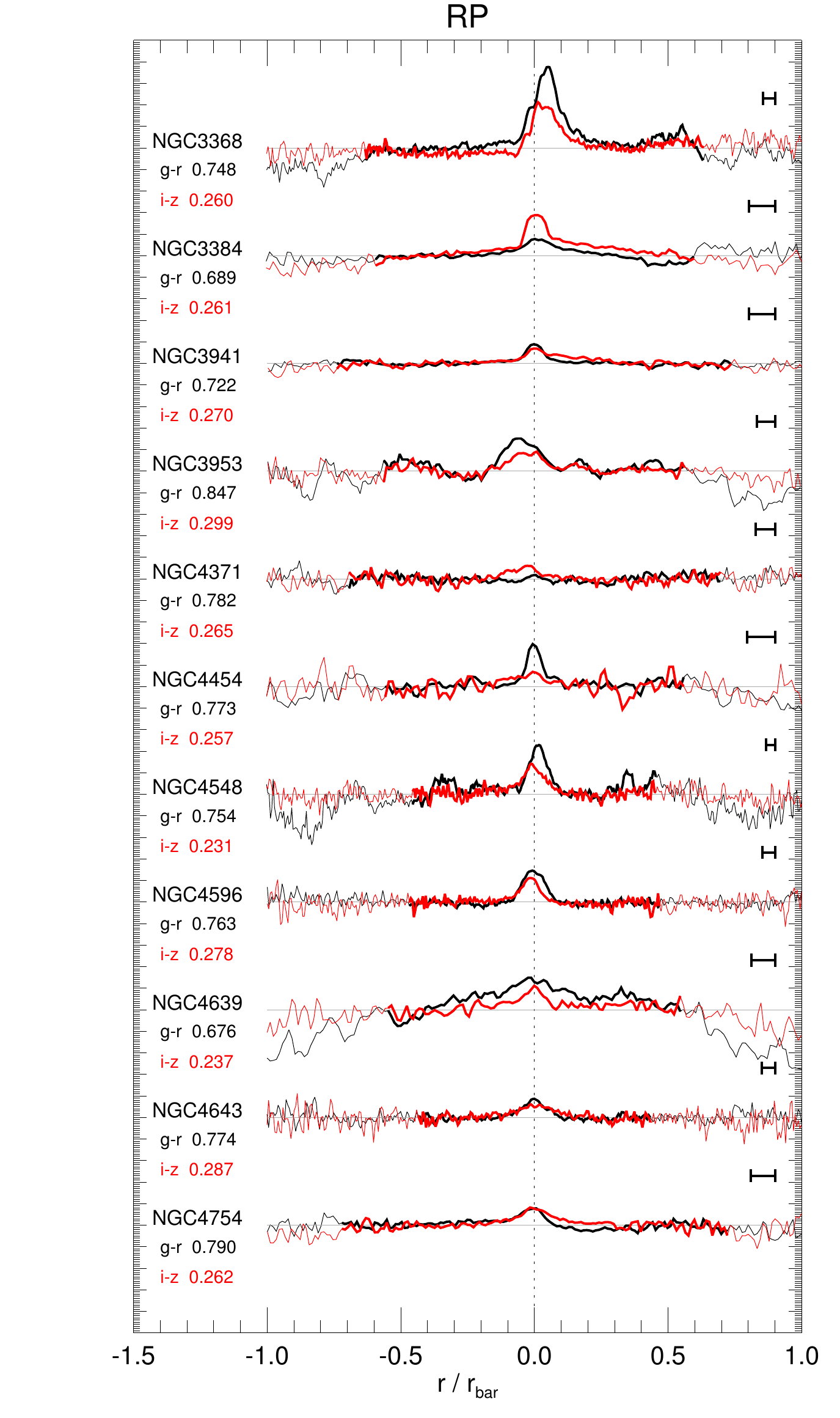}
\includegraphics[scale=.62]{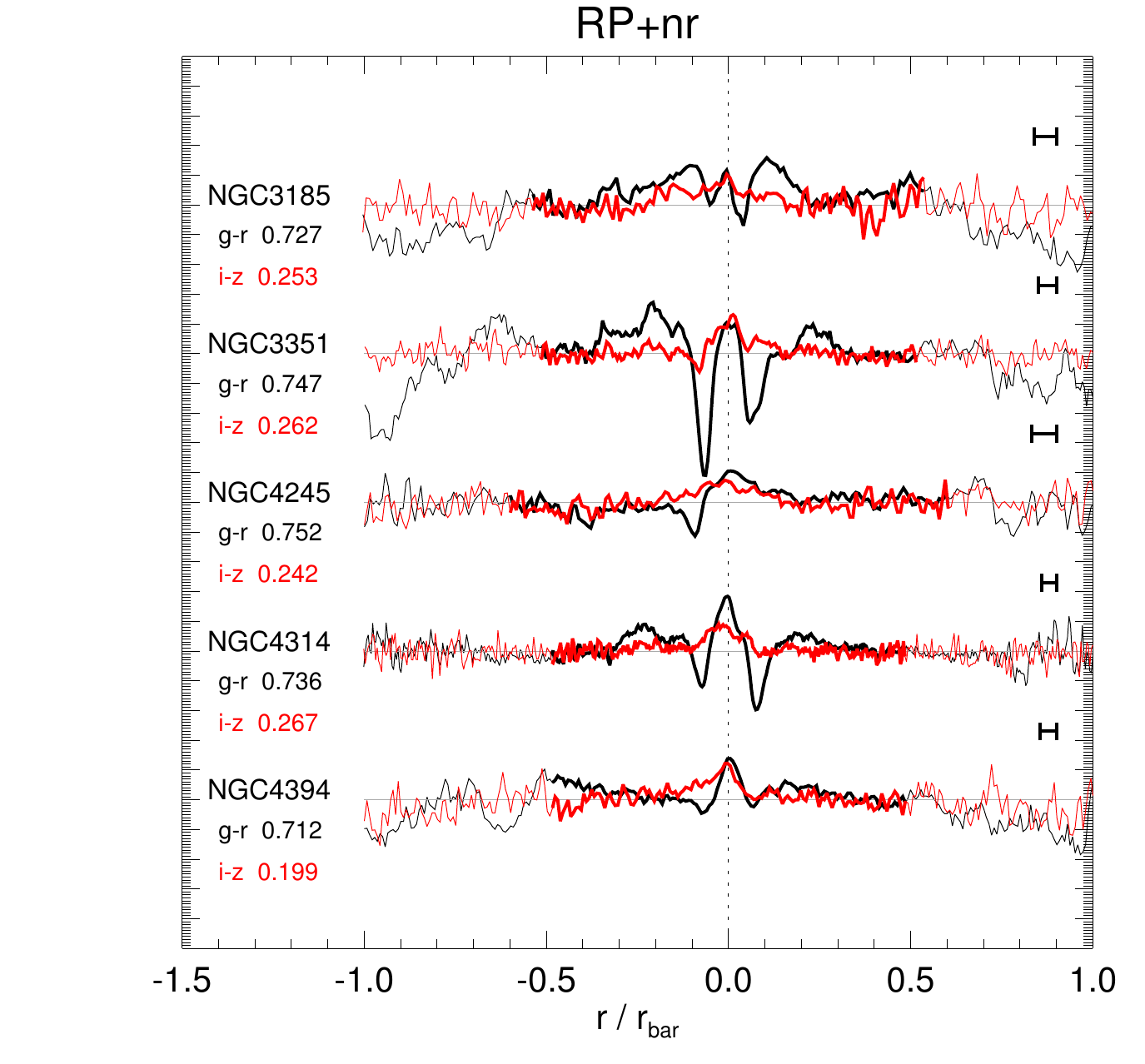}
\includegraphics[scale=.62]{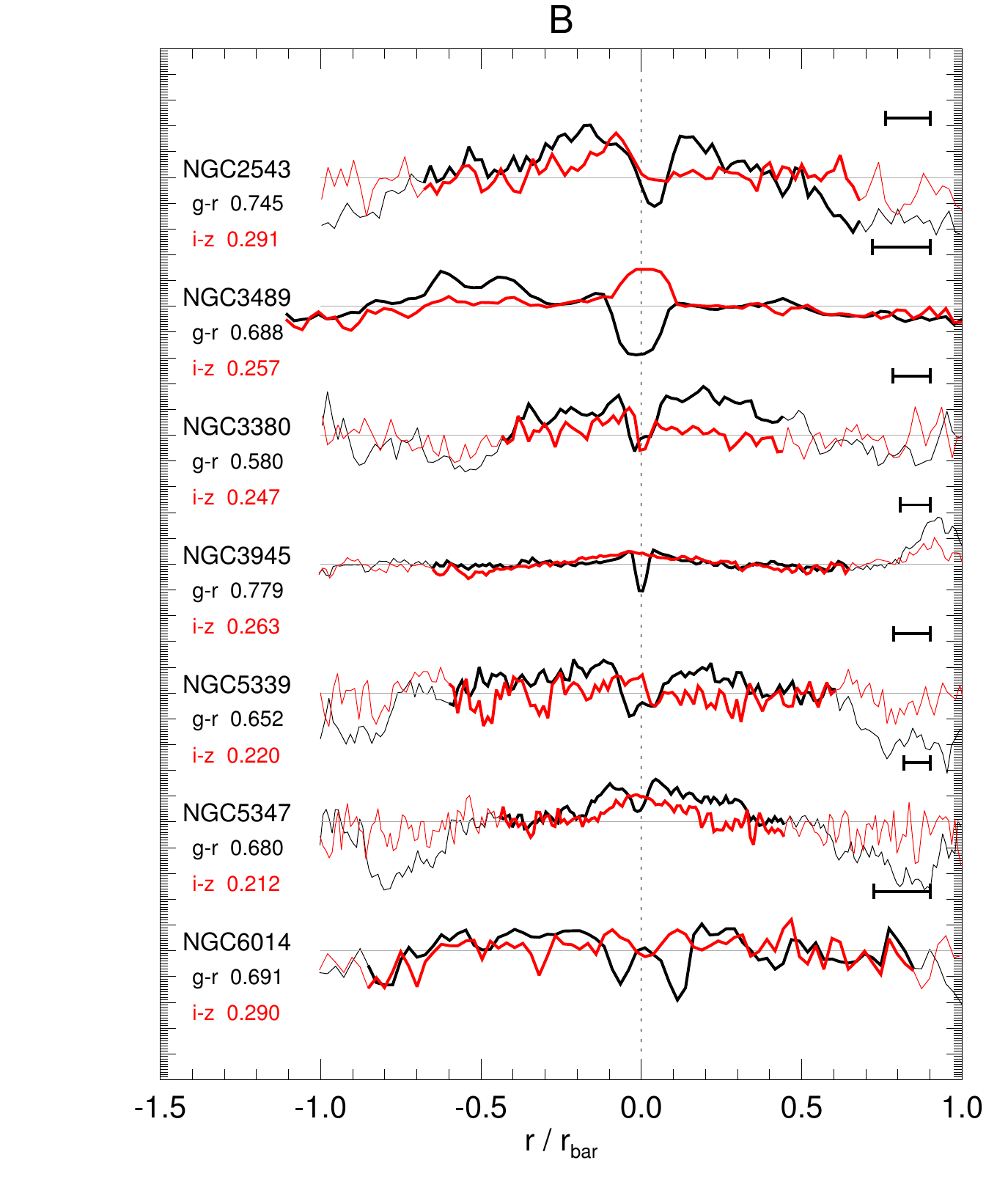}
\includegraphics[scale=.62]{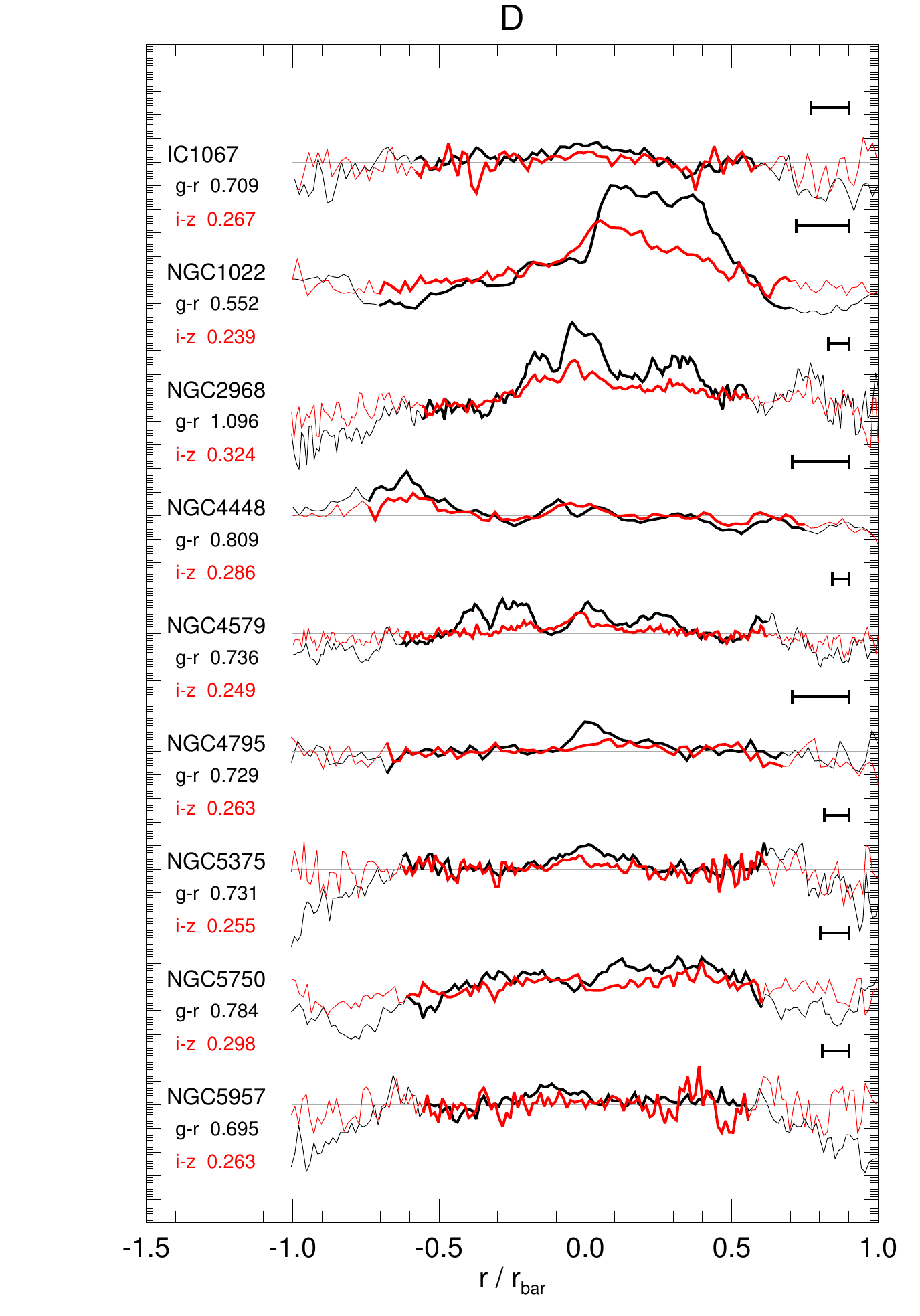}

\includegraphics[scale=.62]{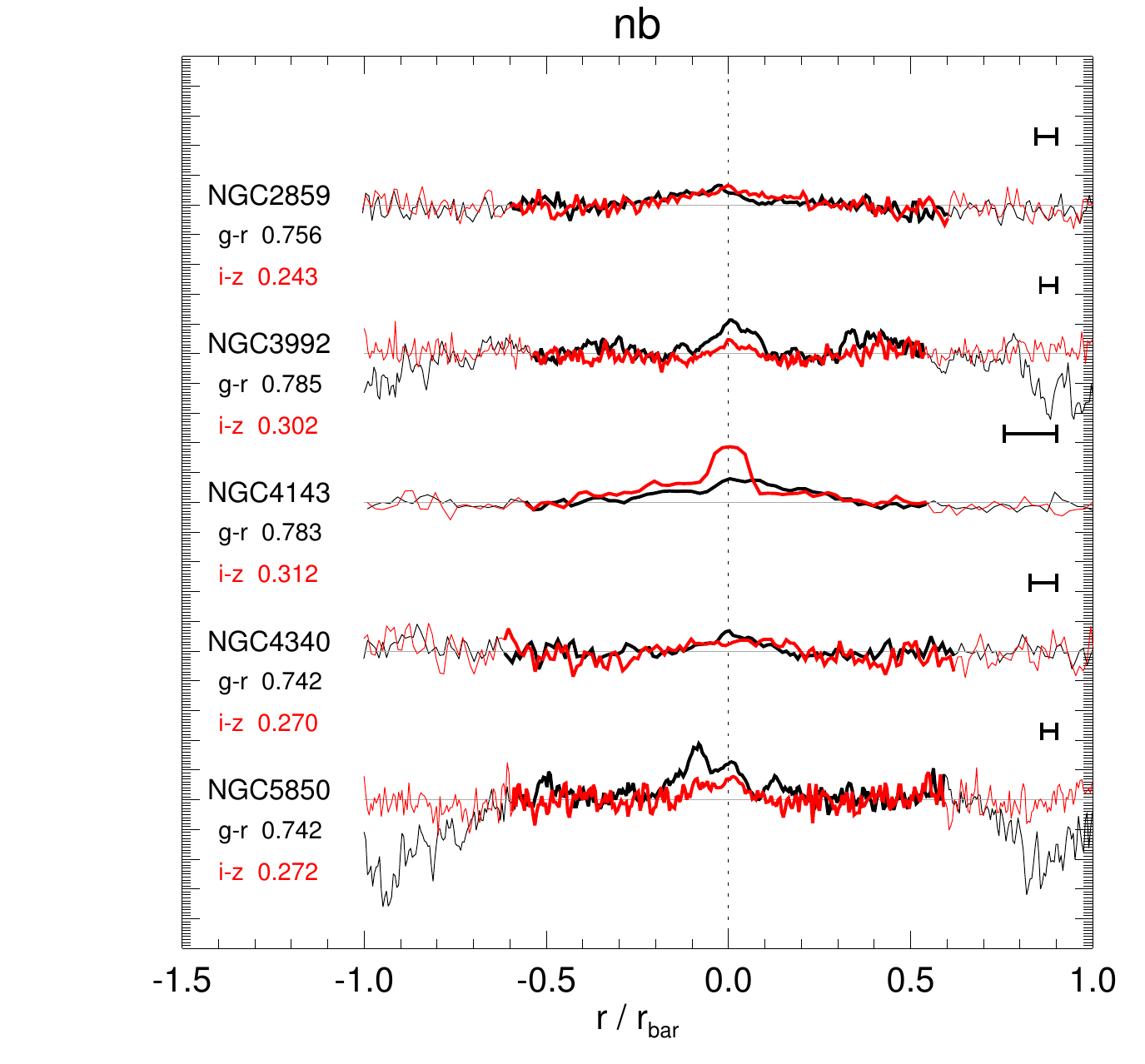}

\clearpage
\onecolumn
\begin{landscape}
\section{Data table}

\setcounter{table}{0}

{\scriptsize
\begin{longtable}{llccccccccccccc}
\caption*{Table 1: Sample of barlens galaxies used in this paper. Shown are the galaxy name, 
the morphological classification, the distance, and the projected properties of 
the disks, bars and barlenses (bl). $i$, PA, r and b/a are for inclination, position angle,
radius and axis ratio, respectively.}\\
\hline 
       &                   &                    &                 & &           &            &              & &             &              &       && &\\
Galaxy & Classification    & i$_{\rm disk}$     & PA$_{\rm disk}$ & &r$_{\rm bl}$ & PA$_{\rm bl}$& (b/a)$_{\rm bl}$ && r$_{\rm bar}$  & PA$_{\rm bar}$ & (b/a)$_{\rm bar}$ && Distance\tablefootmark{a} & Surface brightness\tablefootmark{b} \\
       &                   &  (deg)             & (deg)           & & (arcsec)  & (deg)      & (deg)        & & (arcsec)    & (deg)        &        && (Mpc)  & (mag arcsec$^{-2}$)\\
       &                   &                    &                 & &           &            &              & &             &              &        && &\\
\hline
\endfirsthead
\caption*{Table 1: continued.}\\
\hline
       &                   &                    &                 & &           &            &              & &             &              &       && &\\
Galaxy & Classification    & i$_{\rm disk}$       & PA$_{\rm disk}$   & &r$_{\rm bl}$ & PA$_{\rm bl}$& (b/a)$_{\rm bl}$ && r$_{\rm bar}$  & PA$_{\rm bar}$ & (b/a)$_{\rm bar}$ && Distance\tablefootmark{a} & Surface brightness\tablefootmark{b} \\
       &                   &  (deg)             & (deg)           & & (arcsec)  & (deg)      & (deg)        & & (arcsec)    & (deg)        &        && (Mpc)  & (mag arcsec$^{-2}$)\\
       &                   &                    &                 & &           &            &              & &             &              &        && &\\
\hline
\endhead
\hline
\endfoot
IC 1067\tablefootmark{\ast}  & SB(r,bl)a$\underline{\rm b}$                                                   & 38.3 $\pm$ 2.1 & 117.4 $\pm$  2.0 & & 10.89 $\pm$ 0.14 & 143.2 $\pm$  0.4 & 0.542 $\pm$ 0.008 && 18.66 $\pm$ 0.17 & 151.1 $\pm$ 0.5 & 0.357 $\pm$ 0.003 && 23.3 & 20.4 \\
IC 2051                   & SB($\underline{\rm r}$s,bl)b                                                   & 53.2 $\pm$ 0.7 &  72.0 $\pm$  0.6 & & 11.97 $\pm$ 0.02 &  69.8 $\pm$  0.5 & 0.611 $\pm$ 0.009 && 18.21 $\pm$ 2.10 &   6.4 $\pm$ 3.2 & 0.530 $\pm$ 0.061 && 23.9 & 19.8 \\
NGC  613                  & SB($\underline{\rm r}$s,bl,nr)b                                                & 38.9 $\pm$ 2.0 & 125.9 $\pm$  4.1 & & 39.39 $\pm$ 0.40 & 125.1 $\pm$  0.2 & 0.481 $\pm$ 0.007 && 77.10 $\pm$ 4.03 & 125.6 $\pm$ 1.1 & 0.245 $\pm$ 0.013 && 25.1 & 20.2 \\
NGC  936\tablefootmark{\ast} & (L)SB$_a$($\underline{\rm r}$s,bl)0$^+$                                        & 42.4 $\pm$ 0.5 & 126.6 $\pm$  1.5 & & 22.62 $\pm$ 0.15 & 124.9 $\pm$  0.8 & 0.876 $\pm$ 0.007 && 36.70 $\pm$ 0.40 &  82.3 $\pm$ 0.6 & 0.528 $\pm$ 0.006 && 20.7 & 20.1 \\
NGC 1015\tablefootmark{\ast} & (R$^{\prime}$)SB(r,bl)0/a                                                      & 30.5 $\pm$ 1.0 &  12.5 $\pm$  4.4 & & 10.62 $\pm$ 0.09 &  77.6 $\pm$  8.1 & 0.951 $\pm$ 0.013 && 21.49 $\pm$ 0.24 & 101.4 $\pm$ 0.5 & 0.477 $\pm$ 0.005 && 33.1 & 21.0 \\
NGC 1022\tablefootmark{\ast} & (RL)SAB($\underline{\rm r}$s,bl,ns)$\underline{\rm 0}$/a                       & 21.6 $\pm$ 0.6 & 175.9 $\pm$  4.3 & & 11.99 $\pm$ 0.15 & 135.7 $\pm$  3.5 & 0.644 $\pm$ 0.027 && 16.93 $\pm$ 0.37 & 108.2 $\pm$ 0.3 & 0.602 $\pm$ 0.013 && 18.5 & 19.8 \\
NGC 1079                  & (R$\underline{\rm L}$)S$\underline{\rm A}$B$_a$($\underline{\rm r}$s,bl)0$^+$  & 52.3 $\pm$ 3.2 &  79.1 $\pm$  2.7 & & 19.20 $\pm$ 0.22 &  94.4 $\pm$  1.1 & 0.723 $\pm$ 0.028 && 33.40 $\pm$ 0.80 & 119.6 $\pm$ 3.8 & 0.500 $\pm$ 0.006 && 26.0 & 17.7 \\
NGC 1097                  & (R$^{\prime}$)SB(rs,bl,nr)ab pec                                               & 48.1 $\pm$ 0.4 & 126.5 $\pm$  2.6 & & 55.18 $\pm$ 0.81 & 143.1 $\pm$  0.3 & 0.571 $\pm$ 0.009 && 93.96 $\pm$ 0.58 & 141.1 $\pm$ 2.9 & 0.354 $\pm$ 0.002 && 20.0 & 20.4 \\
NGC 1201                  & SAB$_a$(r$^{\prime}$l,bl,nb)0$^0$                                              & 51.4 $\pm$ 0.2 &   4.9 $\pm$  1.4 & & 16.80 $\pm$ 0.46 &   9.4 $\pm$  0.7 & 0.701 $\pm$ 0.041 && 25.00 $\pm$ 0.00 &  15.0 $\pm$ 0.0 & 0.560 $\pm$ 0.000 && 20.2 & 16.9 \\
NGC 1291                  & (R)SAB(l,bl,nb)0$^+$                                                           &  7.1 $\pm$ 1.9 &  17.7 $\pm$ 36.6 & & 56.01 $\pm$ 0.83 & 168.9 $\pm$  2.2 & 0.892 $\pm$ 0.016 && 97.20 $\pm$ 1.04 & 165.7 $\pm$ 0.3 & 0.594 $\pm$ 0.006 &&  8.6 & 20.3 \\
NGC 1300                  & (R$^{\prime}$)SB(s,bl,nrl)b                                                    & 33.4 $\pm$ 0.0 & 106.1 $\pm$  0.0 & & 44.17 $\pm$ 0.33 &  98.4 $\pm$  0.4 & 0.526 $\pm$ 0.005 && 80.71 $\pm$ 1.87 &  99.7 $\pm$ 1.3 & 0.223 $\pm$ 0.005 && 18.0 & 21.5 \\
NGC 1302                  & (RL,RL)SAB($\underline{\rm r}$l,bl)0$^+$                                       & 15.7 $\pm$ 0.8 &  11.6 $\pm$ 14.2 & & 16.77 $\pm$ 0.19 &  15.9 $\pm$  3.7 & 0.833 $\pm$ 0.020 && 27.15 $\pm$ 0.36 & 172.2 $\pm$ 1.4 & 0.664 $\pm$ 0.009 && 20.0 & 20.0 \\
NGC 1326                  & (R$_1$)SAB$_a$(r,bl,nr)0$^+$                                                   & 37.2 $\pm$ 1.3 &  80.4 $\pm$  3.1 & & 19.16 $\pm$ 0.11 &  55.6 $\pm$  2.0 & 0.858 $\pm$ 0.009 && 28.66 $\pm$ 1.67 &  16.9 $\pm$ 0.9 & 0.583 $\pm$ 0.034 && 17.0 & 20.0 \\
NGC 1350                  & (R)SAB$_a$(r,bl)0/a                                                            & 58.2 $\pm$ 0.7 &   4.7 $\pm$  1.8 & & 32.83 $\pm$ 0.03 &   9.0 $\pm$  0.5 & 0.577 $\pm$ 0.017 && 57.41 $\pm$ 0.84 &  36.3 $\pm$ 0.2 & 0.422 $\pm$ 0.006 && 20.9 & 20.5 \\
NGC 1398                  & (R$^{\prime}$R)SB($\underline{\rm r}$s,bl)a                                    & 42.2 $\pm$ 0.3 &  93.4 $\pm$  0.7 & & 22.02 $\pm$ 0.03 &  90.2 $\pm$  1.5 & 0.886 $\pm$ 0.009 && 36.91 $\pm$ 0.00 &  10.2 $\pm$ 0.0 & ---\tablefootmark{c} && 21.0 & 19.5 \\
NGC 1440                  & (L)SB($\underline{\rm r}$s,bl)0$^0$                                            & 39.5 $\pm$ 0.2 &  23.1 $\pm$  1.4 & & 12.60 $\pm$ 0.02 &  41.8 $\pm$  0.3 & 0.778 $\pm$ 0.010 && 20.80 $\pm$ 0.80 &  52.1 $\pm$ 2.0 & 0.540 $\pm$ 0.208 && 18.2 & 17.1 \\
NGC 1452                  & (RL)SB($\underline{\rm r}$s,bl)0/a                                             & 53.3 $\pm$ 0.0 & 113.7 $\pm$  0.0 & & 14.70 $\pm$ 0.04 & 114.4 $\pm$  1.6 & 0.898 $\pm$ 0.025 && 27.58 $\pm$ 0.13 &  32.5 $\pm$ 0.5 & 0.523 $\pm$ 0.003 && 22.8 & 20.5 \\
NGC 1512                  & (R$\underline{\rm L}$)SB(r,bl,nr)a                                             & 42.7 $\pm$ 1.1 &  74.5 $\pm$  1.9 & & 38.60 $\pm$ 0.62 &  51.6 $\pm$  0.7 & 0.654 $\pm$ 0.015 && 71.38 $\pm$ 1.07 &  42.2 $\pm$ 1.1 & 0.343 $\pm$ 0.005 && 12.3 & 21.0 \\
NGC 1533                  & (RL)SB(bl)0$^o$                                                                & 15.6 $\pm$ 1.7 & 127.0 $\pm$ 12.8 & & 14.71 $\pm$ 0.32 & 138.0 $\pm$  8.9 & 0.906 $\pm$ 0.020 && 26.85 $\pm$ 0.03 & 167.5 $\pm$ 0.5 & 0.631 $\pm$ 0.001 && 18.4 & 19.8 \\
NGC 1640                  & (R$^{\prime}$)SB$_a$(r,bl)$\underline{\rm a}$b                                 & 24.1 $\pm$ 0.0 &  42.8 $\pm$  6.9 & & 14.21 $\pm$ 0.11 &  38.5 $\pm$  1.0 & 0.739 $\pm$ 0.012 && 25.48 $\pm$ 2.40 &  44.0 $\pm$ 1.3 & 0.375 $\pm$ 0.035 && 19.1 & 20.3 \\
NGC 2273                  & (R)SAB(rs,bl,nb)a                                                              & 53.4 $\pm$ 0.2 &  62.0 $\pm$  1.4 & & 10.90 $\pm$ 0.32 &  74.9 $\pm$  1.9 & 0.821 $\pm$ 0.069 && 16.10 $\pm$ 0.70 & 115.5 $\pm$ 2.0 & 0.590 $\pm$ 0.026 && 28.4 & 17.1 \\
NGC 2293                  & SAB$_a$(bl)0/a                                                                 &  0.0 $\pm$ 0.2 &   0.0 $\pm$  1.4 & & 18.30 $\pm$ 1.32 & 108.2 $\pm$  2.6 & 0.804 $\pm$ 0.063 && 25.30 $\pm$ 0.10 & 134.4 $\pm$ 2.0 & 0.450 $\pm$ 0.002 && 32.0 & 17.4 \\
NGC 2543\tablefootmark{\ast} & SAB(s,bl)b                                                                     & 59.9 $\pm$ 1.5 &  40.6 $\pm$  7.1 & & 10.12 $\pm$ 0.06 &  30.3 $\pm$  1.9 & 0.658 $\pm$ 0.029 && 14.88 $\pm$ 1.93 & 105.5 $\pm$ 9.0 & 0.457 $\pm$ 0.059 && 33.6 & 20.4 \\
NGC 2787                  & (L)SB$_a$(r,bl)0$^o$                                                           & 56.2 $\pm$ 0.7 & 108.5 $\pm$  0.8 & & 21.29 $\pm$ 0.02 & 108.2 $\pm$  0.9 & 0.611 $\pm$ 0.012 && 29.81 $\pm$ 0.91 & 166.3 $\pm$ 4.3 & 0.646 $\pm$ 0.020 && 10.2 & 19.5 \\
NGC 2859\tablefootmark{\ast,e} & (R)SAB$_a$(rl,bl,nl,nb)0$^+$                                                 & 37.2 $\pm$ 1.6 &  81.8 $\pm$  1.4 & & 20.65 $\pm$ 0.05 & 150.0 $\pm$ 55.9 & 0.948 $\pm$ 0.008 && 34.40 $\pm$ 0.21 & 169.6 $\pm$ 1.1 & 0.625 $\pm$ 0.004 && 27.3 & 20.6 \\
NGC 2968\tablefootmark{\ast} & (L)SB(s,bl)0$^+$                                                               & 43.1 $\pm$ 0.7 &  67.6 $\pm$  1.0 & & 19.32 $\pm$ 0.62 &  69.7 $\pm$  6.6 & 0.839 $\pm$ 0.030 && 34.69 $\pm$ 0.32 &  38.6 $\pm$ 0.7 & 0.552 $\pm$ 0.005 && 13.9 & 20.6 \\
NGC 2983                  & (L)SB$_a$(s,bl)0$^+$                                                           & 53.6 $\pm$ 0.2 &  90.0 $\pm$  1.4 & & 11.90 $\pm$ 0.29 &  87.4 $\pm$  2.5 & 0.871 $\pm$ 0.031 && 19.30 $\pm$ 0.05 &  41.0 $\pm$ 2.0 & 0.540 $\pm$ 0.001 && 27.4 & 17.4 \\
NGC 3185\tablefootmark{\ast,d} & (RL)SAB$_{ax}$($\underline{\rm r}$s,bl)0/$\underline{\rm a}$                 & 49.5 $\pm$ 0.9 & 136.3 $\pm$  1.2 & & 17.46 $\pm$ 0.05 & 135.3 $\pm$  0.7 & 0.563 $\pm$ 0.011 && 32.42 $\pm$ 1.46 & 106.6 $\pm$ 2.5 & 0.389 $\pm$ 0.018 && 24.7 & 20.9 \\
NGC 3266\tablefootmark{\ast} & (R$\underline{\rm L}$)SB(bl)0$^o$                                              & 25.6 $\pm$ 6.6 &  91.9 $\pm$ 17.7 & &  6.94 $\pm$ 0.02 &  96.0 $\pm$  0.4 & 0.805 $\pm$ 0.017 && 11.17 $\pm$ 0.80 &   7.0 $\pm$ 3.4 & 0.737 $\pm$ 0.053 && 28.0 & 20.0 \\
NGC 3351\tablefootmark{\ast,d} & (R$^{\prime}$)SB(r,bl,nr)a                                                   & 45.0 $\pm$ 0.5 &   8.4 $\pm$  1.1 & & 26.33 $\pm$ 0.19 &   8.7 $\pm$  1.6 & 0.838 $\pm$ 0.010 && 51.43 $\pm$ 1.07 & 112.8 $\pm$ 0.2 & 0.541 $\pm$ 0.011 && 10.1 & 20.1 \\
NGC 3368\tablefootmark{\ast} & ($\underline{\rm R}$L)SAB($\underline{\rm r}$s,bl,nl)0$^+$                     & 48.7 $\pm$ 0.9 & 172.6 $\pm$  2.2 & & 35.08 $\pm$ 0.29 & 163.3 $\pm$  1.3 & 0.700 $\pm$ 0.028 && 55.57 $\pm$ 3.59 & 110.6 $\pm$ 8.1 & 0.574 $\pm$ 0.037 && 10.9 & 19.6 \\
NGC 3380\tablefootmark{\ast} & (R$\underline{\rm L}$)SAB($\underline{\rm r}$s,bl)0/a                          & 20.8 $\pm$ 7.0 &  47.4 $\pm$ 27.2 & &  7.84 $\pm$ 0.12 &  30.0 $\pm$  8.7 & 0.937 $\pm$ 0.018 && 17.99 $\pm$ 0.85 &  18.9 $\pm$ 0.4 & 0.420 $\pm$ 0.020 && 24.3 & 20.2 \\
NGC 3384\tablefootmark{\ast} & (L)SA$\underline{\rm B}$(bl)0$^-$                                              & 60.8 $\pm$ 0.2 &  61.8 $\pm$  0.6 & & 13.42 $\pm$ 0.07 &  42.9 $\pm$  0.3 & 0.689 $\pm$ 0.033 && 22.46 $\pm$ 0.00 & 131.3 $\pm$ 0.0 & ---\tablefootmark{c} && 11.6 & 18.5 \\
NGC 3489\tablefootmark{\ast} & (R)SA$\underline{\rm B}$(r,bl)0$^o$:                                           & 60.1 $\pm$ 0.2 &  69.1 $\pm$  0.1 & & 12.97 $\pm$ 0.03 &  74.0 $\pm$  0.6 & 0.510 $\pm$ 0.006 && 11.69 $\pm$ 2.08 &  14.0 $\pm$ 2.7 & 0.709 $\pm$ 0.126 &&  9.6 & 18.6 \\
NGC 3637\tablefootmark{\ast} & (RL)SB$_a$(r$\underline{\rm l}$,bl)0$^{o+}$                                    & 27.0 $\pm$ 0.9 & 138.1 $\pm$  4.0 & &  9.82 $\pm$ 0.10 & 138.4 $\pm$  2.7 & 0.836 $\pm$ 0.027 && 15.70 $\pm$ 0.46 &  38.1 $\pm$ 1.2 & 0.667 $\pm$ 0.019 && 28.2 & 19.7 \\
NGC 3892                  & (RL)SAB(rl,bl)0$^{o+}$                                                         & 16.5 $\pm$ 4.0 &  99.7 $\pm$ 16.2 & & 13.35 $\pm$ 0.09 & 132.2 $\pm$  3.4 & 0.878 $\pm$ 0.011 && 38.86 $\pm$ 2.45 &  95.3 $\pm$ 0.3 & 0.531 $\pm$ 0.033 && 27.2 & 20.0 \\
NGC 3941\tablefootmark{\ast} & (R)SB$_a$(bl)0$^o$                                                             & 50.8 $\pm$ 0.2 &   8.5 $\pm$  0.3 & & 17.87 $\pm$ 0.18 &   1.9 $\pm$  0.7 & 0.567 $\pm$ 0.007 && 24.18 $\pm$ 1.53 & 171.9 $\pm$ 1.5 & 0.548 $\pm$ 0.035 && 14.2 & 19.0 \\
NGC 3945\tablefootmark{\ast} & (R)SB$_a$(rl,nl,bl)0$^+$                                                       & 47.9 $\pm$ 0.2 & 148.0 $\pm$  1.4 & & 23.30 $\pm$ 0.11 & 148.9 $\pm$  1.4 & 0.924 $\pm$ 0.014 && 35.70 $\pm$ 1.20 &  72.1 $\pm$ 2.0 & 0.700 $\pm$ 0.024 && 22.5 & 17.8 \\
NGC 3953\tablefootmark{\ast} & SB(r,bl)b                                                                      & 58.4 $\pm$ 0.1 &  14.0 $\pm$  0.4 & & 16.91 $\pm$ 0.05 &  16.3 $\pm$  0.6 & 0.658 $\pm$ 0.005 && 30.11 $\pm$ 1.39 &  47.6 $\pm$ 0.1 & 0.466 $\pm$ 0.022 && 18.4 & 19.5 \\
NGC 3992\tablefootmark{\ast,e} & SB(rs,bl,nb)ab                                                               & 55.1 $\pm$ 1.8 &  70.1 $\pm$  3.4 & & 29.78 $\pm$ 0.41 &  73.0 $\pm$  1.5 & 0.561 $\pm$ 0.020 && 55.65 $\pm$ 0.82 &  35.1 $\pm$ 1.7 & 0.383 $\pm$ 0.006 && 24.9 & 20.5 \\
NGC 4143\tablefootmark{\ast,e} & (L,R$^{\prime}$L)SAB$_a$(s,nb,bl)0$^-$                                       & 43.8 $\pm$ 0.2 & 148.7 $\pm$  1.4 & & 10.50 $\pm$ 0.43 & 152.0 $\pm$  2.3 & 0.714 $\pm$ 0.036 && 19.10 $\pm$ 0.70 & 155.5 $\pm$ 2.0 & 0.580 $\pm$ 0.021 && 17.0 & 16.0 \\
NGC 4245\tablefootmark{\ast,d} & (RL)SB(r,bl,n$\underline{\rm r}$l)0$^+$                                      & 33.3 $\pm$ 0.9 &   0.2 $\pm$  2.5 & & 21.86 $\pm$ 0.13 & 145.0 $\pm$  0.5 & 0.678 $\pm$ 0.008 && 36.33 $\pm$ 0.59 & 130.9 $\pm$ 4.9 & 0.485 $\pm$ 0.008 &&  9.7 & 20.5 \\
NGC 4262\tablefootmark{\ast} & (L)SB$_a$(l,bl)0$^{-o}$                                                        & 24.5 $\pm$ 1.6 & 158.1 $\pm$  4.3 & &  8.85 $\pm$ 0.06 & 143.4 $\pm$  1.8 & 0.741 $\pm$ 0.033 && 13.40 $\pm$ 0.14 &  26.5 $\pm$ 3.4 & 0.695 $\pm$ 0.007 && 20.5 & 19.2 \\
NGC 4314\tablefootmark{\ast,d} & (R$_1^{\prime}$)SB(r$\underline{\rm l}$,bl,nr)a                              & 20.4 $\pm$ 0.0 &  51.6 $\pm$  4.3 & & 35.10 $\pm$ 0.39 & 145.6 $\pm$  0.5 & 0.642 $\pm$ 0.008 && 72.58 $\pm$ 2.33 & 146.8 $\pm$ 0.0 & 0.360 $\pm$ 0.012 &&  9.7 & 20.7 \\
NGC 4340\tablefootmark{\ast,e} & SB$_a$(r,nr,nb,bl)0$^+$                                                      & 27.1 $\pm$ 0.2 &  98.1 $\pm$  1.4 & & 23.40 $\pm$ 1.29 & 107.8 $\pm$  0.3 & 0.621 $\pm$ 0.039 && 37.90 $\pm$ 0.20 &  37.6 $\pm$ 2.0 & 0.600 $\pm$ 0.003 && 16.8 & 18.2 \\
NGC 4371\tablefootmark{\ast} & (L)SB$_a$(r,bl,nr)0$^{o+}$                                                     & 59.0 $\pm$ 0.8 &  88.1 $\pm$  0.5 & & 25.30 $\pm$ 0.08 &  88.1 $\pm$  0.4 & 0.714 $\pm$ 0.010 && 36.59 $\pm$ 1.40 & 168.1 $\pm$ 6.2 & 0.746 $\pm$ 0.028 && 16.8 & 20.4 \\
NGC 4394\tablefootmark{\ast,d} & ($\underline{\rm R}$L)SB(rs,bl,nl)0/a                                        & 30.4 $\pm$ 0.0 & 109.8 $\pm$  3.7 & & 20.09 $\pm$ 0.31 & 140.4 $\pm$  0.4 & 0.729 $\pm$ 0.014 && 41.41 $\pm$ 2.15 & 143.4 $\pm$ 0.6 & 0.427 $\pm$ 0.022 && 16.8 & 20.5 \\
NGC 4448\tablefootmark{\ast} & (R)SB(r,bl)0/$\underline{\rm a}$                                               & 71.2 $\pm$ 0.4 &  92.0 $\pm$  0.2 & & 10.23 $\pm$ 0.06 &  93.7 $\pm$  0.6 & 0.691 $\pm$ 0.021 && 13.77 $\pm$ 0.00 & 172.0 $\pm$ 0.0 & ---\tablefootmark{c} && 26.5 & 18.9 \\
NGC 4454\tablefootmark{\ast} & (RL)SAB(r,bl)0/a                                                               & 17.6 $\pm$ 4.2 &  25.9 $\pm$ 25.2 & & 12.43 $\pm$ 0.09 &  42.3 $\pm$  6.4 & 0.941 $\pm$ 0.011 && 22.08 $\pm$ 1.26 &  18.3 $\pm$ 2.6 & 0.487 $\pm$ 0.021 && 35.7 & 20.6 \\
NGC 4548\tablefootmark{\ast} & SB(rs,bl)$\underline{\rm a}$b                                                  & 39.0 $\pm$ 0.3 & 149.2 $\pm$  0.6 & & 27.28 $\pm$ 0.14 & 116.1 $\pm$  5.0 & 0.868 $\pm$ 0.013 && 59.73 $\pm$ 0.88 &  61.6 $\pm$ 0.1 & 0.477 $\pm$ 0.007 && 16.2 & 20.6 \\
NGC 4579\tablefootmark{\ast} & ($\underline{\rm R}$L,R$^{\prime}$)SB(rs,bl)a                                  & 41.6 $\pm$ 0.3 &  92.1 $\pm$  0.8 & & 25.38 $\pm$ 0.16 &  72.1 $\pm$  1.7 & 0.703 $\pm$ 0.011 && 40.73 $\pm$ 0.75 &  53.6 $\pm$ 2.8 & 0.520 $\pm$ 0.010 && 19.6 & 19.5 \\
NGC 4593                  & (R$^{\prime}$)SB(rs,bl,AGN)a                                                   & 33.4 $\pm$ 1.2 & 104.4 $\pm$  4.0 & & 23.03 $\pm$ 0.22 &  71.6 $\pm$  0.2 & 0.734 $\pm$ 0.011 && 49.32 $\pm$ 1.59 &  54.4 $\pm$ 1.5 & 0.398 $\pm$ 0.013 && 33.9 & 20.6 \\
NGC 4596\tablefootmark{\ast} & (L)SB(rs,bl)$\underline{\rm 0}$/a                                              & 35.5 $\pm$ 1.8 & 119.2 $\pm$  3.0 & & 25.07 $\pm$ 0.30 &  89.8 $\pm$  1.2 & 0.902 $\pm$ 0.014 && 53.64 $\pm$ 0.04 &  74.6 $\pm$ 0.3 & 0.474 $\pm$ 0.000 && 16.8 & 20.4 \\
NGC 4608\tablefootmark{\ast} & SB(r,bl)0$^+$                                                                  & 30.2 $\pm$ 0.2 & 103.4 $\pm$  1.4 & & 22.20 $\pm$ 1.30 &  49.2 $\pm$  6.3 & 0.884 $\pm$ 0.068 && 43.80 $\pm$ 0.00 &  25.8 $\pm$ 2.0 & 0.490 $\pm$ 0.000 && 16.8 & 18.3 \\
NGC 4639\tablefootmark{\ast} & (R$^{\prime}$)SA$\underline{\rm B}$(rs,bl)ab                                   & 49.2 $\pm$ 0.0 & 134.5 $\pm$  0.0 & & 13.88 $\pm$ 0.05 & 133.8 $\pm$  1.3 & 0.630 $\pm$ 0.024 && 25.56 $\pm$ 0.16 & 173.0 $\pm$ 2.0 & 0.444 $\pm$ 0.003 && 22.4 & 20.1 \\
NGC 4643\tablefootmark{\ast} & (L)SB($\underline{\rm r}$s,bl,nl)0$^{o+}$                                      & 36.8 $\pm$ 0.8 &  57.2 $\pm$  2.5 & & 21.68 $\pm$ 0.30 & 109.4 $\pm$  7.6 & 0.948 $\pm$ 0.014 && 49.91 $\pm$ 1.18 & 133.3 $\pm$ 0.5 & 0.531 $\pm$ 0.013 && 25.7 & 20.1 \\
NGC 4659\tablefootmark{\ast} & ($\underline{\rm R}$L)SAB(l,bl)0$^o$                                           & 43.5 $\pm$ 1.1 & 173.5 $\pm$  1.4 & &  6.48 $\pm$ 0.04 &  29.0 $\pm$  0.5 & 0.642 $\pm$ 0.025 &&  8.43 $\pm$ 0.52 &  89.4 $\pm$ 8.2 & 0.834 $\pm$ 0.051 &&  7.8 & 19.4 \\
NGC 4754\tablefootmark{\ast} & (L)SB$_a$(bl)0$^{-o}$                                                          & 59.7 $\pm$ 0.4 &  22.4 $\pm$  0.6 & & 19.27 $\pm$ 0.03 &  14.8 $\pm$  1.0 & 0.712 $\pm$ 0.010 && 26.58 $\pm$ 1.75 & 141.4 $\pm$ 8.9 & 0.772 $\pm$ 0.051 && 17.3 & 19.7 \\
NGC 4795\tablefootmark{\ast} & (R$^{\prime}$)SA$\underline{\rm B}_a$(l,bl)a pec                               & 43.7 $\pm$ 0.8 & 114.8 $\pm$  1.8 & &  9.12 $\pm$ 0.05 & 135.1 $\pm$  0.3 & 0.669 $\pm$ 0.035 && 13.63 $\pm$ 1.89 &  29.7 $\pm$ 2.2 & 0.704 $\pm$ 0.098 && 40.4 & 19.8 \\
NGC 4902                  & SB($\underline{\rm r}$s,bl)a$\underline{\rm b}$                                & 21.5 $\pm$ 4.6 & 101.4 $\pm$ 11.9 & & 12.49 $\pm$ 0.27 &  83.5 $\pm$  1.2 & 0.782 $\pm$ 0.023 && 23.94 $\pm$ 1.16 &  64.4 $\pm$ 2.0 & 0.415 $\pm$ 0.020 && 39.2 & 19.8 \\
NGC 4984                  & (R$^{\prime}$R)SAB$_a$(l,bl,nl)0/a                                             & 53.9 $\pm$ 0.4 &  24.6 $\pm$  1.9 & & 19.57 $\pm$ 0.05 &  21.2 $\pm$  2.3 & 0.766 $\pm$ 0.013 && 30.04 $\pm$ 0.74 &  94.1 $\pm$ 0.9 & 0.702 $\pm$ 0.017 && 21.3 & 20.0 \\
NGC 5026                  & (L)SB(rs,nl,bl)a                                                               & 54.5 $\pm$ 0.2 &  60.4 $\pm$  1.4 & & 19.10 $\pm$ 0.23 &  37.6 $\pm$  0.8 & 0.737 $\pm$ 0.016 && 22.20 $\pm$ 1.30 & 170.5 $\pm$ 2.0 & 0.620 $\pm$ 0.036 && 46.3 & 17.5 \\
NGC 5101                  & (R$_1$R$_2^{\prime}$)SB($\underline{\rm r}$s,bl)0/a                            & 22.0 $\pm$ 3.7 & 124.9 $\pm$ 47.8 & & 27.12 $\pm$ 0.11 & 124.7 $\pm$  0.8 & 0.791 $\pm$ 0.007 && 50.12 $\pm$ 0.80 & 122.5 $\pm$ 0.3 & 0.471 $\pm$ 0.008 && 27.4 & 20.2 \\
NGC 5134                  & (R)SAB(rs,bl)a                                                                 & 14.7 $\pm$ 5.0 &  79.0 $\pm$ 34.0 & & 22.16 $\pm$ 0.19 & 150.9 $\pm$  0.2 & 0.672 $\pm$ 0.012 && 45.42 $\pm$ 0.52 & 153.3 $\pm$ 0.7 & 0.385 $\pm$ 0.004 && 10.9 & 20.3 \\
NGC 5337\tablefootmark{\ast} & SB(rs,bl)0/a                                                                   & 62.2 $\pm$ 0.8 &  21.9 $\pm$  0.6 & &  5.20 $\pm$ 0.04 &  35.9 $\pm$  0.3 & 0.529 $\pm$ 0.006 &&  9.12 $\pm$ 0.33 &  34.0 $\pm$ 1.1 & 0.353 $\pm$ 0.013 && 52.2 & 19.4 \\
NGC 5339\tablefootmark{\ast} & SA$\underline{\rm B}$(rs,bl)ab                                                 & 38.8 $\pm$ 1.7 &  47.9 $\pm$  4.8 & & 13.30 $\pm$ 0.05 &  73.4 $\pm$  0.7 & 0.758 $\pm$ 0.005 && 22.19 $\pm$ 0.72 &  86.3 $\pm$ 2.2 & 0.294 $\pm$ 0.010 && 38.8 & 21.0 \\
NGC 5347\tablefootmark{\ast} & SB(rs,bl)a                                                                     & 22.8 $\pm$ 8.6 &  89.4 $\pm$ 35.7 & & 13.86 $\pm$ 0.44 & 106.2 $\pm$  1.7 & 0.741 $\pm$ 0.033 && 31.21 $\pm$ 0.45 &  97.4 $\pm$ 1.7 & 0.411 $\pm$ 0.006 && 27.3 & 21.0 \\
NGC 5375\tablefootmark{\ast} & (R$^{\prime}$)SB$_a$(rs,bl)$\underline{\rm a}$b                                & 29.8 $\pm$ 4.4 & 159.3 $\pm$ 15.9 & & 17.04 $\pm$ 0.31 & 171.0 $\pm$  0.4 & 0.617 $\pm$ 0.013 && 27.23 $\pm$ 0.05 & 171.1 $\pm$ 1.6 & 0.421 $\pm$ 0.001 && 41.9 & 20.9 \\
NGC 5701\tablefootmark{\ast} & (R$_1^{\prime}$)SA$\underline{\rm B}$(rl,bl)0/a                                & 15.2 $\pm$ 3.8 &  50.2 $\pm$ 25.1 & & 20.29 $\pm$ 0.16 &   2.1 $\pm$  0.4 & 0.925 $\pm$ 0.009 && 39.04 $\pm$ 0.37 & 174.9 $\pm$ 1.2 & 0.580 $\pm$ 0.006 && 26.1 & 20.6 \\
NGC 5728                  & (R$_1$)SB($\underline{\rm r}^{\prime}$l,bl,nr,nb)0/a                           & 43.0 $\pm$ 2.3 &   6.7 $\pm$  5.7 & & 23.33 $\pm$ 0.04 &  16.3 $\pm$  1.0 & 0.661 $\pm$ 0.004 && 57.61 $\pm$ 0.38 &  34.3 $\pm$ 0.4 & 0.301 $\pm$ 0.002 && 30.6 & 20.5 \\
NGC 5750\tablefootmark{\ast} & (RL)SAB(r$^{\prime}$l,$\underline{\rm r}$s,bl)0/a                              & 60.2 $\pm$ 0.6 &  62.9 $\pm$  2.1 & & 13.16 $\pm$ 0.08 &  52.2 $\pm$  0.8 & 0.642 $\pm$ 0.014 && 21.82 $\pm$ 0.35 & 109.9 $\pm$ 1.1 & 0.596 $\pm$ 0.010 && 33.6 & 20.0 \\
NGC 5770\tablefootmark{\ast} & SAB($\underline{\rm r}$l,bl)0$^+$                                              & 22.4 $\pm$ 2.0 &  38.4 $\pm$ 10.4 & & 10.90 $\pm$ 0.10 &  79.7 $\pm$  1.3 & 0.970 $\pm$ 0.011 && 22.63 $\pm$ 0.33 & 113.2 $\pm$ 1.5 & 0.566 $\pm$ 0.008 && 22.4 & 20.5 \\
NGC 5850\tablefootmark{\ast,e} & (R$^{\prime}$)SB(r,bl,nr,nb)$\underline{\rm a}$b                             & 36.0 $\pm$ 0.9 & 169.4 $\pm$  2.7 & & 37.04 $\pm$ 0.35 & 121.3 $\pm$  0.5 & 0.655 $\pm$ 0.011 && 62.71 $\pm$ 0.02 & 113.7 $\pm$ 1.5 & 0.379 $\pm$ 0.000 && 23.1 & 21.5 \\
NGC 5957\tablefootmark{\ast} & (R$^{\prime}$)SA$\underline{\rm B}$(rs,bl)$\underline{\rm a}$b                 & 25.4 $\pm$ 4.8 &   2.8 $\pm$ 15.7 & & 12.86 $\pm$ 0.05 &  93.1 $\pm$  1.3 & 0.841 $\pm$ 0.011 && 23.02 $\pm$ 0.98 &  93.9 $\pm$ 1.0 & 0.457 $\pm$ 0.019 && 27.6 & 20.9 \\
NGC 6014\tablefootmark{\ast,d} & SAB(rs,bl)$\underline{\rm 0}$/a                                              & 35.3 $\pm$ 1.2 & 150.6 $\pm$  3.4 & &  9.86 $\pm$ 0.09 &   6.9 $\pm$  0.5 & 0.519 $\pm$ 0.024 && 11.71 $\pm$ 0.46 &  30.2 $\pm$ 0.2 & 0.699 $\pm$ 0.027 && 36.6 & 20.3 \\
NGC 6782                  & (R)SAB(rl,nr$^{\prime}$,nb,bl)0$^+$                                            & 22.0 $\pm$ 0.2 &  20.0 $\pm$  1.4 & & 14.20 $\pm$ 0.39 &   6.1 $\pm$  0.9 & 0.803 $\pm$ 0.029 && 25.10 $\pm$ 0.50 & 178.9 $\pm$ 2.0 & 0.480 $\pm$ 0.010 && 52.6 & 17.8 \\
NGC 7079                  & (L)SA$\underline{\rm B}_a$(s,bl)0$^o$:                                         & 51.3 $\pm$ 2.9 &  75.5 $\pm$  3.4 & &  9.27 $\pm$ 0.06 & 119.7 $\pm$  1.5 & 0.764 $\pm$ 0.026 && 17.21 $\pm$ 0.76 &  46.7 $\pm$ 3.5 & 0.626 $\pm$ 0.028 && 31.8 & 19.5 \\
NGC 7421                  & ($\underline{\rm R}^{\prime}$L)SB(rs,bl)ab                                     & 27.9 $\pm$ 3.5 &  64.6 $\pm$ 19.1 & &  8.99 $\pm$ 0.12 &  97.7 $\pm$  1.9 & 0.779 $\pm$ 0.017 && 21.84 $\pm$ 0.06 &  89.6 $\pm$ 1.7 & 0.344 $\pm$ 0.001 && 23.0 & 20.5 \\
NGC 7552                  & (R$_1^{\prime}$)SB(r$\underline{\rm s}$,bl,nr)a                                & 15.8 $\pm$ 1.3 &  38.5 $\pm$ 16.2 & & 23.13 $\pm$ 0.25 &  96.7 $\pm$  0.2 & 0.648 $\pm$ 0.010 && 52.25 $\pm$ 0.89 &  89.8 $\pm$ 6.2 & 0.374 $\pm$ 0.006 && 17.1 & 19.7 \\  
\end{longtable}
\tablefoot{
The morphological classifications are taken from \citet{buta2015} and \citet{lauri2011}.
The disks and bars parameters are taken from the corresponding source, \citet{lauri2011} or
\citet{salo2015} or \citet{herrera2015} (see text). 
\tablefoottext{\ast}{Galaxy in our color subsample, i.e., with available 
SDSS imaging in {\it u, g, r, i,} and {\it z} bands in \citet{knapen2014}.}
\tablefoottext{a}{The distances are taken from the respective survey data (see text).
These are 
redshift independent distances from NED when available 
and obtained assuming $H_0$ = 71 km s$^{-1}$ Mpc$^{-1}$ otherwise. }
\tablefoottext{b}{Surface brightness of the isophote used to measure the barlens properties.}
\tablefoottext{c}{The value of (b/a)$_{\rm bar}$ is not given in the literature 
because no clear ellipticity maximum was found for this bar.}
\tablefoottext{d}{Galaxy with a star forming nuclear ring observed in the 
{\it (g-r)} color profile (see Appendix B).}
\tablefoottext{e}{Galaxy with a nuclear bar observed in the color profiles 
(see Appendix B).}}
}
\end{landscape}

\end{appendix}

\end{document}